\documentclass{ieeeaccess} 
\usepackage{cite}
\usepackage{amsmath,amssymb,amsfonts}
\usepackage{algorithmic}
\usepackage{graphicx}
\usepackage{textcomp}
\usepackage[dvipsnames,table]{xcolor}
\usepackage{comment}

\usepackage{csquotes}
\usepackage{xpatch}
\usepackage{colortbl}

\newcommand{\tableheadline}{\textbf}

\usepackage{bm}
\makeatletter
\AtBeginDocument{\DeclareMathVersion{bold}
	\SetSymbolFont{operators}{bold}{T1}{times}{b}{n}
	\SetSymbolFont{NewLetters}{bold}{T1}{times}{b}{it}
	\SetMathAlphabet{\mathrm}{bold}{T1}{times}{b}{n}
	\SetMathAlphabet{\mathit}{bold}{T1}{times}{b}{it}
	\SetMathAlphabet{\mathbf}{bold}{T1}{times}{b}{n}
	\SetMathAlphabet{\mathtt}{bold}{OT1}{pcr}{b}{n}
	\SetSymbolFont{symbols}{bold}{OMS}{cmsy}{b}{n}
	\renewcommand\boldmath{\@nomath\boldmath\mathversion{bold}}}
\makeatother

\def\BibTeX{{\rm B\kern-.05em{\sc i\kern-.025em b}\kern-.08em
    T\kern-.1667em\lower.7ex\hbox{E}\kern-.125emX}}

\newcommand{\hlcolor}{black}

\begin{document}
	%\history{Date of publication xxxx 00, 0000, date of current version xxxx 00, 0000.}
	\doi{}

	\title{{Resilience-by-Design in 6G Networks: Literature Review and Novel Enabling Concepts}}
	\author{\uppercase{Ladan Khaloopour}\authorrefmark{1}, 
		\uppercase{Yanpeng Su}\authorrefmark{2}, %\IEEEmembership{Graduate Student Member, IEEE},
		\uppercase{Florian Raskob}\authorrefmark{1}, 
		\uppercase{Tobias Meuser}\authorrefmark{1}, 
		\uppercase{Roland Bless}\authorrefmark{3}, %\IEEEmembership{Member, IEEE},
		\uppercase{Leon \textcolor{\hlcolor}{Janzen}}\authorrefmark{1}, % Würsching
		\uppercase{Kamyar Abedi}\authorrefmark{3}, 
		\uppercase{Marko Andjelkovic}\authorrefmark{4}, 
		\uppercase{Hekma Chaari}\authorrefmark{2},
		\uppercase{Pousali Chakraborty}\authorrefmark{5}, 
		\uppercase{Michael Kreutzer}\authorrefmark{6}, 
		\uppercase{Matthias Hollick}\authorrefmark{1},
		\uppercase{Thorsten Strufe}\authorrefmark{3}, 
		\uppercase{Norman Franchi}\authorrefmark{2}, 
		\uppercase{Vahid Jamali}\authorrefmark{1} %\IEEEmembership{Senior Member, IEEE}
	}
	
	\address[1]{Technical University of Darmstadt (TUDa)}
	\address[2]{Friedrich-Alexander University Erlangen-Nürnberg (FAU)}
	\address[3]{Karlsruhe Institute of Technology (KIT)}
	\address[4]{Leibniz Institute for High Performance Microelectronics (IHP)}%{Leibniz-Institut für innovative Mikroelektronik (IHP)}
	\address[5]{Fraunhofer Institute for Open Communication Systems (FOKUS)}
	\address[6]{Fraunhofer Institute for Secure Information Technology (SIT)}

	\tfootnote{The authors acknowledge the financial support by the German \textit{Federal Ministry for Education and Research (BMBF)} within the project ``Open6GHub'' (Grant Numbers: 16KISK005, 16KISK006, 16KISK010, 16KISK014) and by the LOEWE initiative (Hesse, Germany) within the emergenCITY centre (Grant Number: LOEWE/1/12/519/03/05.001(0016)/72).}
	
	\markboth
	{Khaloopour \headeretal}
	{Khaloopour \headeretal}
	
	\corresp{Corresponding author: Ladan Khaloopour (e-mail: ladan.khaloopour@tu-darmstadt.de).}

\begin{abstract} 
	The sixth generation (6G) mobile communication networks are expected to intelligently integrate into various aspects of modern digital society, including smart cities, homes, health-care, transportation, and factories. While offering a multitude of services, it is likely that societies become increasingly reliant on 6G infrastructure. Any disruption to these digital services, whether due to human or technical failures, natural disasters, or terrorism, would significantly impact citizens’ daily lives. Hence, 6G networks need not only to provide high-performance services but also to be resilient in maintaining essential services in the face of potentially unknown challenges. This paper \textcolor{\hlcolor}{provides a general review of the state of the art on resilient systems, definitions, concepts, and approaches. Moreover, it introduces a comprehensive concept, i.e., resilience-by-design (RBD), in three different levels} for designing resilient 6G communication networks, summarizing our initial studies within the German Open6GHub project. \textcolor{\hlcolor}{First, we outline the general RBD enabling principles and discuss their related sub-categories. Next, adopting an interdisciplinary approach, we propose to embed these principles across all 6G layers/perspectives including} electronics, physical channel, network components and functions, networks, services, and cross-layer and cross-infrastructure considerations \textcolor{\hlcolor}{and discuss their challenges. We further elaborate the RBD principles and their realizations along with several 6G use-cases. The paper is concluded by presenting a comprehensive list of open problems for future research on 6G resilience.}
\end{abstract}


	\titlepgskip=-21pt
	
	\maketitle

\section{Introduction}
\label{sec:introduction}

As the fifth generation (5G) of mobile communication networks are currently being deployed across the world, academia and industry have started to develop a vision for the sixth generation (6G) of communication networks \cite{wang2023road}. To identify 6G application scenarios, key requirements, and enabling technologies, various countries have launched large-scale 6G research projects  including European Union (e.g., projects Hexa-X \cite{Hexa-X}, RISE-6G  \cite{RISE-6G}, REINDEER \cite{REINDEER},  6GStart \cite{6GStart}, 6GTandem \cite{6GTandem}, TERA6G \cite{TERA6G}, etc.), Germany (e.g., projects Open6GHub \cite{Open6GHub}, 6G-RIC \cite{6G-RIC}, 6G-life \cite{6G-life}, 6GEM \cite{6GEM}, and 6G-ANNA \cite{6G-ANNA}), Finland (project 6Genesis  \cite{Oulu}),  USA (e.g., program RINGS  \cite{RINGS}), China (e.g., 6G satellite communication  \cite{China_Project2019}), Japan (e.g., 6G satellite communications  \cite{Space_Japan}), Korea (e.g., 6G quantum cryptographic communication \cite{KT6G}).  
\begin{figure*}[t]
  \centering
  \includegraphics[width=1.0\linewidth]{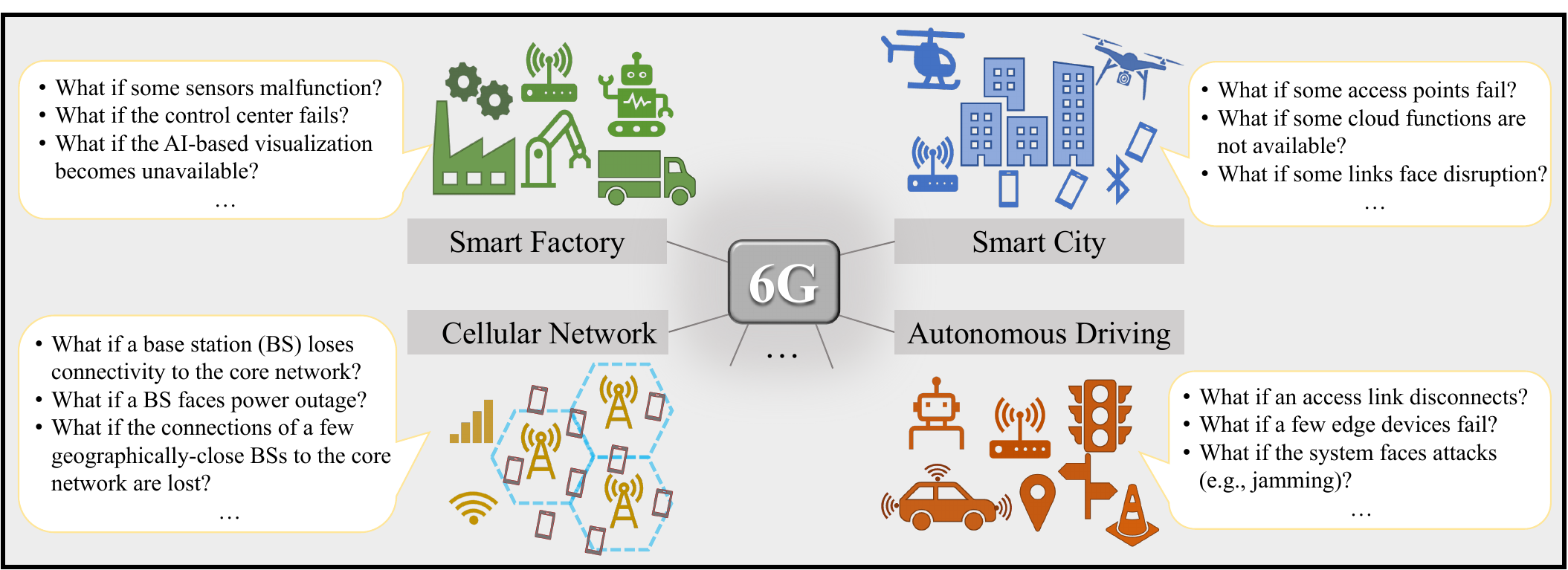}
  \caption{A resilient 6G must not only provide reliable and high-quality services under the normal system operation, but also be able to maintain a minimum service requirement when facing challenges. This figure illustrates four 6G application scenarios, namely smart factory, smart city, autonomous driving, and cellular communication network, and correspondingly a few potential challenges in each scenario.}
  \label{fig: 6G_examples}
\end{figure*}
Recent studies including  \cite{wang2023road,de2021survey,alsabah20216g} have reviewed 6G developments. They envision that 6G will deeply integrate communications, sensing, and computing with enabling technologies such as virtualization, artificial intelligence (AI) and machine learning (ML), THz and visible-light communications, quantum communications, and reconfigurable intelligent surfaces. It will facilitate an intelligent Internet-of-Everything (IoE), offering infrastructures to secure information and communication technology into everyday life. Consequently, this will shape a new, ``smart'' infrastructure of the modern digital society (for instance for smart cities, smart homes, smart healthcare, smart transportation, smart factories, smart grids, smart agriculture, smart radio, etc.).

Our societies will enjoy the multitude of 6G services, and at the same time become increasingly dependent on the communication infrastructure. Any disruption of the provided digital services (due to, e.g., human or technical failures, natural disasters, attacks, or sabotage) will have a significant negative impact on the daily life of the citizens.  Therefore, besides key performance indicators (KPIs) such as data rate, latency, energy efficiency, and so on, the 6G infrastructure is expected to be inherently “resilient”, i.e., capable of providing an acceptable level of service in the face of challenges such as failures and malicious attacks. Some examples of these challenges are provided in Fig.~\ref{fig: 6G_examples}.

\textcolor{\hlcolor}{In this paper, we provide a general review of the relevant previous works on resilient systems and their definitions, concepts, and approaches. Then, we present a comprehensive concept, i.e., resilience-by-design (RBD), for the design of resilient 6G communication networks, which summarizes our initial interdisciplinary studies on this topic within the German Open6GHub project. In fact, we present resilience enabling concepts for 6G networks in three different levels: 
        \\
        \textbf{1) Holistic concepts for resilience:} We outline three main enabling principles of resilience, namely P1) protective design measures, P2) self-awareness capabilities, and P3) reconfiguration capabilities. Moreover, we classify P1--P3 for communication networks into multiple sub-categories with a comprehensive discussion.
        \\
        \textbf{2) 6G layers/perspectives:} Since 6G networks are still a ``standard on the drawing board'' and remain in development, their complexity and evolving architecture mean that the realization of holistic resilience principles will differ depending on the network layer and specific design perspective. Therefore, we present various realizations of these resilience principles for different 6G layers/perspectives including 1) electronics, 2) physical channel, 3) network components and functions, 4) networks, and 5) services perspective. For each layer/perspective, the corresponding challenges and resilience requirements are discussed in detail. Furthermore, we discuss the resilience  cross-layer and cross-infrastructures considerations, aiming to inform the standardization process and ensure resilience is embedded in 6G from the outset.
        \\
        \textbf{3) 6G use-cases:} Finally, since each use-case may demand specific resilience provisions, we further explain the general principles and their realizations in different network layers for several emerging 6G use-cases. These include cloud-based distributed monitoring network, autonomous driving, a production line in a highly automated factory, and virtual and augmented reality.}

In the remainder of this section, we first discuss preliminary concepts, definitions, and approaches for resilience in communication networks. Subsequently, we introduce the proposed resilience-by-design concept for 6G and present the key contributions of this paper.  Furthermore, a detailed discussion on the state-of-the-art research on resilience in communication networks and an overview of the relevant papers are provided at the end of this section.

\subsection{Background on resilient communication networks}

Communication networks may encounter various challenges or failures, including natural disasters, technical issues, and malicious attacks. These challenges may impact physical components, resulting in physical disruptions, or alter the content and quantity of data packets, leading to cyber disruptions \cite{cassottana2023resilience}. Therefore, network components must feature resilience to the most relevant challenges with a large impact. 
Section~\ref{sec:2-resilience} provides an overview and background on resilient communication networks, summarizing the existing definitions of resilience, the proposed resilience-enabling mechanisms, and the distinction of resilience compared to similar terms,  such as availability, reliability, robustness, etc.

In this paper, a \emph{resilient system} is defined as one that is equipped to face and respond arbitrary challenges by resisting, absorbing, recovering, adapting, and learning. 
A detailed discussion on the definition of resilience is provided in Section~\ref{sec:2-resilience}.
Various terms in the literature share similarities with resilience, such as robustness, reliability, availability, fault tolerance, and security. While these terms may initially seem synonymous with resilience, there are significant differences. The primary difference is that the term ``resilience'' encompasses all types of challenges and failures, even those that are rare/unforeseen, and may maintain a minimum service level, in contrast to robust systems that offer normal operation in the face of a failure or fully fail \cite{gribble2001robustness}. 
Another difference is that a resilient system incorporates three crucial properties: \emph{protection}, \emph{detection}, and \emph{recovery} \cite{nistCSF}. 
A detailed exploration of the definitions and distinctions among related terms is provided in Section~\ref{ssec:resilience-differentiation}.

\subsection{Proposed Resilience-by-Design for 6G}

In this paper, we present a general framework for embedding resilience features into the design of 6G communication networks, referred to as \textit{resilient-by-design 6G}. Our approach involves considering different system components across various system layers/perspectives to maintain their functionalities. We identify possible challenges/failures within each layer. Additionally, we account for cross-layer and cross-infrastructure considerations to provide an end-to-end resilience for the overall system.

We propose a holistic framework that consists of three main enabling principles (P1--P3) to realize an RBD concept, known as protective design measures, self-awareness, and reconfiguration capabilities\footnote{The proposed enabling resilience principles are comprehensive but in line with protection, detection, and recovery capabilities of resilient systems proposed in e.g., \cite{nistCSF, smith2011network}. In Section \ref{sec:3-resiliencebydesign}, we discuss different functionalities within each principle and then provide realizations and insights to various relevant layers/perspectives.}. \textit{Protective design measures (P1)} refer to those design mechanisms that are decided during the early design stage of the network and ensure its protection against critical or frequent challenges as well as strategic adversaries. Examples include signing of signalling information, encryption of traffic to prevent potential leaks, isolation of the control plane from the data plane, the design of secure protocols, as well as single-point-of-failure (SPF) free architectures, and stateless designs. 
In addition to these protective design measures, resilient 6G communication networks should be able to identify potential challenges and act accordingly. The \textit{self-awareness (P2)} principle requires that the system should be aware of its state and identify potential challenges and failures. 
It includes local sensing through embedded sensors, 
global monitoring using network data (data acquisition), and utilizing this data for anomaly detection, prediction, and interpretation (i.e., data processing).
Finally, the third principle, \textit{reconfiguration capability (P3)}, means the system should adapt its operation based on its local state, network's global state, and application requirements. This requires embedding multiple operational modes in the design of resilient systems that can be quickly activated in matching failure scenarios to absorb the challenge. Finally, the system will recover its functionality to normal operation, and will evolve over time (i.e., long-term reconfiguration through learning). 

A key part of the proposed RBD concept is to embed RBD principles P1--P3 into all layers/perspectives of 6G communication systems. Therefore, we present the relevant layers/perspectives of 6G communication systems, namely electronics, physical channel, network components and functions, networks, services, and the cross-layer and cross-infrastructure considerations. Moreover, we explain the challenges within each layer and discuss how the proposed RBD principles can be applied to different layers/perspectives to mitigate these challenges.

The design of 6G communication systems following the RBD concept presents trade-offs. 
While using RBD principles can significantly enhance system resilience, they concurrently increase the design complexity and cost. Therefore, an efficient RBD design aims at meeting the required resilience requirements with minimum cost/complexity.  Moreover, various components and layers possess their own constraints \cite{maklachkova2022analysis}. 
For instance, in a cellular network, it may be preferable to add complexity/cost to the design of higher layers, such as core network while keeping mobile user equipment simple and affordable.
Here, we discuss the trade-off between resilience and complexity/cost, resource optimization for end-to-end resilience, and the double-edged sword of reconfigurability for resilience.

Finally, we note that while the design of resilient communication systems and the general requirements have been widely investigated in the literature, see Section~\ref{sec: Introduction_Review papers}, they have not been so far the focal point driving the \emph{end-to-end} system design. Moreover, the consideration of resilience measures is often separately and inhomogeneously treated across different layers. 
However, since the demand for the resilience of communication services grows steadily, 6G  must follow a comprehensive RBD design to realize the end-to-end resilience for critical services, which is the focus of this paper.  

\subsection{Related survey and review papers} \label{sec: Introduction_Review papers}

Resilience is a subject of study across various types of systems, including cyber-physical systems (CPS), IoT networks, vehicular networks, smart grids, optical backbone networks, etc.
In the following, first, we present a literature review on the resilience of these networks that is structured according to their focus topic. 
Then, general design strategies proposed in the literature for enabling the resilience of complex multi-layer systems are discussed. Finally, we conclude this section with a short review of emergency/backup networks for the post-disaster phase.

\textbf{Cyber-physical systems}: 
In \cite{pajic2017design}, security challenges in cyber-physical systems are investigated. Two distinct perspectives are studied from which control-specific CPS security challenges emerge, namely the \emph{physical perspective} (involving attacks on the physical components) and the \emph{control perspective} (concerning attacks on the internal network, including sensors, controllers, and actuators). The focus of \cite{pajic2017design} is on resilience algorithms designed to ensure the performance of control systems in the face of attacks.

Moreover, the work presented in \cite{yang2020adversary} studies Byzantine attacks in cyber-physical systems. This research concentrates on resilience strategies for decentralized and distributed statistical inference, including areas such as resilient distributed detection and distributed estimation to mitigate Byzantine attacks.

\begin{table*}
    \centering 
    \caption{A summary of papers closely related to this paper and a comparison of their research scope.} 
    \begin{tabular}{l|l@{\hspace{.85\tabcolsep}}l@{\hspace{.85\tabcolsep}}l@{\hspace{.85\tabcolsep}}l@{\hspace{.85\tabcolsep}}r@{\hspace{.85\tabcolsep}}l}
        \hline
        \tableheadline{Reference} &
        \tableheadline{Considered failures} &
        \tableheadline{Level/layer} &
        \tableheadline{Structure} &
        \tableheadline{Phase of failure} &
        \tableheadline{Year} &
        \tableheadline{Further notes} \\
        \hline
        \textcolor{\hlcolor}{\cite{kantarci2015resilient}}& \textcolor{\hlcolor}{failures related to SG} & \textcolor{\hlcolor}{two layers} & \textcolor{\hlcolor}{hybrid} & \textcolor{\hlcolor}{after failure} & \textcolor{\hlcolor}{2017} & \textcolor{\hlcolor}{low-latency information acquisition}\\
        \textcolor{\hlcolor}{\cite{sterbenz2017smart}}& \textcolor{\hlcolor}{physical attacks, disasters } & \textcolor{\hlcolor}{multi-level} & \textcolor{\hlcolor}{hybrid} & \textcolor{\hlcolor}{before \& after failure} & \textcolor{\hlcolor}{2017} & \textcolor{\hlcolor}{islands and corridors of resilience}\\
        \cite{hutchison2018architecture} & various challenges  &  multi-level &  decentralized & before, during, \& after failure& 2018 & $\text{D}^2\text{R}^2+\text{D}\text{R}$, routing \& connectivity\\
        \cite{pradhan2018chariot} & device/component failures & three-layer & decentralized & before \& after failure & 2018 & CHARIOT architecture\\
        \cite{benkhelifa2018critical} & security attacks & multi-layer & - & before failure & 2018 &  IDS for IoT technologies\\
        \cite[Ch. 28]{rak2020guide} & failed operator & system-level & centralized & after failure& 2020 & load/traffic redistribution\\
        \cite{khodaei2020scalable} & security attacks & - & decentralized & -& 2020 & resilient CLR distribution\\
        \textcolor{\hlcolor}{\cite{althobaiti2020cybersecurity}}& \textcolor{\hlcolor}{quantum attacks} & \textcolor{\hlcolor}{multi-layer} & \textcolor{\hlcolor}{-} & \textcolor{\hlcolor}{before failure} & \textcolor{\hlcolor}{2020} & \textcolor{\hlcolor}{lattice-based cryptography \& QKD}\\
        \textcolor{\hlcolor}{\cite{saki2021survey}}& \textcolor{\hlcolor}{quantum computers' threats} & \textcolor{\hlcolor}{-} & \textcolor{\hlcolor}{-} & \textcolor{\hlcolor}{before \& during failure} & \textcolor{\hlcolor}{2021} & \textcolor{\hlcolor}{security \& resilience techniques}\\
        \cite{banerjee2021introducing} & natural disasters&  connectivity&  decentralized& after failure& 2021 & SOS\\
        \cite{berger2021survey} & faults, attacks, and failures & different layers & decentralized & before, during, \& after failure & 2021 & IoT resilience mechanisms\\
        \cite{banerjee2022designing} & natural disasters & connectivity & hybrid & after failure & 2022 & SOS-hybrid\\
        \textcolor{\hlcolor}{\cite{hutchison2023importance}}& \textcolor{\hlcolor}{faults/failures, cyber attacks } & \textcolor{\hlcolor}{different layers} & \textcolor{\hlcolor}{-} & \textcolor{\hlcolor}{before, during, \& after failure} & \textcolor{\hlcolor}{2023} & \textcolor{\hlcolor}{networked systems, AI, ML \& SDN}\\
        \cite{cassottana2023resilience} & physical \& cyber disruptions & multi-level & centralized & before \& after failure& 2023 & resilience framework for CPS\\
        This paper & various types of failures  &  different layers &  decentralized & before, during, \& after failure & - & RBD framework for 6G networks\\
        \hline
    \end{tabular}
    \label{tab: comparison}
\end{table*}

\textbf{IoT networks}:
In \cite{berger2021survey}, a survey on the resilience of IoT applications is presented. It provides a classification of resilience and state-of-the-art resilience mechanisms applicable to IoT, and it discusses practical aspects of realizing resilience in this context.
Moreover, \cite{benkhelifa2018critical} discusses security aspects of the IoT network layers, the features of various layers (e.g., physical, link, network, and application layers) that facilitate attacks, and categorizes threats and security attacks based on their types and outcomes. Furthermore, it summarizes intrusion detection systems (IDSs) for IoT technologies.

A decentralized three-layer framework named ``CHARIOT'' was proposed in \cite{pradhan2018chariot} to address failures in devices, nodes, and components within IoT systems. This architecture, designed for autonomous IoT system management, presents a ``sense-plan-act'' loop, which covers failure detection, reconfiguration computation, and system reconfiguration. Additionally, \cite{pradhan2018chariot} includes a case study on smart parking, as a practical application of the proposed framework.

\textbf{Vehicular networks}: 
The authors in \cite{dietzel2010resilient} studied resilience in vehicular networks. Two investigated sources of failures are malicious users (attackers) and malfunctioning remote sensors that may generate untrustworthy or inaccurate information.
Additionally, in \cite{khodaei2020scalable}, privacy challenges and security issues within vehicular communication systems were studied. This paper proposed a framework that is resilient to pollution and denial of service (DoS) attacks, and also to the selfish users. The framework relies on distributing certificate revocation lists (CRLs) based on the trip duration and targeted region of vehicles.

\textbf{Smart grids}: 
The authors in \cite{stefanovic2017resilient} reviewed ``power talk'' in microgrids, which is a communication technique that offers resilient and secure operation against both types of passive and active attacks. 
\cite{liu2017resilient} discusses physical failures for smart grids (SG) and studies resilient (low latency) methods for information acquisition after a failure. It provides an algorithm to dynamically cluster the remaining sensors. 

\textbf{Optical backbone networks}:
A review of research studies on optical networks is presented in \cite{ye2015energy}, which considers the trade-offs between energy efficiency and resilience. This review examines the impact of three key factors on this trade-off, namely long-term traffic predictions, short-term traffic dynamics, and strategies to achieve the requirements of SLAs. In \cite{kantarci2015resilient}, the study focuses on the resilient design of a cloud-based network over an optical backbone. The proposed schemes aim to minimize outage probability and reduce network resource consumption.

\textbf{Multi-layer design}:
Given that communication networks are complex systems with a multi-level or multi-layer structure, it is crucial to ensure resilience at each layer, such as the internet protocol (IP) layer, optical layer, and optical transport network (OTN) layer, as outlined in \cite{ye2015energy}.
Moreover, it is worth noting that the resilience of each layer is essential for the resilience of the above levels and the overall resilience of the network, as highlighted in \cite{hutchison2018architecture}. For instance, the Internet's structure encompasses physical level, routing, autonomous system (AS)-level realm (i.e., policy, mechanism, or a trust boundary), 
and end-to-end connectivity \cite{sterbenz2017smart}. 
The focus of \cite{hutchison2018architecture}, in particular, focuses on the aspects of connectivity and routing, proposing two basic properties for a resilient network.
The first property is survivable connectivity, which requires the utilization of redundant and diverse links to prevent network partitioning in the event of failures. The second property is autonomous isolatability, which necessitates having sufficient local resources for isolated operation in the face of network partitioning.

\textbf{Post-disaster networks}:
Numerous studies in the existing literature have investigated high-impact and very complex failure scenarios that could potentially result in a complete system outage, such as power/system outages or catastrophic natural disasters with significant consequences \cite{rak2020guide, banerjee2022designing, banerjee2021introducing, cassottana2023resilience}. Some proposed solutions involve the use of backup networks, including unmanned aerial vehicle (UAV) networks 
\cite{deepak2019overview,  kaleem2019uav, naqvi2018drone, banerjee2022designing, hu2021building, miranda2016survey}, 
device-to-device (D2D) networks 
\cite{deepak2019overview, kaleem2019uav, naqvi2018drone, banerjee2022designing, banerjee2021introducing}, 
free-space backhaul links 
\cite{miranda2016survey, hu2021building, zhang2016self},  
etc. These backup networks are designed to ensure the provision of essential services until the original communication infrastructure is restored. They are often referred to as post-disaster, emergency, or ad-hoc networks. We refer the readers to \cite{deepak2019overview, tyson2014beyond, miranda2016survey, sakano2013disaster, kaleem2019uav, naqvi2018drone, hu2021building, zhang2016self, banerjee2022designing} for various emergency networks, their service objectives, and the associated research challenges.
A temporary decentralized network for the post-disaster period is proposed in \cite{banerjee2021introducing}. It was argued that people (citizens) can adapt to the disrupted situation during this period. The ``bottom-up'' approach proposed in \cite{banerjee2021introducing} employs self-organization for survival (SOS) using component roles specified based on the available battery charge in a distributed manner. In a follow-up work by the authors, a new design was proposed in \cite{banerjee2022designing} for a hybrid temporary network during the post-disaster period. This design combines two approaches, namely ``bottom-up'' (SOS temporary network) and ``top-down'' (utilizing external equipment such as UAVs, WiFi access points, and radio relays.).

Table \ref{tab: comparison} presents a list of selected papers on the resilience of communication networks that are closely related to this paper as well as their key focuses in comparison with this paper.

\section{Resilience} 
\label{sec:2-resilience}

The characterization of resilience as a property of the (communication) system is an important aspect of resilience research, which sets the objectives (resilience goals) for the design of resilient systems and guides us to develop strategies (resilience enablers) to achieve these objectives. In this section, we present a brief review of the resilience definitions and enablers proposed in the literature and discuss how the term resilience may be differentiated from other related terms such as availability, reliability, robustness, etc.   

\subsection{Resilience definitions}
\label{ssec:resilience-def}

The term ``resilience'' is described as the ``\emph{ability to recover from or adjust easily to misfortune or change}'' according to the Merriam-Webster dictionary. Looking into the scholarly literature, various studies have provided their definitions of resilience \cite{sterbenz2010resilience,thompson2016new,witti2018secure,vugrin2011resilience,avizienis2001fundamental,avizienis2004basic,brinkmeier09methods,bishop2011resilience,erdene2012new,khan2017trust,modarresi2017multilevel,laprie1985dependable,laprie2008dependability,delic2016resilience,sterbenz2010resilience,pradhan2018chariot,us2010dhs}.  
In some studies, resilience is defined as ``\emph{the ability of the network to provide and maintain an acceptable level of service in the face of various faults and challenges to normal operation}'' \cite{rak2020guide, sterbenz2010resilience, sterbenz2014redundancy}, or ``\emph{the ability of the network to provide and maintain an acceptable level of security service in case some nodes are compromised}'' \cite{chen09sensor}. 
Hence, it was suggested that future networks require service level agreements (SLAs) to specify resilience requirements and objectives based on the specific application considered  \cite{rak2020guide}.  
The authors in \cite{avizienis2004basic} defined resilience as a synonym for fault tolerance, self-repair, and self-healing. In \cite{thompson2016new}, resilience is defined from the viewpoint of reactive security: ``\emph{Resilience is achieved if and only if a security breach is detected, contained and resolved}''.   
The recent survey \cite{berger2021survey} studies 41 papers that define resilience in the context of the Internet-of-Things (IoT) and afterwards defines resilience as ``\emph{the property of preserving the dependability and security of a system when the system encounters changes, thus withstanding or recovering from impairments}''.
Moreover, they refer to all means working towards the resilience of a system as ``resilience mechanisms''. Furthermore, formal definitions of resilience have been provided by governmental entities, as well. For example, the United States Department of Homeland Security (US DHS) has defined resilience as the
``\emph{ability to resist, absorb, recover from or successfully adapt to adversity or a change in
conditions}'' \cite{us2010dhs}. The US Computer Security Resource Center (CSRC) gives the following definition for the resilience of information systems \cite{csrc2024resilience}: ``\emph{The ability of an information system to continue to: (i) operate under adverse conditions or stress, even if in a degraded or debilitated state, while maintaining essential operational capabilities; and (ii) recover to an effective operational posture in a time frame consistent with mission needs.}''
Last but not least, in the 2020 EU Strategic Foresight Report, resilience is defined in a more general sense as ``\emph{the ability not only to withstand and cope with challenges but also to undergo transitions, in a sustainable, fair, and democratic manner}''~\cite{EU2020resilience}.

Despite the various definitions of resilience in the literature, most can be assigned to one of the following two categories:   
(\emph{i}) \emph{Service-oriented} definitions which focus on a system's service and that a specific level of service is provided even when the system is faced with a threat/challenge    
\cite{sterbenz2010resilience, thompson2016new, witti2018secure, vugrin2011resilience, avizienis2001fundamental, avizienis2004basic, bishop2011resilience, erdene2012new, khan2017trust, modarresi2017multilevel, laprie1985dependable, laprie2008dependability}.     
This includes temporary service degradation but requires that the system finally provides the expected service.
In this context, some definitions characterize resilience with the trust that can be put on systems to achieve and maintain the expected service \cite{laprie1985dependable, laprie2008dependability, avizienis2001fundamental, avizienis2004basic};
and (\emph{ii})~\emph{Capability-based} definitions which describe the abilities of a resilient system to maintain normal operation, e.g., preparing for a challenge, withstanding it, and recovering from it back to normal operation \cite{delic2016resilience, sterbenz2010resilience, pradhan2018chariot, us2010dhs}. 

In the following, we provide a definition for resilience that is inline and/or based on the definitions reviewed above and is used throughout this paper. Here, we adopt the term \emph{challenge} as an abstraction from the terms used in related works to describe the circumstances a system faces, e.g., threats, crises, disturbances, failures, faults, incidents, or adversarial attacks.
This convention yields a general resilience definition that applies to all types of systems in communication networks from different perspectives.

\textbf{Resilience:} 
\emph{A resilient system is prepared to face challenges, withstand them, and prevent most from causing performance degradation. It can also absorb the impact of significant challenges, ensuring essential functionalities or a minimum service level. Moreover, it can recover (i.e., short-term coping, bouncing back), adapt (long-term coping, bouncing forward), and evolve based on the experiences learned during this process.}

In the next subsection, we address these questions: 
How to realize a resilient system and what are the resilience enablers?

\subsection{Resilience enablers}
\label{ssec:resilience-enablers}

A resilient system generally requires specific properties to achieve resilience objectives, as discussed in previous studies \cite{sterbenz2010resilience, hutchison2018architecture}. These properties include the ability to overcome different challenges, monitor both its status and surroundings, protect against repetitive challenges, predict potential ones, and take proactive measures. Moreover, it should detect challenges in the early stages, absorb, recover, and learn from past failures, and evolve over time \cite{pradhan2018chariot, berger2021survey, sterbenz2010resilience, sterbenz2014redundancy, rak2020guide, hutchison2018architecture, cassottana2023resilience}. The design and implementation of a resilient system is a critical and challenging issue. In the following, we review the enabling capabilities mentioned in the previous works to realize a resilient system.

A common resilient strategy, described in \cite{sterbenz2010resilience, sterbenz2014redundancy, rak2020guide, hutchison2018architecture}, is known as \emph{$\text{D}^2\text{R}^2+\text{D}\text{R}$}. This strategy consists of two loops: an inner loop and an outer loop. The inner loop involves ``defend, detect, remediate, and recover'' and enables rapid adaptation from severely and partially degraded states to normal operation. In contrast, the outer loop, which includes ``diagnose and refine'', facilitates long-term evolution. 
A resilient system should be able to avoid the failures, manage the failures and its operation \cite{pradhan2018chariot}. Thus, detection and diagnose, identification of the appropriate reconfiguration plan, and system reconfiguration are the capabilities proposed in the ``CHARIOT'' method to realize a resilient system \cite{pradhan2018chariot}.

The authors in \cite{berger2021survey} specified enabling means to implement a resilient system, for example, means to discover and fix the faults, prevent their occurrence, protect the service, forecast the faults, and estimate their consequences.
They also studied different resilience mechanisms for IoT systems to provide a practical view and classified them into four main categories: (\emph{i}) \emph{redundancy mechanisms}, such as data redundancy, redundant network links, and auto-scaling to compensate for or absorb failures; (\emph{ii}) \emph{monitoring mechanisms}, including general surveillance, anomaly detection, and intrusion detection systems to collect, store, aggregate data, and trigger alerts when necessary; (\emph{iii}) \emph{protection mechanisms}, such as encryption, verification, authentication, and sensor fusion, to shield and protect the system from external and malicious harms; and (\emph{iv}) \emph{recovery mechanisms}, such as rollback, rollforward, and checkpointing, to restore the system to its functional state. The recovered state may represent a new operational state. The mentioned mechanisms can be combined and used in different layers of IoT systems architecture \cite{berger2021survey}.

In \cite{cassottana2023resilience}, the models and methods for resilience in cyber-physical systems are reviewed and a framework based on three steps is proposed. The first one is the \emph{system description}, where data and information about the system, its components, interdependencies, and processes are collected to identify the components and define their performance measures over time.
In the second step, \emph{disruption scenario}, data and information about possible challenges/failures are gathered or created. In the third step, \emph{resilience strategy}, the resilience strategies are applied to prevent or mitigate challenges and restore system performance. Proactive strategies aim to prevent challenges before they occur, but reactive strategies aim to restore system performance after a challenge and mitigate losses. Some of the most regular strategies are mentioned which are hardening components, redundant components, component restoration, early warning systems, intrusion tolerance systems, and authentication.

In some literature \cite{nistCSF, smith2011network}, the properties/capabilities of the resilient systems are classified into three main categories, {protection}, {detection}, and {recovery}. 
In this context, \textit{protection} denotes the intrinsic capability of the system to proactively resist challenges, while \textit{detection} and \textit{recovery} serve as reactive measures, with detection ensuring that challenges are quickly identified and recovery guaranteeing adaptive responses to restore the system.

Our goal is to combine the mentioned resilience enablers into the design of 6G communication networks. \textcolor{\hlcolor}{The challenge is the network’s heterogeneity and complexity, involving diverse systems, interconnections, interdependencies, and different layers/perspectives.} Consequently, we conducted a thorough investigation of three primary enabling principles, as described in Section \ref{sec:3-resiliencebydesign}. 
\textcolor{\hlcolor}{We focus on different layers of 6G networks and present corresponding realizations for each enabling principle within each layer/perspective. Moreover, we discuss the cross-layer and cross-infrastructure considerations.}

\subsection{Differentiation with other similar terms}
\label{ssec:resilience-differentiation}

While resilience is meant to enable the system to remain functional when it faces a challenge, there are other terms 
in the literature that aim at capturing similar capability.
An overview of the different dimensions of resilience and their relations is provided in \cite{sterbenz2010resilience}.
Some terms, e.g., reliability, availability, robustness, security, etc., 
seem to be interchangeable to resilience at first glance; however, there are important differences present.  
While resilience combines these classical terms, it goes beyond them by requiring the ability to account for unforeseen causes of performance degradation and long-term improvement.
In the following, we discuss a few important terms 
that are closely related to resilience and highlight the differences between these terms and resilience.

\begin{itemize}
    \item\textbf{Availability:}
    It captures the readiness for service/usage \cite{maklachkova2022analysis, sterbenz2010resilience, berger2021survey}. 
    So, availability is related to the duration for which the system components remain operational, and it is a part of resilience. Availability traditionally is the prime objective in robustness, and an important objective of information security; two fields that aim to achieve availability by various means.

    \item\textbf{Integrity:}
    It ensures that services and data are not modified without proper authorization 
    \cite{landwehr2001computer}.
    Integrity at its core refers to the ability to detect if network traffic has been modified. 
    Integrity of signalling information between intermediate devices is vital, as adversaries otherwise maybe able to interfere with the proper operation of the network (cf. secure interdomain routing (RFC 6480), DNS security (RFC 9364) and others).  
    Considering service integrity, resilience acknowledges that disruptions may occur and prepares the system to withstand, recover from, or adapt to them and provide its services potentially at a reduced level.

    \item\textbf{Reliability:}
    It often defines continuity of service \cite{sterbenz2010resilience}, and refers to an object's ability to sustain its capacity over time, ensuring the achievement of necessary functions within specified modes and conditions \cite{maklachkova2022analysis, avizienis2004basic}.  
    Reliability is often used with respect to commonly existing perturbations which are typically accounted for in the design stage (such as noise, interference, fading, etc.). Examples are reliable communication strategies over noisy channel with fading. In this context, reliability is generally measured over a short time horizon and does not consider unforeseen events, significantly differentiating it from resilience.

    \item\textbf{Robustness:}
    Robustness is used with different meanings across different disciplines. In some literature, robustness is used very broadly, making together with security, the key component of resilient systems \cite{berger2021survey, hutchison2018architecture}. 
    In most literature, however, robustness is rather employed to describe the ability of the design to cope with perturbations that are not accounted for \cite{sterbenz2010resilience}, e.g., a design based on perfect channel state information (CSI) can be evaluated with respect to its robustness to CSI estimation error. Here, the communication strategies can be developed based on the robust-by-design approach, e.g., the potential CSI estimation error is modeled and accounted for in the design stage. In this context, the term robust design is often used rather than resilient design.
    A robust system continues ``normal'' operation in the face of a challenge (or fully fails) \cite{gribble2001robustness}, while a resilient system is capable of addressing even rare challenges and it is prepared to switch to a different operational mode and offer a minimum level of service.

    \item\textbf{Fault-tolerance:}
    It is defined as the ability of a system to tolerate faults such that service failures do not result.
    Although self-awareness can also exist in a fault-tolerant system, resilience systems provide full-scale monitoring in the sense that both hardware and software faults as well as security attacks are considered. Thus, the main difference to a fault-tolerant system is the degree to which causes of performance degradation can be detected.
    Self-adaptation/configuration is also possible in standard fault-tolerant systems \cite{cassottana2023resilience}, but not all fault-tolerant system support adaptation. As mentioned, a resilient system provides a cross-layer adaptation, and reconfiguration at multiple abstraction levels (e.g., it could be possible to reconfigure the processor, either entirely or partially, to switch on or off different sensors). 

    \item\textbf{Confidentiality:} 
    Beyond availability and integrity, information technology (IT) security additionally requires confidentiality of data. This requires that only authorized parties may get access to information that is stored, transmitted, or processed within the network. 
    An obvious need for confidential communication is the exchange of keying material between intermediate systems, which is required to achieve integrity as described above.
    Confidentiality of signalling information can also prevent adversaries from learning structural information and defense strategies, which would put them in the position of adapting and improving their attacks on availability.
\item\textbf{Accountability and controlled access}: It is imperative to verify the identity of communication peers, and to establish the ability to identify the entity responsible for a specific event, such as the use of service.
Access to services and data has to be restricted to authorized entities.
Strategic adversaries may otherwise impersonate entities with privileged permissions and interfere with the expected operation of the networking infrastructure.
    
\end{itemize}

In conclusion, resilience emerges as a comprehensive term, unifying related concepts and directing attention not only towards the most probable failures but also covering rare and unforeseen challenges \cite{maklachkova2022analysis, sterbenz2010resilience}.

%%%%%%%%%%%%%%%%%%%%%%%%%%%%%%%%%%%%%%%%%%%%%%%
\section{Resilience-by-Design Concepts for 6G}
\label{sec:3-resiliencebydesign}
%%%%%%%%%%%%%%%%%%%%%%%%%%%%%%%%%%%%%%%%%%%%%%%
In Section \ref{sec:2-resilience}, we reviewed the definitions of resilience and provide our definition. The resilience enablers were summarized from previous works, and similar terms related to resilience were discussed. Nevertheless, the specific realization of a resilient 6G communication network remained unexplored.
To achieve resilience, some requirements have to be considered, appropriate measures have to be undertaken, and corresponding hardware/software resources need to be implemented. These actions contribute to an increase in system costs and complexity. Thus, designing resilient systems is based on trade-offs between different design objectives. 
This section provides a comprehensive exploration of the design phase for resilient communication systems, with a particular focus on those tailored for the upcoming generation of mobile communication networks, i.e., 6G. First, we introduce the concept of ``resilience-by-design'', exploring the relevant requirements and functionalities. We then explore different layers and perspectives within this framework. Finally, we discuss the resilience performance vs. cost/complexity trade-offs in achieving RBD.

%%%%%%%%%%%%%%%%%%%%%%%%%%%%%%%%%%%%
\subsection{Proposed RBD definition}
 
Section~\ref{ssec:resilience-def} outlined the main properties of a resilient system, namely resistance, absorption, and recovery. 
Achieving these features requires the integration of a specific set of resilience capabilities into the systems design. This comprehensive design approach is referred to as \emph{RBD concept} in this paper.

First, we establish the concept of ``end-to-end resilience'', which is a core part of RBD. A system is called end-to-end resilient if it maintains functionality despite challenges occurring at various parts within the system and across different layers. It covers the entire life-cycle of a system, from design to operation and evolution, with the ultimate goal of ensuring uninterrupted delivery of intended services or outputs even in the face of challenges.
The concept of designing for resilience means that in every phase of the system design, the resilience aspects should be considered together with other standard design requirements (e.g., performance, power consumption). Thus, we define the term resilience-by-design as follows.

\noindent
\textbf{Resilience-by-design:} \textit{A system design follows an RBD concept if resilience capabilities/features at different system layers/perspectives have been integrated into the system, enabling it to cope with the potential challenges and provide end-to-end resilience.}

The key terms in the above definition are \textit{resilience capabilities/features}, \textit{system layers/perspectives}, and \textit{potential challenges}. Obviously, the identification of potential challenges that the system may face is central to the RBD concept, which has not been a key focal point in conventional design approaches. However, the communication system is quite complex consisting of various interconnected and interdependent components. Therefore, it is susceptible to various cascading failures in addition to the single points of failure, and thus, identifying all failure states for the entire system is not a scalable approach, if not infeasible at all. Furthermore, policies and solution approaches required for realizing resilience capabilities/features across various system layers (electronics, physical channel, network components and functions, networks, and services) are often quite diverse and their development demands domain-specific knowledge. In order to provide ``end-to-end'' resilience, i.e., across all system layers, we will study the principles of RBD for different layers in Section~\ref{subsec:RBD_perspectives}.

%%%%%%%%%%%%%%%%%%%%%%%%%%%%%%%%%%%%%%%%%%%
\subsection{Relevant capabilities/features} 
\label{subsec:relevant-capabilities-features}

To realize the RBD concept, a set of resilience capabilities/features are required that include protection, detection, prediction, interpretation,  absorption, countermeasure, recovery, adaptation, learning,  sustainability, etc. \cite{sterbenz2010resilience}. In the following, we introduce three holistic enabling principles of RBD which include the aforementioned resilience capabilities/features as special cases.

\textbf{Holistic RBD enabling principles:} Before presenting 6G resilience enablers at different layers and/or from different perspectives, we introduce three holistic enabling principles P1--P3. These principles are based on the resilience enablers: protection, detection, and recovery discussed in Section~\ref{ssec:resilience-enablers} and expand them.

\textbf{P1 -- Protective design measures:} 
A resilient system should incorporate some mechanisms to \textit{inherently} ensure its protection against various types of challenges. This protection should be continuous, even during the system's \textit{normal} operation. Such protective measures may be considered at different layers of the system design. For instance, they include strict authentication and isolation of any signalling information,  choices of system topology, the network management strategy, protocol design, and provisions down to the choice of a specific component. Protective measures are specifically designed for application to challenges of high frequency or significant importance.  Four main categories of these challenges are malicious security attacks, single points of failure (SPF), perturbations, and complex state management.
\begin{itemize} 
      \item \textbf{Isolation and authentication:} RBD implies remote configuration, reconfiguration, and collaboration between distributed components. Leveraging machine-learned models (`AI')---which would most likely be off-loaded to \emph{virtualized} resources---is increasingly suggested, at many different levels. This in turn implies exposure of interfaces for remote access. Impersonating a controller, or a seemingly cooperating peer is one of the most common attacks in legacy network protocols.
      The prime security objective of resilient systems hence has to be strict authentication of communicating parties and data origin, as well as strict isolation of any management, potentially also user plane traffic, against arbitrary interference (forgery, denial of service).
      \item \textbf{SPF-free architectures:} Single point of failures are components within a system, the failure of which can lead to the failure of the entire system. Often, centralized systems may encounter single points of failure (which pose a threat to \emph{availability}) whereas distributed systems are less susceptible to crashes if they avoid full reliance on individual components and implement some local adaptation. Redundancy hence helps achieve a minimum level of protection as part of the system design.   
      \item \textbf{Perturbation-resistant components:} Components have different degrees of resistance to perturbations. For instance, misordered protocol messages or bits within a protocol message, significant variations in temperature, and other perturbations in general may negatively impact the performance of the affected component (which pose a threat to \emph{robustness}). Here, implementing integrity schemes and investing in high-quality components that inherently are able to withstand large temperature variations is beneficial from the resilience perspective. However, this generally implies a higher cost and hence its applicability to most system components is limited.
      \item \textbf{Statelessness and soft states:} Stateless designs should be used wherever possible to avoid complex state management such as state 
      synchronization and state replication. Stateless entities can be easily
      replaced by others in case they fail.  If state management is required, 
      soft-state concepts should be used as they are more robust due to their 
      self-healing property: soft-states get deleted automatically unless they
      are refreshed by corresponding events or messages, and they get created
      automatically when needed.
\end{itemize}
Since the above protection measures concern the normal operation of the system, they increase the design cost and complexity and may even yield a reduction in the overall performance of the system. Security measures frequently incur a cost of performance, due to inherent computation and communication overheads. For instance, while a distributed algorithm implemented on several low-capacity computing edge nodes is not subjected to a single point of failure, it may have lower performance  compared to a centralized algorithm implemented on a high-capacity computing cloud server. Therefore, there is a trade-off between cost/complexity and performance w.r.t. resilience, which is discussed in detail in Section~\ref{subsec:RBD_tradeoffs}. Furthermore, this trade-off motivates the design of strategies that are not used as a normal operational mode of the system, but only come into play when the system faces a challenge. To this end, we introduce two additional enabling principles for RBD.

 \textbf{P2 -- Self-awareness capability:} Sensing/monitoring mechanisms are crucial for the awareness of the system state, identifying potential failure states, and deciding about the proper operational modes. Self-awareness capability of a RBD System is divided in two part which are \emph{data acquisition} and \emph{data processing}.
 The sensing information can be employed for anomaly detection, interpretation, and/or prediction. The following resilient capabilities/features related to self-awareness include: 
 \begin{itemize} 
      \item \textbf{Local sensing:} By embedding physical sensors, relevant information about the local system state, environmental/operating parameters  (e.g., temperature, supply voltage, and clock frequency), or hardware faults (e.g., soft error, aging, and electromagnetic interference sensors) can be collected \cite{rai2007temperature, kuzubasoglu2020flexible, catterall2010ion, rajachandrasekar2012monitoring, amouri2011, anik2020, rahmanikia2017}. 
      \item \textbf{Global monitoring:} In addition to the local information collected by sensors, monitoring of the global network data can be useful for the identification of failures in the network and decision on the appropriate counter-action. This data includes, for example, software errors (e.g., critical tasks deadline monitors, application constraints monitors, and operative systems monitors) and communication status (e.g., data transmission and jamming monitors) \cite{keedy1979structuring,xu2019indoor,cai2017proactive, hu2017proactive}.
      \item \textbf{Anomaly detection:} 
      Using the sensing/monitoring data, the current system state/anomalies/failures can be detected. Here, one may distinguish model-based anomaly detection \cite{housh2018model, zhang2013time}, 
      which is applicable when the system behavior lends itself to a mathematical characterization, and data-based anomaly detection \cite{uwagbole2017applied, silva2014prbs},  
      which is relevant for more complex network dynamics. 
      \item \textbf{Anomaly prediction/anticipation:} When sufficient sensing/monitoring data is available, it is possible to learn the dynamic behavior of the system, which can then be used for the prediction of future system state/failures \cite{cassottana2023resilience, uwagbole2017applied, zhan2015predicting}.  
      Although,  in principle, both model- and data-based models can be adopted for prediction, cloud-based or AI-based processing methods, such as neural networks, are often more suitable for processing a large database of collected data.
      \item \textbf{Interpretation:} 
      The ability of the system to infer the position, extent, impact, etc. of the failures is referred to as the interpretation capability \cite{lundberg2006resilience}. This feature is crucial for planning efficient system response as discussed in the following.
 \end{itemize}

\begin{figure}[t]
  \centering
  \includegraphics[width = 1.0 \linewidth]{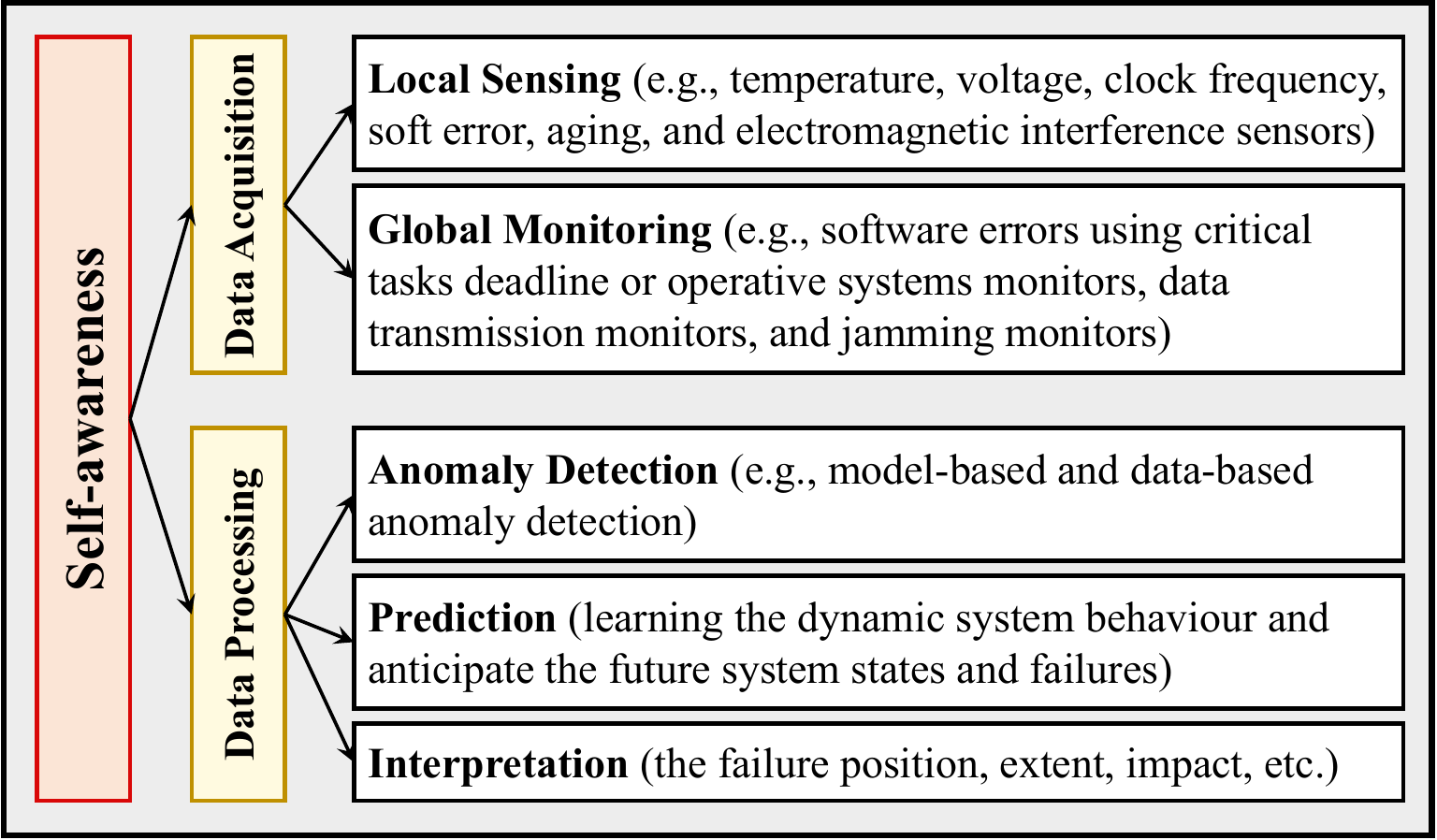}
  \caption{Self-awareness capability of RBD systems.}
  \label{fig: self-awareness}
\end{figure}

 The different features related to self-awareness capability are illustrated in Fig.~\ref{fig: self-awareness}.

\begin{figure*}[t]
  \centering
  \includegraphics[width=1.0\linewidth] {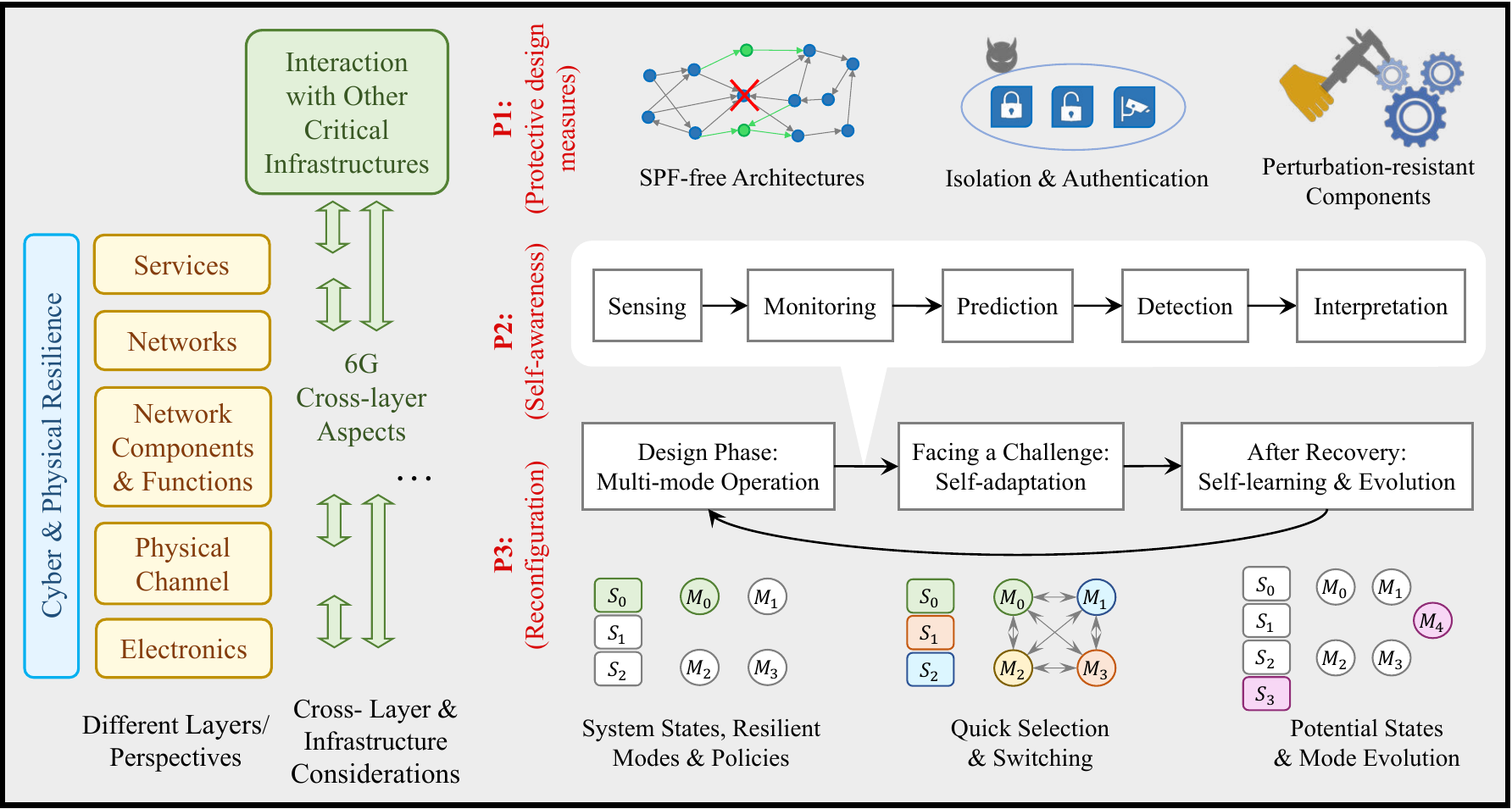}
  \caption{Proposed RBD concept for 6G illustrating (from left to right) cyber vs. physical resilience, resilience for relevant different layers, cross-layer and cross-infrastructure considerations, and key principle features of 6G resilience, namely protective design measures, self-awareness and reconfiguration. The local system states $S_i$ and the available operational modes $M_j$ are shown. In the normal state, $S_0$, the system operates on the normal mode $M_0$. During a challenge, the system state is changed to $S_1$ or $S_2$, where the system selects a new operational mode, i.e., $M_1$, $M_2$, or $M_3$. It can also switch among modes to recover from the challenge. The system will learn potential system states, such as $S_3$, using the knowledge obtained during the challenge, and it will evolve by developing a new operational mode, denoted as $M_4$.}
  \label{fig: Resilience by design}
\end{figure*}

 \textbf{P3 -- Reconfiguration capability:} Three steps are required for full reconfiguration functionality in the RBD concept:
  \begin{itemize} 

        \item \textbf{Design phase -- Resilience operational modes:} After the occurrence of a challenge, the communication system may no longer be able to maintain a minimum level of service using its normal operating mode even if the challenge is identified. Therefore, resilient operational modes have to be developed and embedded in the system during the design phase (i.e., RBD concept) such that they can be activated once a challenge state is detected and identified, supplementing the conventional mode designed for normal scenarios. Among straightforward options are redundant components.  However, such approach is not always feasible and cost-efficient, and hence given the rare occurrence of some challenges, cannot be afforded in most systems. 
        Therefore, policies should be developed for re-purposing the existing system components when facing a challenge. For instance, a specific communication hardware component (e.g., a phased array of the transmitter) may be designed for a given nominal range of temperatures. This phased array is by design resilient to abnormal (e.g., high) temperatures by not only deploying a thermometer sensor for the detection of temperature fluctuations (P2 capabilities) but also with new beamforming policies that adapt themselves to the change in the phase-shifting characteristics at high temperatures (P3 capabilities).
        \item \textbf{Facing a challenge -- Self-adaptation:} After predicting a challenging state through the analysis of sensing or monitoring data (i.e., P2 capabilities), the system must promptly initiate proactive measures by adjusting its operational mode. This adjustment aims to either prevent the predicted challenge if avoidable or mitigate its potential effects if unavoidable. In the latter scenario, where the challenge is inevitable, the system needs to promptly detect its occurrence. Subsequently, the system has to quickly decide to which of the available resilience operational modes it has to switch to in order to achieve an efficient absorption of the challenge, and eventually, recovery to a functional operation. Self-adaptation determines the system's functionality to dynamically transition among the available operational modes given the evolving system state. 
        \item \textbf{After recovery -- Self-learning:} Depending on the considered layer, failure events can be rare, especially for e.g., higher layers of the communication system. The knowledge gained after the recovery from a failure can be exploited to learn potential failure states and efficient adaptation decisions leading to a constant improvement of resilience features.
 \end{itemize}
The following resilient functions/attributes require capabilities P1--P3 for their realization:	
 \begin{itemize}
      \item Absorption (requires both quick detection of a challenge and quick transition to a suitable resilient mode)
      \item Recovery (requires continuous monitoring of the system states after absorption and transition to intermediate resilient modes until the system is fully restored)
      \item Prevention (efficient handling of challenge states also implies that compared to non-resilient systems, the larger cascading failures are prevented in large complex systems under the RBD approach)
      \item Sustainability (intelligent transition among the available operational modes can be employed to ensure long-term sustainability of the system)
 \end{itemize}

The capabilities/features mentioned for RBD concept are summarized in Fig.~\ref{fig: Resilience by design}. It is worth noting that considering interdependencies is crucial when dealing with complex systems. The aim is to establish a comprehensive RBD methodology for every layer/component to adhere to. In Section \ref{sec:4-usecases}, we present several specific 6G use-cases that facilitate a more detailed elaboration of the overall RBD concept.

%%%%%%%%%%%%%%%%%%%%%%%%%%%%%%%%%%%%%%%%%%
\subsection{Different layers/perspectives}
\label{subsec:RBD_perspectives}
As discussed in the previous section, RBD offers a holistic approach to resilience that aims at embedding resilience into all layers/perspectives of the 6G communication systems. In this section, we present various relevant layers/perspectives (see Fig.~\ref{fig: Resilience by design}) and elaborate on the measures that can be taken to increase the resilience of the overall system (i.e., end-to-end resilience). To this end, we divide the system and its environment into different categories, which are not directly linked to, e.g., the layers of the ISO/OSI model. These categories are electronics, physical channels, network components and functions, networks (e.g., RAN, core, backbone), services, and cross-layer and cross-infrastructure considerations. In the following, we provide insights into each of these perspectives/layers.

\subsubsection{Electronics}
Electronic devices, such as microprocessors, memories, transceivers, etc., form the lowest (physical) level of the overall system and build up the network components. 
They are responsible for a partial functionality of the network components. 
As such, they have a pivotal role in improving the resilience of these components.
While network components often incorporate redundant electronic components, there are already measures possible to improve the resilience of the electronic components themselves, which will be highlighted in the following.
Another important aspect when considering electronics is the required software, e.g., firmware to operate micro-controllers/processors, since this software may also be prone to faults.

\paragraph{Challenges}
To enhance the resilience of network components, electronic systems have to be robust against challenges like cold, heat, and power disruptions.
Depending on the environmental conditions, a reduction of the load on this component is necessary, which influences the performance of the component and, consequently, of the surrounding network component.
The degree of load reduction necessary is an open challenge for the electronic perspective.
Another dimension of load management is the fact that the aging process of electronics is accelerated when running at high temperatures for a longer period of time. Aging is a progressive process of the degradation of electronic components, occurring while the system operates, and it is the result of various wearout effects such as Bias Temperature Instability (BTI), Hot Carrier Injection (HCI), Random Telegraph Noise (RTN) and Time Dependent Dielectric Breakdown (TDDB). Aging has become highly critical in modern nanoscale CMOS technologies \cite{amrouch2014}.
In fact aging poses certain problems like security risks or higher risk of cascading failure. 
Besides aging, another very critical reliability issue in modern electronic systems are the soft errors, i.e., bit-flips in memory elements \cite{baumann2005}. These effects may occur due to a hit of a high-energy particle, which is a serious concern for space and avionics. In addition, soft errors may be caused by electromagnetic interference (EMI), which may occur as a result of external noise or due to noise generated inside the system.
In case the software running on the electronics is faulty, electronic components may fail or their lifespan could be reduced.
Furthermore, malfunctioning software may result in security risks, in case electronics do not behave as expected.
It is important to note that under real operation scenarios any electronic system may be exposed to the joint impact of multiple types of faults or fault sources  \cite{Goerl2019}. For example, electronics may be exposed simultaneously to EMI and high temperatures. 
A matter frequently overlooked is the threat of backdoors in devices, or hardware trojans \cite{sethumadhavan15trustworthy,moriam18protecting}. 
Especially for increasingly deployed multi-processor systems-on-chip solutions, the design process has become so complex that it increasingly relies on the integration of components from third parties, which may be untrustworthy.

\paragraph{Requirements}
We discuss three main categories of the requirements for electronic components: temperature management, resistance to the environment, and state management.
The \textit{temperature management} of an electronic component can be improved using appropriate materials (Section~\ref{subsec:relevant-capabilities-features}: P1).
In case of overheating (sensed by thermometer P2), may it be due to environmental or load reasons, the load should proactively be reduced to lower the temperature (P3).
The \textit{resistance} of electronics to environmental challenges should be assessed through testing, involving the exposure of respective components to extreme temperatures or different power supply scenarios.
Moreover, it is also important to have protection against overvoltage when power is restored (P1).
This can be accomplished, e.g., using surge protectors.
\textit{State management} is especially important if components are either connected to non-reliable power sources or may be powerless after a disaster.
In case of such a power outage, it is important that electronics are either stateless or are capable of retaining state, such that their functionality is not decreased after such a power outage (P1).
The design in addition should imply breach of the trust domain directly on the hardware level. Protective measures or peer-monitoring hence are desirably already on the level of electronics.

\paragraph{Discussion}
There are several factors that influence the resilience of an electronic component.
For example, monitoring the temperature of electronics helps detecting potential failures in advance since high temperatures can indicate malfunctioning (P2).
As the environmental conditions pose a significant challenge to the electronics, monitoring of the environmental conditions and the electronic component, e.g., through temperature and humidity sensors, can further improve adaptation to the environment, increasing the lifetime of the individual components. 
Here, we have to also pay attention to the accuracy of sensing data and the corresponding cost of design. 
The supply voltage at which the electronic components operate also affect the fault tolerance. It is known that the reduction in supply voltage increases the likelihood of soft errors and timing errors.
To evaluate the quality of software running on electronics software testing is also an important factor.

\subsubsection{Physical Channel}

In communication networks, data is transmitted over physical channels, which can be either wired mediums like optical fiber connections or wireless mediums such as sub-6GHz or mmWave access links. However, these communication mediums themselves are susceptible to various threats, including man-in-the-middle attacks, eavesdropping, and jamming, which can severely disrupt the network's functionality. For example, in the context of autonomous driving, unmitigated jamming could take down the communication links among vehicles or between the vehicles and the 6G infrastructure, thereby compromising critical autonomous driving services.

Security in the face of an omnipotent adversary is impossible, and protecting communications always implies assumptions that bound the adversaries in some sense.
The accepted approach in cryptography is to bound the computational resources of an adversary, and to analyse the security of cryptographic constructions against adversaries that are assumed to be bound in probabilistic polynomial time (PPT) or, given the possible implementation of sufficiently powerful quantum computers of sufficient size, in bounded-error quantum polynomial time (BQP). If the parameters of the
cryptographic construction are sufficiently large (for example, if the
key is long enough), the schemes are proven to be secure by
demonstrating that a worst-case attack would take longer than the
lifespan of the sun to succeed. 

Physical layer security (PLS), and quantum-based security for that matter, make an alternative assumption: they do not necessarily bound the adversary in their computation, but rather consider limitations in the communication or sensing capabilities of potential adversaries.
Assumptions on the transmission and reception capabilities of adversaries are always optimistic in nature. 
In addition, many of the implied assumptions in PLS have been shown to be broken \cite{walther18improving, walther19blind, walther21ray}.

The domain of PLS, however, remains an interesting field for research, especially when considering quantum mechanics. 
\textcolor{\hlcolor}{The quantum attacks can be distinguished into two types: 1) attacks using quantum computing technology, which results in eavesdropping challenges addressed in the next subsection, and 2) attacks aimed at quantum computers, which include security/privacy issues (because of noises, errors, cross-talk, etc.) and several adversary threats including input tampering, Trojan insertion, program mis-allocation, reverse engineering, fault injection, and cloning \cite{saki2021survey}.}
Therefore, an RBD 6G demands the provision of relevant measures to ensure the resilience of the underlying physical channels.

\paragraph{Challenges}

Examples of common challenges that are directly associated with the physical communication channels:
\emph{Jamming:} A malicious attacker tries to impair signal reception at the legitimate user \cite{mpitziopoulos2009survey}. Jamming is often a concern in wireless communication systems where due to the broadcast nature of the wireless medium, the attack can be launched at any location in a reasonable vicinity.
\emph{Eavesdropping:} An illegitimate user tries to intercept the confidential information exchanged between legitimate users \cite{pandey2022security, kapetanovic2015physical}. Significant progress has been made in developing encryption methods and PLS schemes. However, these schemes are being challenged in the face of the emerging quantum computing and communication era \textcolor{\hlcolor}{with possible quantum computing attacks threatening public-key encryption and hash function security in blockchains, and increasing eavesdropping issues in the face of an attacker with quantum computing resources. Therefore, new solutions are required to guarantee post-quantum security such as quantum key distribution (QKD), quantum machine learning, lattice-based cryptosystems \cite{faruk2022review, althobaiti2020cybersecurity, nejatollahi2019post}.} 
\textit{Unfavourable channel conditions:} Over the past decades, various techniques (e.g., channel coding, multiple-antenna communications, etc.) have been developed to cope with the inherently random and time-varying nature of wireless channels. However, despite this progress, in many scenarios, these approaches may be ineffective for unfavorable conditions of the wireless channels. For instance, communication cannot be established if the user is located in a blockage area or typical PLS schemes cannot offer a non-zero secure rate if the eavesdropper's channel is better than the channel of the legitimate user.

\paragraph{Requirements}
The above-mentioned challenges associated with the physical channel have been widely studied in the literature; see \cite{pirayesh2022jamming, grover2014jamming, shakiba2021physical}.  For example, protective mechanisms such as channel hopping, spectrum spreading, and coding techniques have been proposed to mitigate jamming attacks \cite{jeung2011adaptive, pinola2012experimental} (P1). A promising research direction for designing jamming-resilient systems involves interference mitigation using learning techniques or multiple-input multiple-output (MIMO) techniques for detection and decoding data packets in the face of jamming signals \cite{davaslioglu2019deepwifi, yan2014mimo, yan2016jamming} (P2 \& P3).
Strategies to overcome eavesdropping and to increase the secrecy capacity include secret coding and encryption \cite{rezki2017secret}, reciprocity-based key establishment \cite{wang2015survey}, QKD schemes based on quantum physics, ensuring secrecy irrespective of the attacker's computing power \cite{alleaume2014using, huang2021quantum, amer2021introduction}, and using ML to detect active eavesdroppers \cite{kapetanovic2015physical, hoang2021physical} (P1 \& P2).
Additionally, in recent years, intelligent reconfigurable surfaces (RISs) have been extensively studied as a means to cope with the inherent random nature of the wireless channel by directly manipulating/shaping it \cite{renzo2019smart,wu2019towards,yu2021smart}. It has been shown that RISs can be used to extend coverage, improve the rank of the channel, and enable secure communications in scenarios, in which without RISs this may have been not feasible \cite{ji2020secret, yu2020robust, ji2021random} (P3).

\paragraph{Discussion}
Important factors from the physical channel perspective include availability, reliability, integrity, and confidentiality. A jamming attacker has the potential to weaken the channel quality, decrease the Signal-to-Interference plus Noise Ratio (SINR), and interfere with synchronization. As a result, it can compromise both availability and integrity, as discussed in \cite{pirayesh2022jamming}. Furthermore, it can employ deceptive techniques, sending meaningful but misleading radio signals, which poses a threat to the reliability of communication \cite{pirayesh2022jamming}. Additionally, an eavesdropper attacker may decode a portion of the signal based on its channel quality, posing a threat to confidentiality and, consequently, the overall security of communication across the network \cite{shakiba2021physical}.

\subsubsection{Network Components and Functions}
The network, which is discussed in the next subsection, consists of various network components.
These are either hardware components or virtual network functions (VNFs).
For hardware components, they are built of electronics, which were discussed in the previous section.
In addition to that software is also a key part of the network, since even very simple network hardware components need some software/firmware to run.
VNFs, on the other hand, are software-only implementations of network components running on ordinary servers.
As it is easy to update software, the concept of VNFs provides significant reconfiguration capabilities (P3).
Regarding resilience it is crucial to take network components into account, since they provide the functionality for the overall network, including monitoring, control, and management functions that are part of the overall system resilience (e.g., P2, P3).

\paragraph{Challenges}
From the perspective of network components, they should be designed to handle several types of disruptions.
As for electronics this includes power outages as well as failures of single electronic parts and security attacks.
In case of a power outage, network components that are stateful themselves like for example firewalls or databases are responsible for keeping the overall network state and therefore should not lose their states.
Network components consist of electronic parts and rely on them.
Since these parts can fail, network components have to be capable of absorbing the failure of single electronic parts.
Attacks may try to exploit vulnerabilities of network components with the goal of decreasing their functionality, changing their behavior or extracting information.
Protective measures hence have to be put in place, protocol messages and their origin be authenticated, and repudiation be prevented \cite{abedi24improving, osman21mitigating,li19infas}.
This explicitly includes the control and management access to the network components, since many of the network components have to be managed before functioning properly or even during runtime (unless they are functioning fully autonomously).
For example, this involves receiving policy requests, a different configuration, or reporting monitoring data.

\paragraph{Requirements}
Considering the challenges physical network components may face, they are required to do state management, be robust against partial failures in electronics and secure against attacks.
Stateless design should be used wherever possible, however, if network components require to have state, \textit{state management} can help to avoid them losing their state in case of disruptions.
Hence, components should be designed in a way, that enables to store their state in backups and recover it after facing challenges that corrupt or delete current states.
Furthermore, chips that can maintain state even in case of power loss to continue operation later on shall be used. 
The \textit{robustness} of network components relies to a great extent on the electronics they are build of, but can further be extended by redundantly deployed chips to compensate for potential failures (P1).
To increase the \textit{security} of network components, \textit{adaptability} and \textit{upgradability} play an important role, since they allow modifying components based on past challenges increasing their security for the future (P3).
This can be achieved by using programmable parts like FPGAs.

Considering VNFs, there has to be an infrastructure allowing for updating the software.
Moreover, software should be designed such that it is easy to update without reprogramming large parts of it (P3).
Last but not least, the zero-trust (ZT) paradigm has to be applied, which is a security model.
It resolves perimeter security issues by never making any assumptions about trustworthiness, assuming that no part of the network can be trusted, and adversaries can be within any perimeter of the network.
Additionally, ZT architecture assumes all communication over the network to be potentially compromised, hence ZT architecture stresses the crucial point that no communicating entity should be trusted by default, whether in terms of their identity or their authorizations.

\paragraph{Discussion}
Regarding resilience, testing the components against potential challenges provides an indication about how robust they are against them.
Furthermore, repeated testing over time allows to observe how the robustness of components evolves, which is a central aspect of resilience (P1).
In case of a crisis multiple components of the network may fail, which leads to failure of other components relying on them.
To avoid this, the interdependence of network components shall be minimized and especially circular dependencies must be avoided by design.
Moreover, this allows easy replacement and therefore accelerates the recovery of the overall network.

For VNFs, as well as for hardware components, the quality of software is directly related to the resilience of the overall system.
The difference, however, is that VNFs can be restarted, replaced, or scaled out much easier than the software running on hardware components.
Hence, software testing is a central way to guarantee quality of the components and to determine how well components cope with potential challenges (P2).

Finally, all components need to implement strict access control and authentication measures. The control plane and virtualization infrastructures for management of VNFs need to be isolated (physically or at least logically resource-wise) from the user plane, to reduce attack surfaces and prevent cross-plane effects, for instance to sabotage the system \cite{schuchard10losing}.

\subsubsection{Networks}
A network consists of a combination of network components and functions, and should be constructed such that failures of links and/or components do not lead to unnecessary service disruptions.
For the failures, one should consider both failures caused by the environment and attacks.
While there are different types of networks present to form a fully-functional and resilient 6G network (RAN, core, transport, backbone), they all fundamentally benefit from the same principles/ideas.

The connectivity between the network components is an essential aspect for each of these networks \cite{freitas2022graph, hutchison2018architecture}.
Therefore, its resilience is a central factor to consider.
This connectivity can be further subdivided into control plane and data plane connectivity, while the latter is typically not available without a proper working control plane.

\paragraph{Challenges}
Continuously maintaining the end-to-end connectivity of any two devices in the network is a challenge:
Even if network components and the contained electronic components are built in a resilient manner, these components may fail or may get disconnected.
When wired connections are used, like in the 6G core network, the wires may be broken or get disconnected, leading to a connection failure.
When wireless connections are used, environmental influences like the weather or jamming attacks can drastically degrade the performance of the wireless connection, up to a complete disconnection of the components.
These disconnections may then lead to network splits or partitioning whereby essential functions or services may not be reachable anymore, because they are residing in the other partition.
Another challenge is that the load in the network can be highly unpredictable and is expected to surge at disasters, where the network capacity is at its lowest.

Malicious interference with the protocols exacerbates this situation. Simple attacks on routing have proven disastrous on the level of intra-domain routing \cite{cohen15small}. 
Considering the increasing complexity of service provision that comes with dependencies on other protocols (like, for instance, DNS and its known vulnerabilities \cite{cert97bind,alexiou10formal,herzberg12security,man20dns}), secure provisioning, management, and operation are continuously at risk.

\paragraph{Requirements}
To design a resilient network withstanding the aforementioned challenges, different enablers for resilience should be considered while building networks, e.g., using diverse network components, deploying additional resources (overcapacity and redundancy P1), collecting network state information (P2), and enabling the reprogramming of network components to alleviate the impact of failures (P3). 
Diversity of network components is especially important for dealing with dependent failures leading to outages of certain areas, communication paths, or devices~\cite{BORBOR201996,STERBENZ20101245}. 
As the network needs to deal with failures of network components and links, deploying additional network components and links is pivotal for such resilient networks.
In addition, the increasing programmability of network components enables utilizing network components in a previously unintended way, e.g., by shifting functionality from the 6G core to the 6G RAN to enable basic connectivity services even during a disconnect between the RAN and core.
As last requirement, the network needs to be able to provide basic functionality even when the load is highly fluctuating or surging, e.g., by prioritizing important traffic (P3).

Most importantly, configuration interfaces need to be hardened by strict authentication of communication peers, data, and data origins.
These measures need to be founded in the ZT paradigm, and have to include protection of secondary services, the networks rely on.

\paragraph{Discussion}
Important factors for the resilience of a network are the survivability and adaptability of this network.
Similar as mentioned in the network components, adaptability is pivotal for resilient networks.
While the focus of adaptability of network components is on the adaptation to load and other environmental conditions, adaptable networks need to be able to dynamically route content along available paths, recover from link and node failures quickly, and prioritize important traffic such that a basic service can be provided at all times~(P3).

Survivability captures the ability of a network to provide  a minimal level of functionality even during crises.
For example, if the core network is unreachable for the RAN, the RAN should be able to allow communication between UEs within local areas (P3).
This may include the RAN to provide core functionalities at the network edge. In addition to that, 3D networks can be utilized to further enhance the survivability.

Strict authentication, access-control, data-integrity measures and non-repudiation services must be implemented. Trust has to be established based on either direct knowledge, or chains of trust rooted in trustworthy authorities.

\subsubsection{Services}

From a service perspective, the performance of applications and services running over the network is pivotal.
We can differentiate between two types of services, services offered internally from the 6G network and services offered externally only utilizing the 6G network.
Integrated services are expected to be more prominent in future 6G networks.
Those services will be more complex and support, for example, environmental monitoring via joint communication and sensing (JCAS).
This environmental monitoring can be used for environmental awareness of autonomous agents like autonomous vehicles, or by the network to optimize the performance of the radio channel or the mobility management.
On the other hand, there will be external services which utilize the 6G network to transport data.
These services can be resilient by themselves, e.g., having a certain tolerance for connectivity loss or the possibility of a temporary offline operation.
Overall, this perspective deals mostly with increasing the resilience of these applications and services given a partially reliable network underneath, which includes the consideration of trust between entities in the network.

\paragraph{Challenges}
Since applications and services running over the network heavily rely on the underlying network, their functionality may be significantly impaired if parts of the network fail.
In the worst case, the service is fully reliant on the underlying network and fails if there are severe impairments of the network, leading to high costs or other problems.
From the services perspectives, the different priorities of applications, especially during shortages, must be considered.
For instance, during a disaster, the communication between disaster response teams is more important than the functioning of streaming platforms.
Thus, information about the priority of services needs to be provided to the underlying network.

Another important aspect is linked to the increasing virtualization of the 6G network, potentially extending its reach into shared cloud infrastructures.
In this case, the traditional approach of relying on perimeter security becomes futile.
As network functions transform into software services, the line between the control plane within the network core and external networks becomes less defined.
Trust assumptions in components are no longer secure, as adversaries can gain control over processes within the virtualization infrastructure.
This not only poses a risk of unauthorized access to various services but also the potential for impersonating seemingly legitimate entities.
However, there might be the need of a trade-off in disaster situations, in which the ability to validate trust can be limited \cite{rescue2022}.
Only in those scenarios, the previously established trust might be sufficient to temporarily make exceptions on the strict execution of trust.

\begin{table*}[t]
    \centering
    \caption{A summary of the relevant layers/perspectives of 6G communication systems.}
    \begin{tabular}{|>{\centering\arraybackslash}p{1.1cm}|p{6cm}|p{9cm}|}
    \hline
    \textbf{Layers} & \textbf{Challenges and failure types}& \textbf{Requirements} \\
    \hline
    Electronics & 
    •	Cold, heat, and power disruptions
    
    •	Aging process due to BTI, HCI, RTN, and TDDB 
    
    •	Component degradation \& failures, security risks, and cascading failures
    
    •	Internal \& external noises, EMI, soft errors, faulty software, and security risks  
    & •	Temperature management of an electronic component improvement (P1).

    •	Proactive load reduction to lower the temperature (P3).
 
    •	Resistance to environmental challenges \& protection against overvoltage when power is restored (P1).

    •	State management, stateless or state retaining electronics (P1).
 \\
 \hline
    Physical channel & 
    •	Man-in-the-middle attacks
    
    •	Jamming and eavesdropping with new bounds on the adversary's resources, e.g., sufficiently powerful quantum computers
    
    •	PLS in the case of adversaries with unlimited computing resource and limited communication \& sensing capabilities
    
    •	Unfavorable channel condition, e.g., signal blockage
    &  

    •	Protective mechanisms, e.g.,
    channel hopping, spectrum spreading, and coding techniques (P1).
    
    •	Interference mitigation using learning \& MIMO techniques, and jamming-resistant data packets detection and decoding (P2- P3).
    
    •	Strategies to overcome eavesdropping, e.g., secret coding, encryption, reciprocity-based key establishment, and QKD, and to increase the secrecy capacity using ML to detect active eavesdroppers (P1-P2).

    •	Using RISs to extend coverage, improve the rank of the channel, and enable secure communications (P3).
 \\
 \hline
    Network components \& functions & 
    •	Disruptions such as power outages, failures of single electronic parts, and security attacks

    •	Control and management access to the network components to manage policy requests, different configuration, or reporting monitoring data
 & 
    •	Stateless design or state management to avoid state losing in case of disruptions, e.g., designing components that store their state in backups and recover it (P1).
    
    •	Adaptability \& upgradability, i.e., modifying components based on past challenges (P3).
 
    •	Easy to update software  without reprogramming large parts of it (P3).

    •	ZT paradigm, where each communication needs to be authenticated, all access controlled, authorizations strictly adhered to, and repudiation be prevented (P3).
 \\
    \hline
    Networks & 
    •	Network splitting/partitioning due to broken wired connections and environmentally influenced or jammed wireless connections

    •	Highly unpredictable load surge in the low-capacity network at disasters
    
    •	End-to-end connectivity between network components, i.e., control plane and data plane connectivity

    •	Malicious interference and routing attacks
 & 
    •	Enhance and secure provisioning, management,
and operation.

    •	Deploying diverse components \& additional resources (P1) and collecting network state information (P2).

    •	Enabling the reprogramming of network components to alleviate the impact of failures (P3).
 
    •	Providing basic functionality in highly fluctuating/surging load, by prioritizing important traffic (P3).

    •	Hardening configuration interfaces and protection of secondary services by strict authentication using ZT paradigm (P1).
 \\
    \hline
    Services & 
    •	Impairments of the network and disaster situations
 
    •	Lack of information in the priority of different services

    •	Adversary's control within the virtualized  infrastructure, risk of unauthorized access to services, and impersonating seemingly legitimate entities

 & 
 •	Adaptability regarding service placement (P3).
 
 •	Prioritization profiles and inter-service dependencies for various potential situations in the face of resource shortages (P2).

 •	Robustness of network services against failures by making use of monitoring data to pre-train the ML models (P1, P2).
 
 •	Online-learning and automatic adaptation (P3).
 
•	ZT architecture implementations are required (P3).
 \\
    \hline
    \end{tabular}
    \label{tab: layers}
\end{table*}

\paragraph{Requirements}
When facing partial outages of the underlying network, \emph{adaptability} regarding service placement is needed, i.e., for shifting services from remote locations (e.g., the cloud) back to edge-servers or even local devices.
To allow fast decisions regarding the prioritization of services when facing resource shortages, different \emph{prioritization profiles} for various potential situations are required.
Thus, services with lower priority can be shut down, e.g., during a disaster scenario and the limited resources can be used to still provide the most important services.
As services might rely on other services, prioritization also includes the consideration of inter-service dependencies.
To make services more resilient themselves, they have to be \emph{robust} against potential network failures or reconfiguration, i.e., they have to be designed such that they can be restarted or moved to different locations within the network without having their functionality impaired.
As the 6G network architecture is quite complex and produces a lot of monitoring data, ML-based models can be envisioned for automatically analyzing this data and improving the system.
The utilized ML algorithms need to be pre-trained and constantly improved.
For that purpose, generalization and online learning of ML algorithms are decisive, such that network services can automatically adapt and learn from new situations.
These properties are pivotal for services that are resilient by design.

As mentioned in the previous section, due to the virtualized infrastructure of 6G networks, trust assumptions in components are no longer secure.
A solution to this problem might be the ZT architecture~\cite{rose_zero_2020}.
From a service perspective, that is, no assumptions about trustworthiness of the services provided externally or internally by the 6G network, are made.
Moreover, all communication among or with these services is not trusted.

To realize ZT in 6G networks, authentication and access control are two key requirements.
All communication must be authenticated before granting access to the resources, and the requested access to the resources must be verified for effective permissions and enforcement of access control. 
This approach provides multiple layers of defense against an insider attacker who has compromised security within the network perimeter and prevents attackers from escalating their privileges or accessing sensitive resources.

\paragraph{Discussion}
The time it takes to detect a challenge, remediate and recover from it, is one key factor to determine resilience.
Therefore, the mean time between the occurrence of a service failure and the point at which the failure is detected is the \textit{Mean Time to Detect (MTTD)}.
The MTTD provides a metric on how well services are monitored and on how good and fast the analysis of the monitoring data is.
The \textit{Mean Time to Remediate (MTTRm)} is the mean time a service needs after facing a challenge until it reaches a minimal level of acceptable functionality.
Hereby, it is important to consider, that a service might consist of multiple services, which is especially the case regarding the one internally provided by the 6G network.
Regarding different operation levels, the MTTRm can be defined as the time that a service needs to go from one operation level to the next one.
A service is recovered when it reaches its original level of functionality.
Hence, the \textit{Mean Time to Recover} is the mean time from the occurrence of a challenge until a service is at its initial operation level.
Refinement is also an essential part of resilience.
Therefore, the \textit{Time to Adapt}, i.e., the time a service needs to adapt based on what was learned from a challenge, is also of importance.
After a service has adapted, its resilience is typically considered to be higher than it was initially.

In addition to that, \textit{survivability} describes how much a service is impaired after a challenge.
The \textit{availability} of a service can be quantified by measuring its up-time and relating it to its down-time.

A summary of the discussed layers/perspectives of the 6G communication systems, their related challenges and requirements is provided in Table \ref{tab: layers}.

%%%%%%%%%%%%%%%%%%%%%%%%%%%%%%%%%%%%%%%%%
\subsection{Cross-Layer and Cross-Infrastructure Considerations} 
\label{sec:Cross-layer-Considerations}

In the following, we discuss cross-layer aspects within 6G networks as well as the interaction of 6G networks with other critical infrastructures.

\subsubsection{6G Cross-layer Aspects}
The 6G network infrastructure will consist of many interdependent components and layers
(e.g., see Figure 12 in \cite{Organic6G-IEEEAccess:2023}). Mastering the complexity of
these highly interdependent sub-systems and services to achieve a highly resilient overall
system is difficult. For example, the use of virtualized 6G components seems to provide the 
necessary flexibility for scalability as well as for reconfiguration as response to challenges (P3). 
However, in case virtual resources need to be added or exchanged in response to a challenge, the 
virtual infrastructure manager entity must work correctly and be able to access the physical and 
virtual resources without problems. This requires a properly working \emph{Control Plane (CP)} connectivity.
It may be realized out-of-band, i.e., by using a separate dedicated network that must be configured and 
managed as well, thereby introducing another dependency. Alternatively, CP connectivity may be realized 
in-band, i.e., using the same links and components as the user's data packets and thus largely sharing 
the same failure fate with data plane components. A combination of both approaches is also possible and 
used in practice. CP access and its connectivity can be seen as cross-layer issue. Without
CP connectivity, many higher layer resilience mechanisms will not work anymore, e.g., monitoring
data from individual components cannot reach responsible controllers anymore (e.g., affecting monitoring and anomaly detection of P2) and they cannot exert necessary reconfiguration actions to perform a suitable adaptation 
(e.g., affecting self-adaptation of P3) when facing a challenge. So one conclusion is to maintain CP 
connectivity for the various components at the different layers.
Ideally, this CP connectivity is provided in a zero-touch manner (P1), i.e., it does not depend on any 
manual configuration and it is self-adapting (P2 and P3), in order to avoid losing control over the 6G infrastructure. 
A zero-touch connectivity solution that has no dependencies (e.g., like KIRA \cite{BlessZitterbartDespotovic2022_1000148953}) 
cannot fail by misconfiguration of underlying components.

\subsubsection{Interaction with Other Critical Infrastructures}

The 6G communication networks interface with various critical infrastructures such as power distribution, water management, transportation networks, governmental operations, and financial systems. This interconnectedness significantly impacts the mutual resilience of these systems, as any disruption or malicious attack targeting one infrastructure may cause cascading effects across others. For concreteness, in the following, we discuss the interplay of 6G communication networks with the power distribution network.

The operational 6G communication systems rely on specific requirements within its deployment environment such as power supply and cooling systems. Given these prerequisites, the resilience of 6G communication systems is interconnected with the resilience of the power distribution network. Various solutions have been proposed in the literature for enhancing the resilience of the power network, e.g., distributed power generators based on renewable energy sources. We refer interested readers to \cite{bhusal2020power, blaabjerg2017distributed, bie2017battling}, as a full discussion on this topic is out of the scope of this paper. Nonetheless, RBD measures can be embedded in the 6G communication network to prepare them for potential power supply disruptions. Example strategies include the provision of low-energy operational modes or the reconfiguration of the remaining functional part of the 6G communication infrastructure to compensate for the failure of any subsystem (e.g., a BS) due to a power outage.

%%%%%%%%%%%%%%%%%%%%%%%%%%%%%%
\subsection{Design trade-offs} 
\label{subsec:RBD_tradeoffs}

In this section, we investigate design trade-offs to achieve resilience in a network. 

\subsubsection{Resilience-complexity/cost Trade-off:} RBD designs (see principles P1--P3) introduce various provisions to ensure minimum service requirements in the face of potential challenges. These provisions increase the complexity and cost of installing and operating the 6G communication systems and often do not (at least significantly) improve the performance of their normal operations. For instance, deploying redundant components/links does improve the system resilience but at the expense of a considerable additional cost. \textit{Therefore,  the art of an RBD design is to enhance the system's resilience with a minimum addition to its complexity and/or cost.} Straightforward solutions such as adding redundancy and diversity, while being effective for parts of the 6G communication networks (e.g., redundant routing paths), are not always the most efficient approaches (e.g., devices cannot afford duplicates of all their components). Moreover, improving the resilience of an individual component/layer does not necessarily imply a considerable increase in the end-to-end resilience of a given service since other components/layers of the 6G network may be the bottleneck.  Hence, it is crucial to characterize the impact of the resilience enablers discussed in this section on the end-to-end resilience of the 6G systems for offering a given service. Such characterization reveals the trade-off between resilience and complexity/cost and guides the system designer using 6G services in choosing the right setting for any given application.

\subsubsection{Resource Optimization for End-to-end Resilience:} The different components and layers of 6G communication networks possess their own constraints \cite{maklachkova2022analysis}. For instance, in a cellular network, adding high complexity and cost to the design of mobile user equipment is not always feasible. However, it is more tolerable in higher levels/layers, such as the core network. For example, the infrastructure should incorporate potential resources for different scenarios, necessitating considerations of complexity, investment costs, and running costs (including energy consumption and maintenance). 
\textcolor{\hlcolor}{The optimum strategy for resource/cost management may depend on the considered application. For instance, processing data at the edge is often more cost-effective, especially for applications that generate substantial amounts of data. Transmitting and processing such data within the core network can be expensive, whereas edge devices can conduct initial data processing, filtering, and aggregation to reduce the load on the core network \cite{yu2017survey, pan2017future}.} 
Moreover, not all applications require the same level of resilience. For instance, remote surgery demands a very high level of resilience, while in a monitoring network, lower levels of resilience remain acceptable.
Thus, we encounter limited resources at each layer for resilience improvement and have potentially different resilience requirements for different applications. In addition to the characterization of the resilience vs. complexity/cost trade-off, another pivotal question becomes: \textit{How to maximize the end-to-end resilience given the resource constraints?} and in which layer is applying the resilience enablers most efficient to meet the requirement of a given service? 
These questions pose a resource allocation problem aiming to maximize resilience while adhering to design constraints at different levels.

\subsubsection{Reconfigurability -- A Double-edged Sword for Resilience:}
As discussed before, resilient 6G systems have to be able to reconfigure themselves in order to cope with the challenges (attacks, failures, etc.), which is why the reconfiguration capability was presented as a key resilience enabler. Emerging communication technologies like software defined networking (SDN) and network functions virtualization (NFV)\textcolor{\hlcolor}{,
and AI provide possibilities of realizing the reconfiguration capability \cite{hutchison2023importance}. However, enhancing the system reconfiguration also could imply that the ways that a system can potentially fail or be attacked would increase, too (e.g., a larger attack surface). Example of AI-driven threats include adversarial attacks, privacy threats, and loss of control in highly autonomous/complex environments.} 
In addition, the improper management of the reconfiguration itself may be a cause of the system failure. In other words, managing the \textit{complexity of the entire communication network} is a crucial challenge that must be addressed when implementing resilience strategies. This is primarily because the higher the network complexity, the more susceptible it is to complex failures. Finding a balance between the system reconfigurability and complexity constitutes a challenging trade-off for the design of 6G communication systems. 

\section{6G Use-Cases}
\label{sec:4-usecases}

The general framework for RBD is application-dependent. In the following, we provide 6G use-cases and discuss the proposed concept of RBD for these systems.

\subsection{Cloud-based distributed monitoring network}

\begin{figure}
	\centering
	\includegraphics[clip, width=1.0\linewidth]{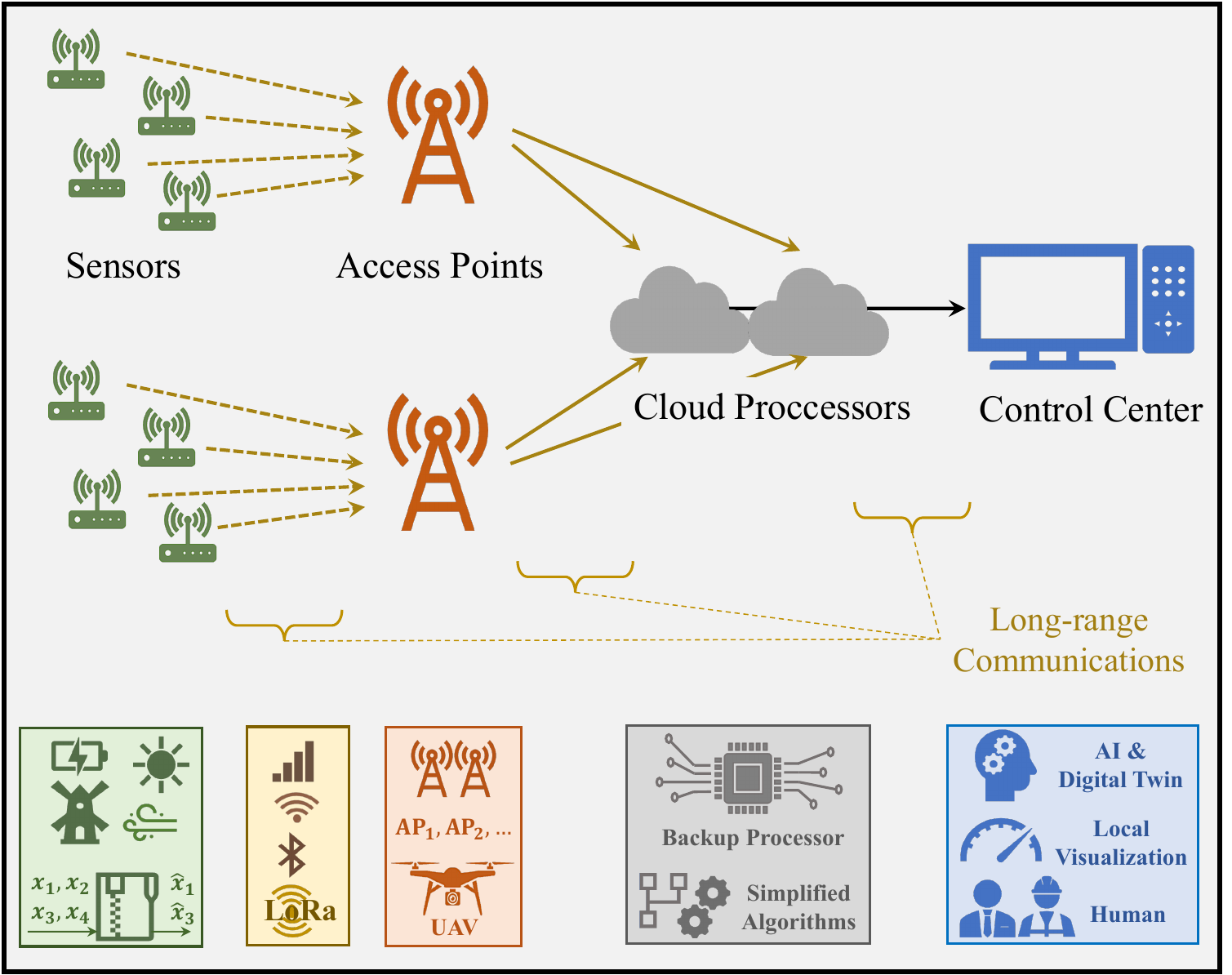}
	\caption{Cloud-based distributed monitoring network as a 6G use-case.}
	\label{Fig: usecase_Cloud_Based_Monitoring}
\end{figure}

With recent advances in sensing, communication, and computing technologies, 6G-enabled distributed monitoring and processing are expected to play a crucial role in the digitalization of smart cities \textcolor{\hlcolor}{\cite{javed2022future, band2022smart}}. 
\textcolor{\hlcolor}{A distributed monitoring network includes a unit that visualize the network (e.g., using a digital twin (DT)) and making decisions. The resilience of this network depends on the updated decisions and the monitoring should be performed continuously (without interruption) \textcolor{\hlcolor}{\cite{band2022smart}}, possibly with a reduced level of service, i.e., the critical parameters are updating and the non-critical parameters are ignored. Therefore, the set of desired resilience metrics could involve accuracy, updating rate of the DT, and age of information (AoI).}
In a typical setting, numerous sensors are deployed across the city to gather relevant sensory data (sensing aspect). The collected observations are then transmitted to a limited number of access points (communication aspect), which forward the data to some cloud-based processors for computation and analysis (computation aspect). The resulting insights are subsequently sent to a control center (e.g., city administration or network operator) for visualization and monitoring. A schematic illustration of such a distributed monitoring network is depicted in Fig.~\ref{Fig: usecase_Cloud_Based_Monitoring}. 

Existing design approaches primarily aim at efficiently exploiting the benefits of the abundant sensing data, powerful processing resources, and high-quality monitoring which can be practical in normal scenarios. However, whether such system designs are resilient in the face of failures in crises has remained an unexplored question. There are many possibilities for things to go wrong. 
Any partial failure may result in a problem in the end-to-end performance of this network.
For instance, in potential failure situations, certain components might become inaccessible, or specific links/connections could be disrupted. To showcase how the proposed RBD concept can be applied to distributed cloud-based monitoring, we analyze a few different states of the system (due to e.g., potential challenges/failures) and offer relevant strategies for mitigating these challenges from different perspectives (i.e., sensing, communication, computation, and visualization).

\textbf{System state 0 (normal system state):}
In this state, there is no failure. The sensing and monitoring mechanisms do not report any failures. Every subsystem (component) operates in its normal mode, similar to the conventional system. Thus, there is no need to consider any resilience measures.

\textbf{System state 1 (sensor failure):} 
A sensor encounters power constraints (challenge), which are reported by its local energy monitors (P2). The sensor has the capability to activate various designed energy sources (e.g., electricity lines, battery, solar panels), and its sampling rate can be reduced (P3 at sensing perspective).
The sensor's transmission power is adjustable, allowing for a decrease in power when necessary (P3 at communication perspective).
The cloud determines the data processing approach using simpler algorithms (P3 at computation perspective), while the control center specifies how the data should be visualized (P3 at visualization perspective).

\textbf{System state 2 (link failure):} 
The sensor-to-access point link is down (challenge). Consequently, the link quality monitor is unable to receive any acknowledgment or feedback from the access point, leading to the detection of a link disruption (P2).
A diverse (or backup) link can be activated, such as LoRa, WiFi, or Bluetooth technologies in addition to the primary cellular technologies (P3 at communication perspective)
A portion of computations can be executed locally at the sensor, and the resulting processed data, which has a reduced rate, can be transmitted (P3 at computation perspective).
The sensor decides which data should be sent (P3 at sensing perspective) and how the preprocessing should be applied (P3 at computation perspective). 

\textbf{System state 3 (processing failure):} 
A cloud processor is inaccessible due to a disruption in the link from the access point or a malfunction in the access point itself (challenge). 
The access points and control center detect the unavailability of the cloud by monitoring the quality of their connections to and from it (P2).
Backup local processors can be deployed at the access points to execute a portion of computational tasks locally. A new (less complex) algorithm can be adopted/developed for computation (P3 at computation perspective). The processed data can then be transmitted to the control center (P3 at communication perspective). Reliable links between the new processors and the control center are essential for seamless communication.
The control center decides how the data should be visualized (P3 at visualization perspective). 

\textbf{System state 4 (control center failure):} 
This center receives processed data from the cloud, visualizes/interprets the data, and then makes decisions based on this information. Let us focus on two types of failures (challenges) for this center: (\emph{i}) a failure in data visualization/interpretation, and (\emph{ii}) a failure in the decision-making algorithm.

6G will enable maximum exploitation of the processed data of the sensor network at the control center by interpreting it using AI and Big Data (e.g., measurement history, data from other correlated sources, etc.) \cite{letaief2021edge, mahmood2022comprehensive}, and integrating it into visualization interfaces such as a digital twin \cite{khan2022digital}. The algorithms realizing these powerful features are also subject to attack/failure and are often not locally available (e.g., due to the associated cost). The visualization/interpretation unit at the control center follows an RBD concept if it is equipped with multiple operational modes (P3, reconfiguration), where for example, if the powerful remote AI algorithm used for visualization fails, a backup and perhaps less powerful local unit takes over the visualization task.  

The existence of a control center and ``human-in-the-loop'' may be essential for various monitoring-based decision-making settings. However, the control center itself may constitute an SPF in this system. To ensure an RBD system, in the case that the control center cannot be reached, an autonomous AI-based control can be provisioned, which detects such events (P2, self-awareness) and makes a transition to an autonomous policy-based algorithm or adopts a semi-autonomous human-in-the-loop strategy. The latter involves human intervention to enhance validation and handling, constituting a form of reconfiguration (P3).    

As illustrated in the aforementioned examples of failure, a malfunction in one component or link affects the optimal operational mode of other components. Consequently, all components adjust their strategies to absorb the failure and guarantee end-to-end resilience, ensuring a minimum level of visualization for the end-user.

While typically systems are designed for their normal state, an RBD system accounts for relevant \textit{abnormal} states, too. Obviously, the larger the set of challenges/failures considered during design, the more resilient the system can be but the higher the system complexity and cost. The art of RDB design is to achieve the resilience requirement with minimum cost/complexity, see Section~\ref{subsec:RBD_tradeoffs} for our discussion on this trade-off. 

\subsection{Autonomous driving}

\begin{figure}
	\centering
	\includegraphics[clip, width=1.0\linewidth]{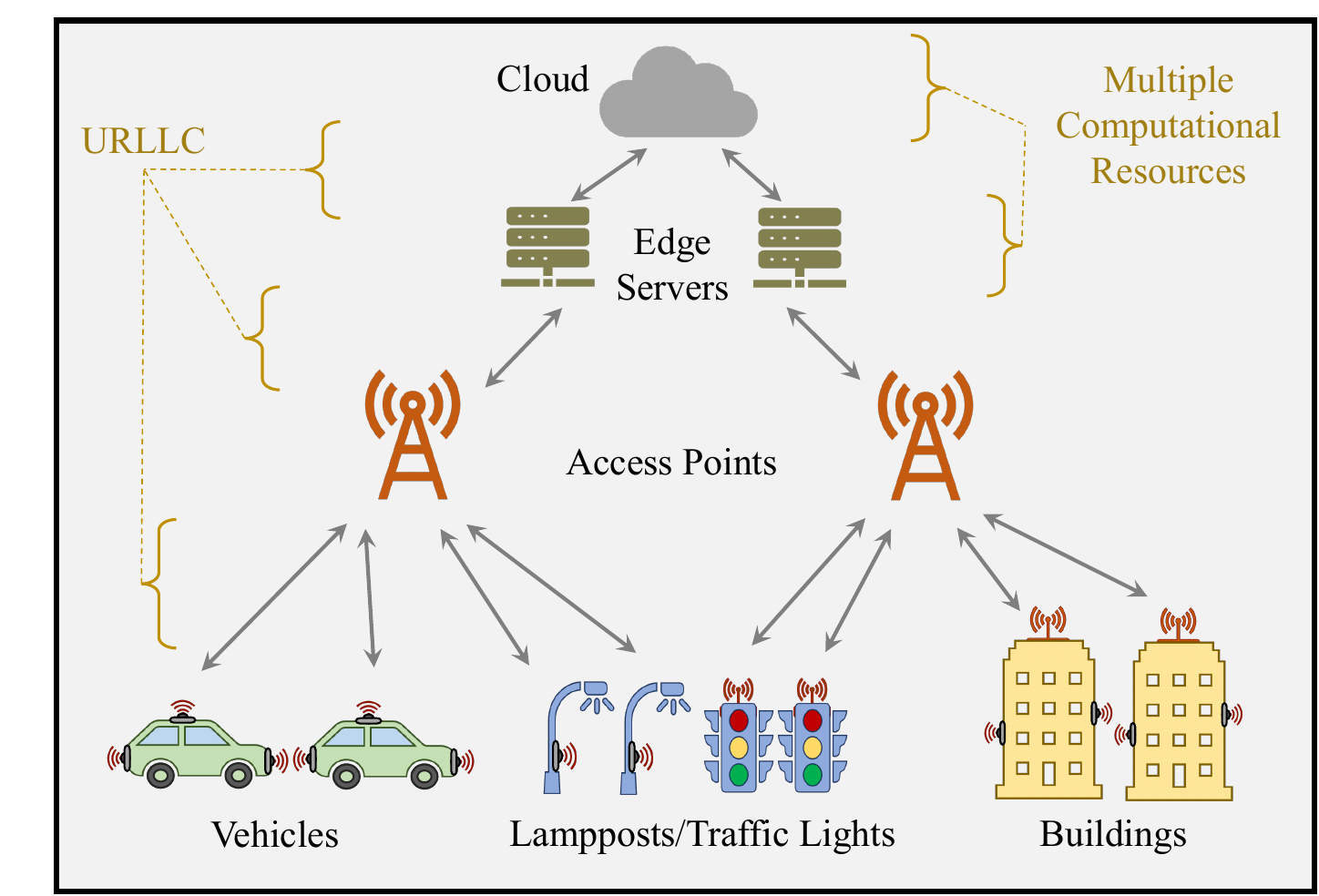}
	\caption{Autonomous driving as a 6G use-case.}
	\label{Fig: usecase_autonomous_driving}
\end{figure}

With edge computing and Internet of Everything (IoE), 6G will enable  massive enhancements regarding sensing and ultra reliable low latency communication (URLLC) \textcolor{\hlcolor}{\cite{serodio20236g}}.
Thus, it paves the way for autonomous and even remote-controlled driving, as outlined in Fig.~\ref{Fig: usecase_autonomous_driving}.
Vehicles are equipped with sensors allowing them to collect observation data about their environment, while additional sensors are placed at environmental objects, e.g., lampposts or buildings.
The data collected from both the vehicles and the environmental sensors is then transmitted via the access points to the edge servers.
By processing and analyzing this data, a model of the environment is created at the edge, through which both the edge server and the vehicles gain an overview of the overall traffic situation.
However, as the resources of an edge server tends to be limited, some of the processing, especially non-time-critical processing, might need to be performed in the cloud.
Examples for low-latency applications are the remote control of vehicles and the provisioning of safety-critical sensor data \textcolor{\hlcolor}{\cite{mohammed2024enabling}}.
For remote-controlled vehicles, control information will be shared with the vehicles in real-time.
But even the performance of non-remote-controlled autonomous vehicles can drastically be increased if a high-quality environmental model is available in real-time.
For less time-sensitive tasks like finding a route to drive from A to B; collecting data on the global traffic situation; or making long-term analyses of traffic, respective data is also transmitted to the cloud, which then stores and processes it. 
This data can then later be used in the edge to make vehicle control-related decisions.

Since driving errors may lead to severe consequences, the 6G network has to be ultra-reliable and able to send controlling commands to the vehicle with a guaranteed low latency \textcolor{\hlcolor}{\cite{dressler2022v, aguiar2023inter, pannu2023data, yonis2023performance}}.
\textcolor{\hlcolor}{Therefore, the relevant set of resilience metrics for this use-case includes accuracy and end-to-end operating speed (low latency).}
In the following, we describe how the RBD concept is capable to guarantee this by going over some potential challenges.

\textbf{System state 1 (sensor failure):}
When it comes to failing sensors it has to be distinguished between two cases: a failure of the vehicular sensors and a failure of the environmental sensors.
If the vehicular sensors fail, vehicles should be designed such that they can identify the failing sensor (P2), compensate for the failing sensor using redundant sensors (P1), and infer the required data from other sensors (P3).
Moreover, the overall system can, to a certain degree, compensate for failing sensors by not only utilizing the observational data provided by the vehicles, but also information provided by other vehicles or sensors placed on environmental objects (P1).
However, in this case, the requirements for the 6G network increase drastically, as the vehicle's movement relies solely on external sensors.
This could, for example, enable the vehicles to return to a safe state and come to a stop safely (P3).
Failures of environmental sensors can be compensated for, and vice versa, by making use of data provided by the vehicles~(P1).

\textbf{System state 2 (link failure):}
In case the link between a vehicle and a base station is impaired, there are multiple ways this challenge can be addressed by the 6G system.
If the link still exists, but the bandwidth is restricted, sensor information may be prioritized over other non-critical data traffic.
Furthermore, the sampling rate of sensors can be reduced to minimize the overall link load (P3).
Another alternative is to make use of vehicle to everything (V2X) communication, which commonly relies either on 6G-based Device-to-Device communication or on WiFi ad-hoc networks (P1).
V2X allows to share data in a peer-to-peer fashion between vehicles, other road participants, and potentially available road side units, which then can relay the information to the next available edge server (P3).

{The link may be interrupted when the base station is under a jamming attack. Generally, the jammer aims to perform a DoS attack by transmitting interference signals to the receiver. It is foreseeable that at the age of 6G, the jammer will not only impact the communications service but also threaten the sensing function. Issues on jamming and anti-jamming strategies in communication networks have been widely discussed \cite{pirayesh2022jamming,7751164,5473884}. However, research on the impact of jamming on sensing and the corresponding anti-jamming technologies is just emerging. Simple methods like controlling transmission power and increasing radar processing gain via collecting more samples or choosing alternate waveforms with better anti-interference performance can more or less improve the system behavior under jamming. Furthermore, coordinated multi-point (CoMP) transmission is regarded as a solution against jamming, as multiple transmitters or receivers are involved in the transmission process. If one transmission/reception point is down due to jamming, the others can still provide services. }

\textbf{System state 3 (processing failure):}
Services provided by an edge node might be impaired due to server outages or overloading.
In such a case the system has to activate additional resources.
This might be possible by simply utilizing backup servers, which are available for such a scenario.
However, if there are no additional backup resources, services with lower requirements, e.g., regarding latency, or lower priority might be moved to the cloud or deactivated until the challenge has been overcome (P2, P3).
For the use-case of autonomous driving, that is, controlling the vehicles has the highest priority, while, e.g., preventing traffic jams is of lower priority.
As last resort, vehicles might be put into a safe state, eventually slowing down traffic drastically or even stopping the vehicles completely.

As discussed for this use-case, autonomous driving works best, if the whole system is running without any impairments.
However, it also became clear that autonomous driving is possible even while the system faces challenges.

\subsection{A production line in a highly automated factory}

\begin{figure}
	\centering
	\includegraphics[clip, width=1.0\linewidth]{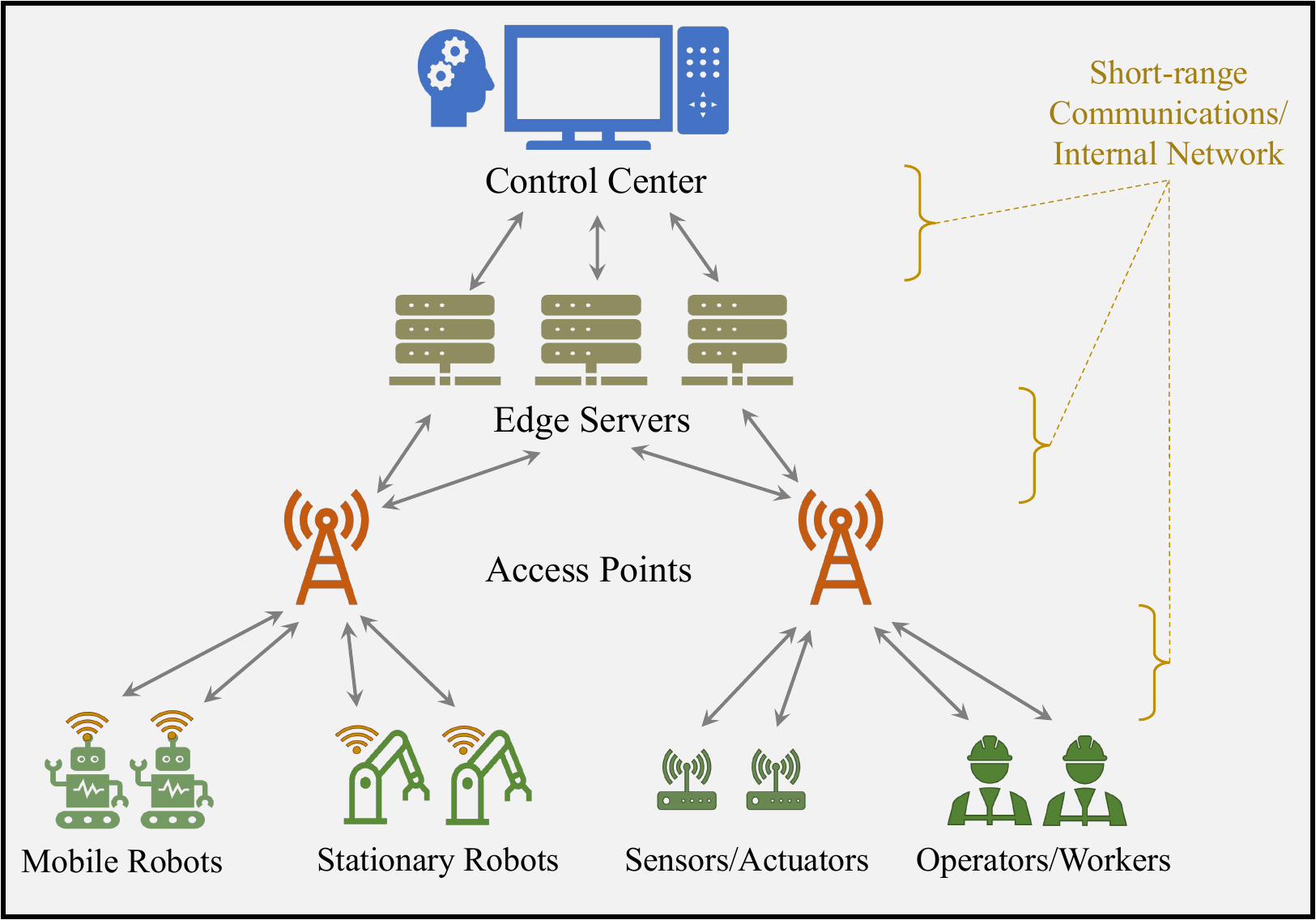} 
	\caption{Automated factory as a 6G use-case.}
	\label{Fig: usecase_smart_factory}
\end{figure}

Modern factories are increasingly based on automated production processes, thereby increasing the overall production efficiency. In general, an automated factory can be centralized (all production activities are taking place at a single site), or distributed (there are multiple production sites) \textcolor{\hlcolor}{\cite{boccella2020evaluating}}. For simplicity, we will consider the case of a single production site. An automated production line may involve a large number of stationary or mobile machines/robots, a number of sensors and actuators, and a limited number of humans (operators/workers). \textcolor{\hlcolor}{The set of relevant resilience metrics for an automated production line involves production efficiency, low latency, safety, low cost, and sustainability to achieve a real-time control \cite{fortoul2023smart}.}
One of the trends in automated (smart) factories is the use of wireless communication between different process elements, allowing seamless remote control of the production process. A general concept of an automated factory as a 6G use-case is illustrated in Fig.~\ref{Fig: usecase_smart_factory}.

The 6G technology has the potential to revolutionize manufacturing by enabling highly interconnected, intelligent, and agile production systems in smart factories. The URLLC technology would provide reliable and low-latency communication, thus enabling real-time control of complex manufacturing equipment \textcolor{\hlcolor}{\cite{peng2022resource}}. The massive machine-type communication (mMTC) technology would support massive connectivity, allowing seamless integration and control of a huge number of sensors required in an automated factory. With envisioned AI support in 6G networks, intelligent decision making based on massive sensor information would be more reliable in automated factories. As 6G networks will support higher bandwidths, the application of augmented and virtual reality, as well as digital twins,  would allow operators in factories to visualize complex processes and perform virtual testing and troubleshooting.

One of the key requirements for automated factories is the resilience, because any error in the production system may cause expensive damage or even the loss of lives. A production line in an automated factory may be subjected to various challenges that may compromise the reliability of the production process. In many manufacturing processes high temperatures may be employed, which leads to increased aging rate and soft error rate of electronics. Another important issue is the electromagnetic noise, which may result in soft errors. The risk of security attacks on wireless communication networks should be also taken into account. In the following, we discuss how the proposed RBD concept can be applied to a smart factory.      

\textbf{System state 1 (sensor failure):}
In an automated factory, three main classes of sensors may be employed: sensors for monitoring the environment, sensors for monitoring the machines/robots, and sensors for monitoring the quality of the final product. Sensors for monitoring the factory environment may involve sensors for monitoring temperature, humidity, pressure, energy consumption, presence/movement of humans and mobile robots, etc. All machines/robots may be equipped with sensors for monitoring their performance, providing the information when the maintenance may be required. Special sensors are also used to monitor the quality of materials used in the production process and to test the final product. Critical sensors may be duplicated (P1) to ensure that sensing functions are preserved even if one sensor fails. All sensors may be equipped with the self-checking features to enable timely detection of erroneous sensor information. If an error is detected in a sensor, it will be switched to test mode to identify the error and possibly repair the sensor, while the redundant sensor will be activated to continue the measurement (P3). 

\textbf{System state 2 (link failure):} The wireless communication in an automated factory may be affected by failures in the access points or disruptions in links between the access points and actors in the production process (robots and humans). A typical approach for mitigating the effects of link failure in smart factories is through redundant communication links based on a different technology. Redundant communication links may be implemented (P1) in the form of wired communication protocol such as process field bus (PROFIBUS) and process field network (PROFINET), or proprietary wireless communication such as industry local area network (LAN). Continuous monitoring of the link status (P2) will ensure timely detection of link disruptions, and redundant links will be immediately activated (P3). 

\textbf{System state 3 (processing failure):}
Apart from control center, data processing may be performed by machines/robots and edge servers. Each machine/robot may be equipped with processing resources for local processing, necessary for immediate actions. Data from robots is collected by the edge servers for further processing. In order to reduce the likelihood of failures, redundant processing elements are used. In the case of robots, reconfigurable multi-core processing platforms may be used (P3). The reconfigurability allows changing the operating modes in real time, thus enabling to achieve a trade-off between performance, power consumption and reliability. That is crucial because robots are powered with batteries having a limited lifetime. In the case of edge servers, multiple servers may be used, such that if one server fails another could take over the processing tasks.  

\textbf{System state 4 (control center failure):}
The control center in an automated factory monitors the overall production process and makes high-level decisions on the required changes. The resilience threats in this case are similar to those in a distributed monitoring system, and thus similar measures may be applied. 

\textcolor{\hlcolor}{\subsection{Virtual and augmented reality}}
% Scenario
\textcolor{\hlcolor}{Extended realities (XR), encompassing augmented reality (AR) and virtual reality (VR), are emerging technologies with diverse applications in fields of networking, gaming, healthcare, and education \textcolor{\hlcolor}{\cite{fu2022systematic, akyildiz2022wireless}}.}

\begin{figure}
	\centering
	\includegraphics[clip, width=1.0\linewidth]{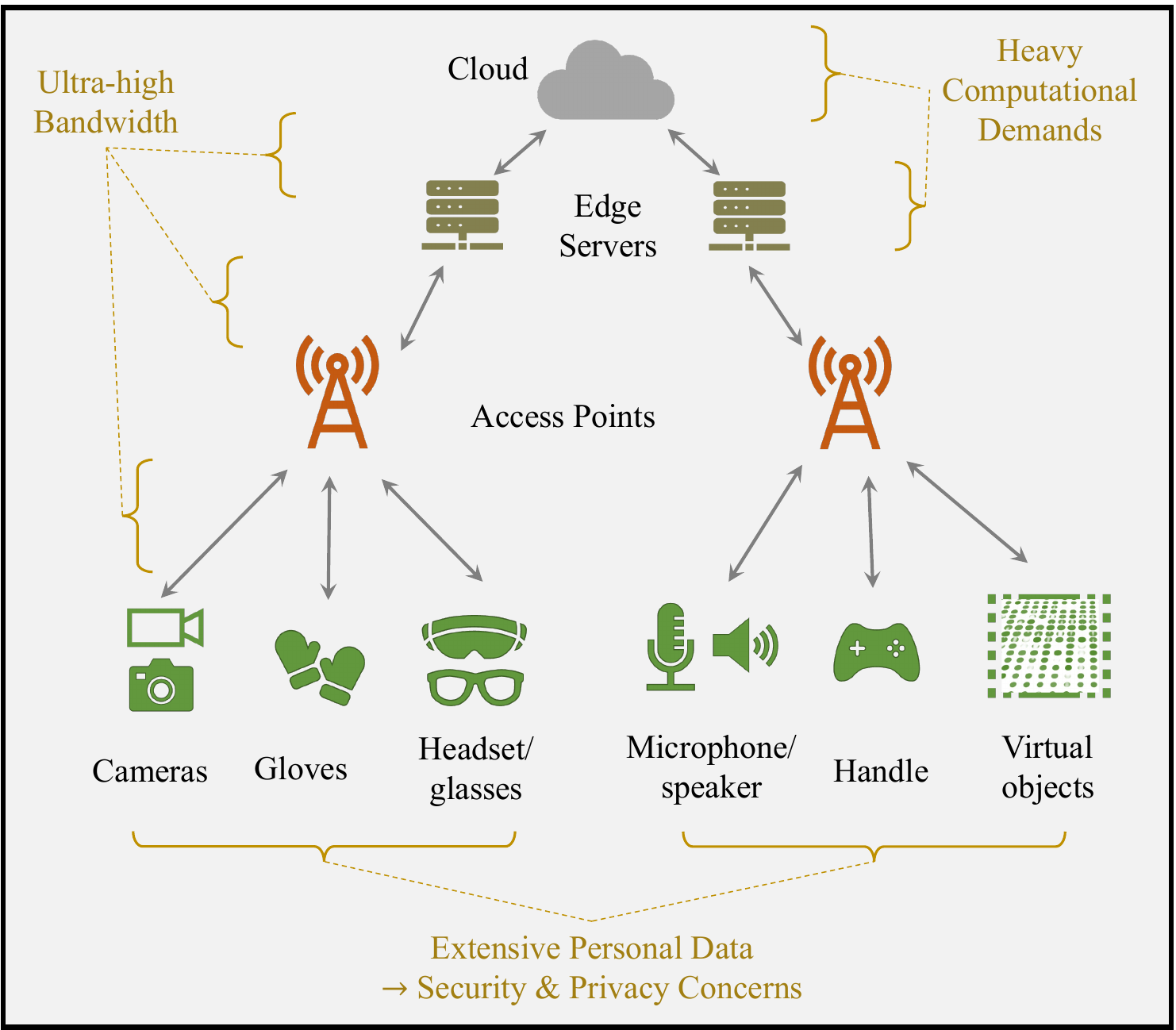}
	\caption{Virtual and augmented reality as a 6G use-case.}
	\label{Fig: usecase_AR_VR}
 \vspace{-1.0 em}
\end{figure}

\textcolor{\hlcolor}{XR systems are envisioned as immersive platforms that support scalable, trustworthy, and persistent interactions through advanced communication and collaboration technologies (see Fig. \ref{Fig: usecase_AR_VR}). A key challenge in enabling scalable XR interactions is efficiently managing cloud/edge computing resources and transmitting XR camera and sensory inputs over wireless networks, while meeting quality of service (QoS), quality of experience (QoE), and data consistency standards.}

% Performance Requirement
\textcolor{\hlcolor}{Since XR transforms the physical world into a virtual shared experience, interactions are often limited to localized environments due to the complexity of maintaining consistent spatial and semantic associations in diverse deployment scenarios. Additionally, XR experiences require real-time synchronization and tracking of digital assets across multiple users, imposing scalability challenges \textcolor{\hlcolor}{\cite{akyildiz2022wireless}}. This is particularly relevant in scenarios like large classrooms and collaborative learning, where increased latency and intensive rendering computations strain network and computational resources, which must be dynamically adjusted based on scene complexity and user load. In the following, we discuss a few challenges specific to XR applications and the corresponding potential resilience provisions based on principles P1--P3.}

% Special requirements and resilience challenges
\textcolor{\hlcolor}{\textbf{System state 1 (processing challenge):}
Modern XR systems involve heterogeneous devices and envision offloading to cloud/edge servers with varying computational capabilities. Achieving full immersion is challenging due to the heavy computational demands and large required communication bandwidth, especially when 3D scene models require significant data storage and processing. Interaction among numerous users further exacerbates these demands. 
These challenges can be detected at the users by monitoring the delayed feedback from the cloud/edge serves (P2). One possible solution is to deploy local processors at the users to perform a portion of the required computations in a decreased level. For example, the local processors can provide a few 2D scene models instead of a 3D model to achieve the virtual shared experiences in a minimum level (P3).} 

\textcolor{\hlcolor}{\textbf{System state 2 (communication challenge):}
XR imposes substantial demands on communication and networking infrastructure, requiring ultra-low latency, ultra-high bandwidth, and precise synchronization of multiple data streams. Holographic communication, for instance, demands bandwidths ranging from 100 Gb/s to 1 Tb/s, far exceeding current capabilities \textcolor{\hlcolor}{\cite{hu2020cellular}} (P1). Additionally, ultra-low latency is critical to prevent cyber-physical sickness, particularly in scenarios involving haptic feedback and head-mounted displays (P1). XR use-cases also necessitate stringent synchronization of various data streams, as misaligned data can degrade the quality of rendered images and videos, complicating data communication, networking, and fusion processes. The diverse modalities involved (audio, video, haptic) further challenge data multiplexing and resource allocation. 
In case of bandwidth limitations, the link quality monitors at the users detect this challenge based on the delayed acknowledgements from the clouds (P2). The local processors at the users can perform preprocessing algorithms and the results with a reduced rate can be transmitted over the available bandwidth (P3). New/less complex preprocessing algorithms should be performed by the local processors, and diverse communication links can be deployed between users and clouds (P3). Also, the cloud can apply new strategies to process the preprocessed data (P3).}

\textcolor{\hlcolor}{\textbf{System state 3 (security and privacy challenges):}
Moreover, XR introduces new security and privacy concerns. AR and VR devices collect extensive personal data, including physical movements, eye tracking, and voice recordings, which can be used to track, identify, or infer private user attributes. The integration of multiple modalities in XR applications heightens security and privacy risks. Given XR's reliance on wireless communication, protecting against eavesdropping, data breaches, and malicious attacks is particularly challenging. 
Since the security challenges are very critical, a minimum security level must be ensured using coding and encryption techniques (P1). Additional secure and resilient operational modes can be activated when an active eavesdropper is detected using ML methods (P2). In future, adapted privacy-enhancing technologies and post-quantum crypto schemes shall be developed and integrated to ensure security and privacy for XR applications and their users (P3).}

\section{Future Research Directions}
\label{sec:future-research-directions}
By introducing the RBD concept, our work provides a framework for building resilient 6G networks, thus establishing a basis for future 6G development and standardization.
To apply our RBD concept to 6G networks, future work must investigate ways of implementing each of the three main principles (P1--P3) of the RBD framework.
That is, resilience should be considered at an early design phase such that the network is protected against critical or frequent challenges, as well as strategic adversaries.
Moreover, techniques to make the network self-aware and enable it to reconfigure itself have to be explored.
For example, this might be the development of a monitoring framework for 6G networks.
Regarding the ability of the network to reconfigure itself, multiple operation modes might be defined for the system, which can be activated depending on the overall system state.
However, reconfiguration and changes can often only be exerted by a working control infrastructure (e.g., controllers that coordinate reconfiguration actions and communicate their decisions as commands to their controlled resources). 
Future networked systems need autonomic management \cite{hutchison2023importance}, zero or at least
automated configuration to be more resilient and  avoiding configuration
mistakes introduced by human operators. Providing a resilient zero-touch control plane 
connectivity as base for autonomic management and control of a 6G network together with the necessary
security is challenging though. For instance, security often requires some form of configuration,
e.g., provisioning of certificates, which can only be automated to a certain extent.
Consequently, all future work has to consider the trade-offs that come with a resilient design, e.g., employing an additional monitoring system for self-awareness results in increased overhead and various dependencies on it. In the following, we discuss some of these future research challenges in more details.

\textbf{CAD tools for resilient electronic systems:} Currently, there is no appropriate support for the computer-aided design (CAD) tools for resilient electronics systems. Existing design tools for integrated circuits are focused on achieving a compromise between chip area, performance, and power consumption. Some commercial simulation tools can analyze fault effects, but they consider only individual fault types (transient or permanent). In order to enable a cost-effective and time-efficient resilience-by-design process, new software design tools that will consider resilient requirements from the early design phases are required. The main challenge in design of CAD tools is to achieve low runtime (fast analysis) for very complex designs consisting of millions of components. Another challenge is to take into account the physical effects of faults, which requires to adopt a cross-layer simulation approach.

{\textbf{Strategic threats and security services:} Resilience towards strategic threats is impossible without prior protection of each component, and the cooperating system as a whole.
The protection mechanisms have to be tailored to each component and on every layer as part of the initial design, as security is not merely a feature but a fundamental property. 
Achieving the different security objectives, mainly availability but as secondary goals also integrity, confidentiality, and controlled access, will require both design and implementation of suitable security services, including authentication and access control as well as investigations about their compositions when considering the overall system as a whole.
Currently, a pressing security concern is the complexity of upcoming 6G networks, which involve assemblies of intricate components.
These are drastically expanding the attack surface over simpler, traditional networks.
Data origin and peer authentication, but also comprehensive access control will have to be designed and developed, to prevent malicious interference in the increasingly open and distributed infrastructures.} 

\textbf{Anomaly detection in 6G:}
Anomaly detection in 6G will probably have to leave familiar paths. Due to the real-time requirements and highly dynamic conditions, a constant operating mode cannot be assumed.  Anomaly detection will therefore be a tightrope walk: false positives are unacceptable, while at the same time some \emph{normal states} will hardly last long enough to be suitable as a reference. Currently, ML-based research approaches that rely on prediction and that can independently change abstraction levels appear promising.

\textcolor{\hlcolor}{\textbf{Integrating resilient modes into networks:}}  %Design of the resilient operational modes: 
One of the principles for the proposed RBD framework is reconfiguration capability, which includes designing resilient multi-operational modes, specifically to address possible challenges and improve the metric of system resilience in the face of such challenges. These modes can evolve or be discontinued based on updated failure scenarios and system metrics. For instance, when a more general mode is introduced through system evolution, older modes may be eliminated. Key questions for future research include: How should operational modes be designed? How many modes are necessary to achieve a desired level of resilience? Which modes can evolve over time, and which ones should be discontinued? To address these questions, the application requirements should be taken into account. \textcolor{\hlcolor}{This approach is particularly relevant as we transition from 5G to 6G. In this evolution, not all components of the 6G network will provide the same level of resilience. Therefore, we must achieve resilience across the network, even if some components are not inherently resilient. This transition naturally involves upgrading certain parts of the network to meet higher resilience standards, while other parts may remain less resilient.}

\textcolor{\hlcolor}{\textbf{Scalability and modularization:} The 6G communication network is very complex including various interconnected and interdependent components. Therefore, the set of possible failures, attacks, and other disturbances that could affect this network is also complex. This makes the scalability of RBD approaches quite challenging. In this paper, our approach is to simplify the complex failure scenarios by considering possible failures within different layers and providing realizations of enabling principles P1--P3 for each layer. This partially addresses scalability challenge as the failure sets relevant for each layer is more limited. Nonetheless, due to interdependency of different layers scalability of RBD 6G remains an open challenge. 
One possible solution to overcome this issue is based on functional modularization, where the 6G network is broken down into a system of subsystems (i.e., modules) based on their functionalities, where the modules operate relatively independent of each other.
For example, computational resources can be a functional module which, regardless of the failures/threads it is affected by, can operate in several functional states (e.g., high, medium, low). Note that each functional module may still span multiple network layer; however, due to their relative functional independence, other modules need to adapt to the functional status of the modules and not to the specific failures that they are affected by.
Furthermore, the modules can be empowered with additional operational modes (including the the enabling principles proposed in this paper) to allow reconfigurability in the face of possible failure scenarios. The proper level of modularization, the design of resilient modes for each module, the distributed and automatized reconfiguration of different modules, and the characterization of end-to-end resilience are open research problems for future research.}

\textbf{Measuring end-to-end resilience:} 
The proposed solutions in the previous studies to enhance the resilience of systems, usually focus on one or more metrics such as availability, reliability, security, robustness, etc. We clarified the differences between resilience and other related terms and investigated a general framework for designing resilient communication networks. However, there are other questions that need to be discussed in future studies, e.g., what is the set of relevant resilience metrics for the end-to-end system? Can a general set of metrics be proposed, or does it depend on the application of the considered system? How can we maximize these metrics? Thus, a potential future research direction could focus on measuring the resilience of systems.

\textbf{Long term availability of the network services:} 
To have resilience in the network, the network should be able to mitigate failures and continue to provide services to the end users. From the software core network perspective in 6G, the long-term availability of the network services is very essential. The resource usage changes depending on the operation load in the system.
In the face of insufficient and unevenly distributed resources among network components, they may fail or perform poorly. 
Finding the optimal allocation of resources for the network services based on the requirements is very challenging. Also in case of insufficient resources for the network operations, how to provision resources dynamically in an efficient way is an open question. Defining system stages with strange attractors and how close the system is to the attractors that predict failures, and taking required actions can enhance the system reliability.

\textbf{Emergency/backup communication networks:} 
It is important to emphasize that in RBD, we design systems that can continue functioning despite partial challenges/failures. However, it is impossible to forecast and account for every possible challenge. When a challenge is very complex and has a wide-ranging impact on the communication system, it could potentially lead to a complete system outage. 
Different studies in the literature have discussed these complex challenges and proposed emergency/backup networks (see Section \ref{sec: Introduction_Review papers}). For concreteness, we did not explicitly discuss these backup networks in this paper and refer interested readers to the provided references. However, we note that the emergency/backup networks can be explained as resilience-enabling ``operational modes'' of the entire communication networks that are activated when the existing network is extremely damaged. Moreover, the design of an interface between emergency networks and the existing (still functioning part of the) infrastructure is naturally a part of the proposed RBD concept.

\textbf{Social and legal aspects of resilient design:} 
We studied various technical aspects to embed resilience into communication system design. We discussed resilience principles (P1--P3), including protective design measures (default considerations in the design), self-awareness capability (including sensing, monitoring, prediction, etc.), and reconfiguration capability (including the design of multi-operational modes). However, determining the legal and social aspects of protective measures considered default in the design, as well as the type of data legally permitted to be sensed and monitored, introduce challenges that require regulation or legislation. Furthermore, the introduction of new operational modes requires alignment with specific standards and regulations. Hence, the social and legal aspects of this resilient design are as essential as the technical aspects and require further studies.

\textbf{Resilience-by-default versus RBD:}
In the recent literature of security, a distinction has been made between security-by-design and security-by-default \cite{chattopadhyay2020autonomous, casola2016security, simoiu2020secure}. While security-by-design means that the security configurations should be embedded in system design, security-by-default means that the minimum security levels should be provided by default without additional configuration. A similar analogy is expected for resilience, i.e., RBD, which is studied in this paper and resilience-by-default, where any parts of the 6G system should have resilience features enabled by default such that no action is required to benefit from the resilience capabilities of the system. However, given the complexity/cost trade-off, not all the services require the same degrees of resilience. A potential solution would be to establish a few levels of resilience where services assigned to each level will ``by default'' feature a minimum predefined resilience guaranteed by the 6G communication network.

\section{Conclusion}
Communication networks constitute a critical infrastructure that is the basis for many critical services in a digital society. Therefore, resilience has been considered a driving force in the design of 6G communication networks. This paper has introduced a comprehensive framework embedding resilience concepts into the design of 6G communication networks. After reviewing the background on resilience concepts, definitions, and approaches, we have introduced the proposed holistic RBD concept for 6G communication networks. This concept accounts for physical and cyber resilience across all layers of the communication systems, addressing electronics, physical channel, network components and functions, networks, and services. Additionally, it considers cross-layer and cross-infrastructure interactions. Three different enabling principles, namely protective design measures, self-awareness capabilities, and reconfiguration capabilities are discussed in detail and further exemplified from various perspectives and layers in 6G networks. In addition, the trade-off between complexity, efficiency, and resilience is discussed. The proposed RBD concepts are explained along several concrete 6G use-cases. This paper is concluded by identifying   
various open problems for future research on 6G resilience.

\section*{Acknowledgment}
The authors would like to thank Prof. Stefan Götz, Prof. Anja Klein, Prof. Milos Krstic, Prof. Thomas Magedanz, Prof. Steffen Paul, Prof. Marius Pesavento, Prof. Björn Scheuermann, Prof. Hans Schotten, Prof. Haya Shulman, Prof. Ralf Steinmetz, Dr. Markus Ulbricht, Dr. Fabian Vargas, Prof. Michael Waidner, and Dr. Lara Wimmer for their suggestions and comments.

%\printbibliography

\begin{IEEEbiography}[{\includegraphics[width=1in,height=1.25in,clip,keepaspectratio]{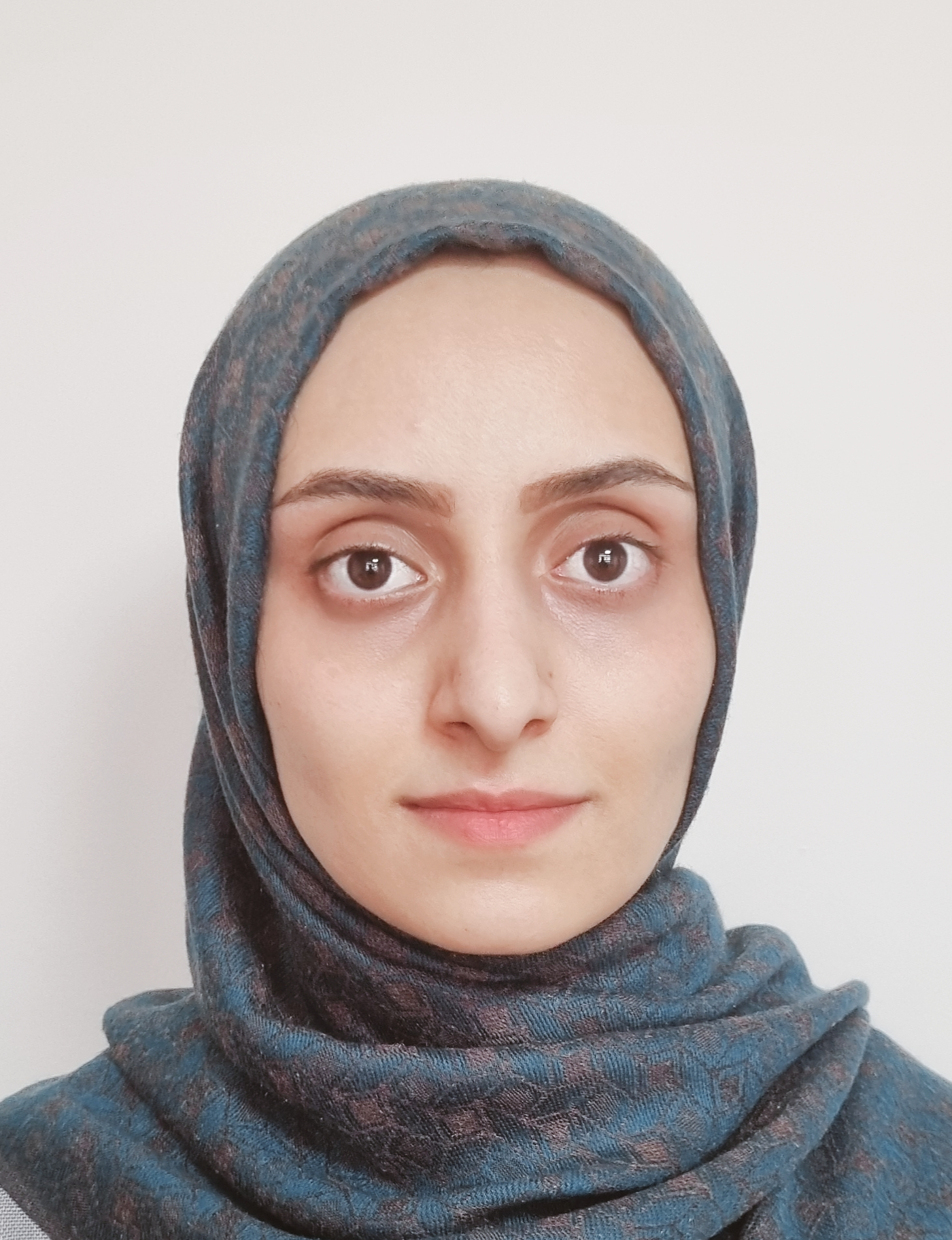}}]{Ladan Khaloopour}
received the B.Sc., M.Sc., and Ph.D. degrees in electrical engineering from Sharif University of Technology (SUT), Tehran, Iran, in 2015, 2017, and 2022, respectively. In 2022, she joined the Resilient Communication Systems Group at the Institute of Telecommunications, Technical University of Darmstadt (TUDa) as a researcher.
Her research interests include molecular communications, information theory, and resilient communication networks.
\end{IEEEbiography}
\begin{IEEEbiography}[{\includegraphics[width=1in,height=1.25in,clip,keepaspectratio]{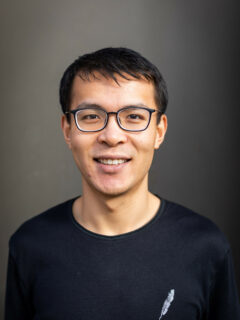}}]{Yanpeng Su}
(S'22) received the B.Sc. degree in electrical engineering and automation from the China University of Petroleum (East China) (UPC), Qingdao, China, in 2018 and the M.Sc. degree in electrical engineering from the Friedrich-Alexander-Universität Erlangen-Nürnberg (FAU), Erlangen, Germany, in 2022, respectively. In 2022, he joined the Chair of Electrical Smart City Systems, FAU, as a research assistant. His current research interests include channel modeling and waveform design for joint communications and radar sensing and coordinated multipoint radar sensing system design.
\end{IEEEbiography}
\begin{IEEEbiography}[{\includegraphics[width=1in,height=1.25in,trim={2cm 0 2cm 0},clip,keepaspectratio]{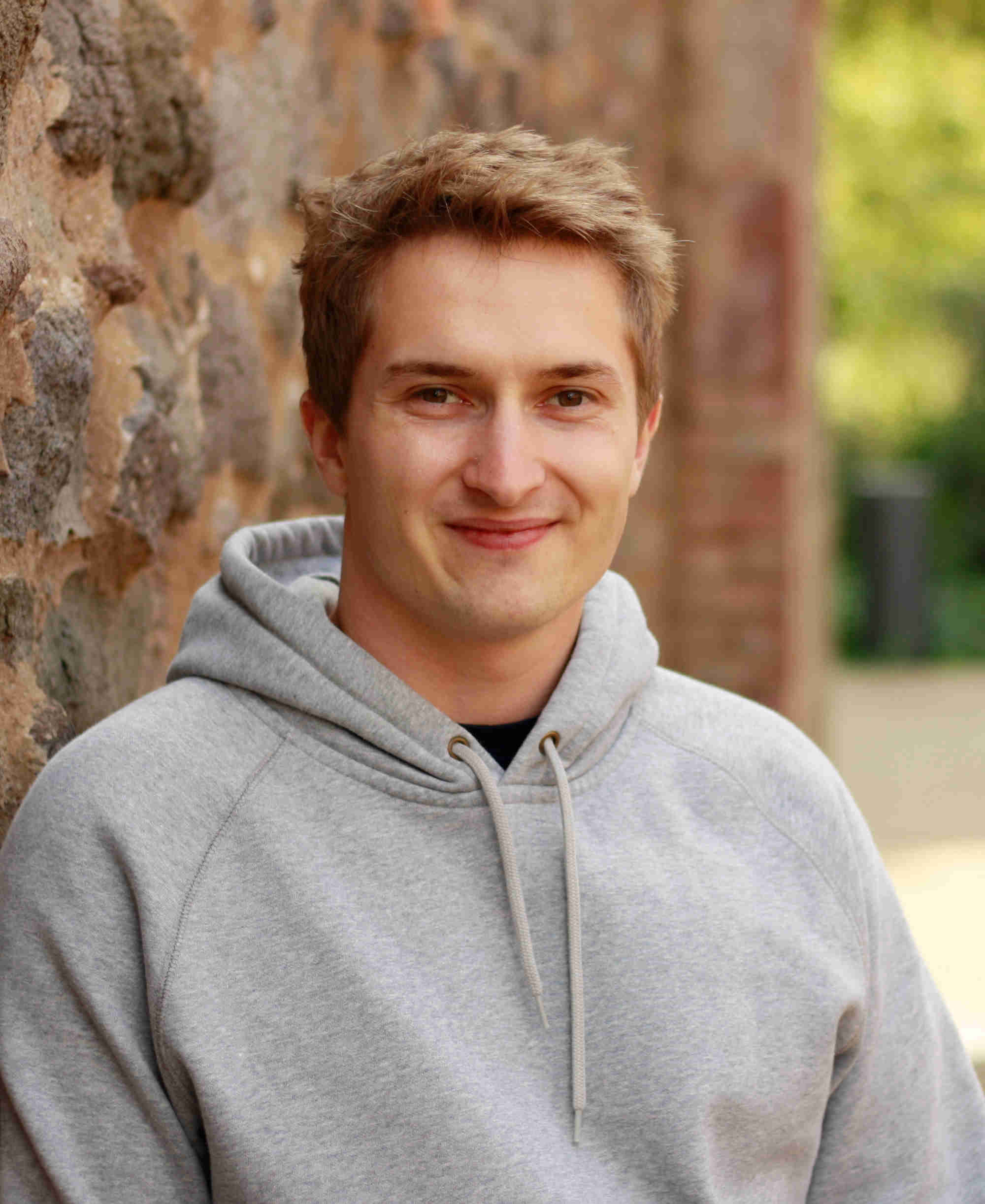}}]{Florian Raskob}
received his B.Sc. in "Cognitive Informatics" from the University of Bielefeld in 2019 and his M.Sc. degree in "Computer Science" with focus on model-based system development from the Humboldt University of Berlin (HU Berlin) in 2022. Since 2023, he is working on his Ph.D. degree with focus on the performance of 6G networks at the Communication Networks Lab at Technical University of Darmstadt (TUDa).
\end{IEEEbiography}
\begin{IEEEbiography}[{\includegraphics[width=1in,height=1.25in,trim={14cm 2cm 12cm 5cm},clip,keepaspectratio]{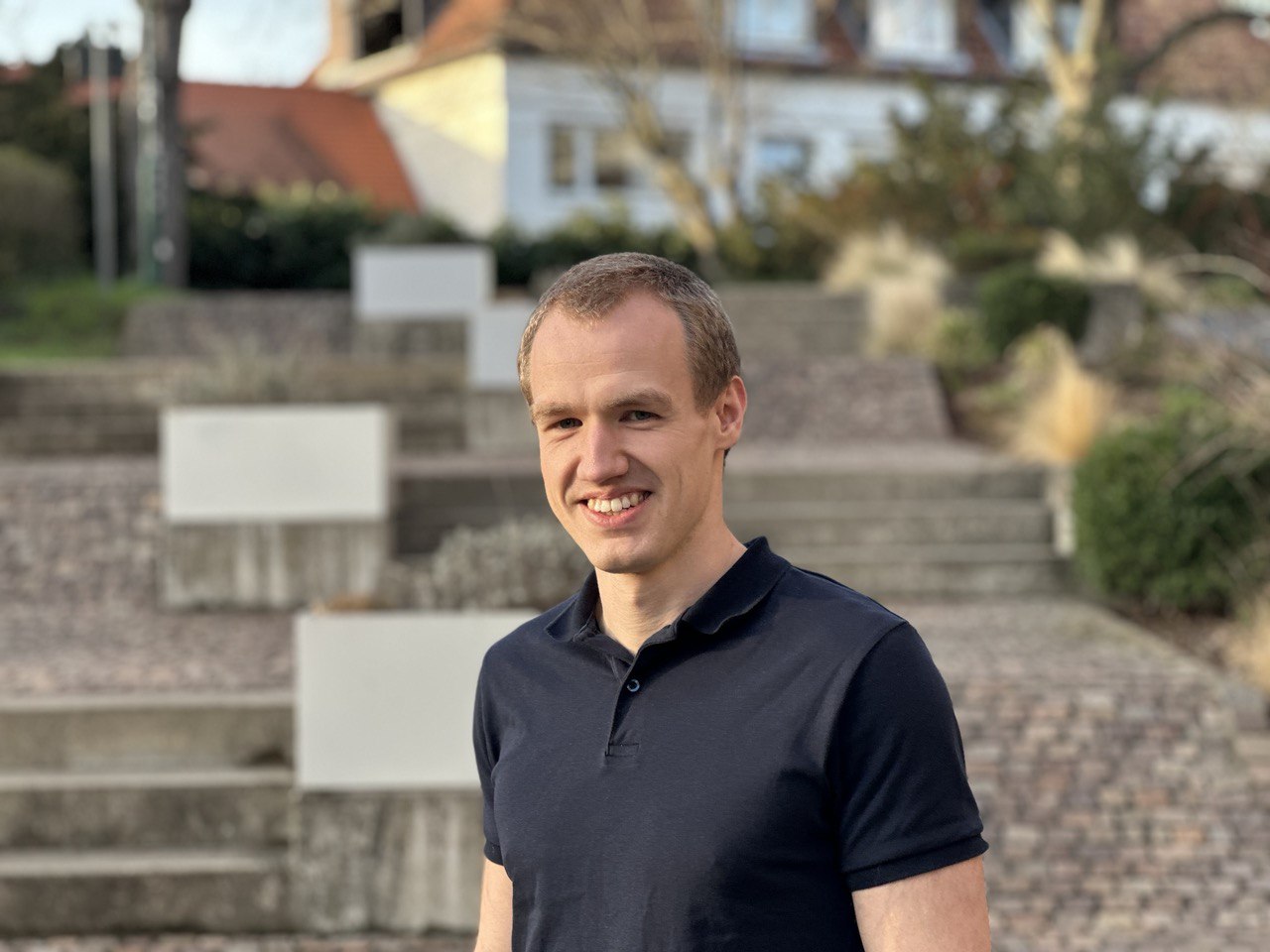}}]{Tobias Meuser} 
graduated with his Ph.D. in electrical engineering and information technology at the beginning of 2020, after which he became the head of the "Adaptive Communication Systems" group within the Multimedia Communications Lab at TU Darmstadt. With the start of the third phase of the CRC MAKI, he became a principal investigator in project B1 - Monitoring and Analysis and later became part of the Athene Young Investigator Program for independent researchers at TU Darmstadt. Since mid 2023, he is head of the "Adaptive Communication Systems" group within the chair of communication networks at TU Darmstadt. His current research interest is centered around approximation techniques for communication networks to increase the resilience of these networks, especially for the mobile communication standard 5G. 
\end{IEEEbiography}
\begin{IEEEbiography}[{\includegraphics[width=1in,height=1.25in,clip,keepaspectratio]{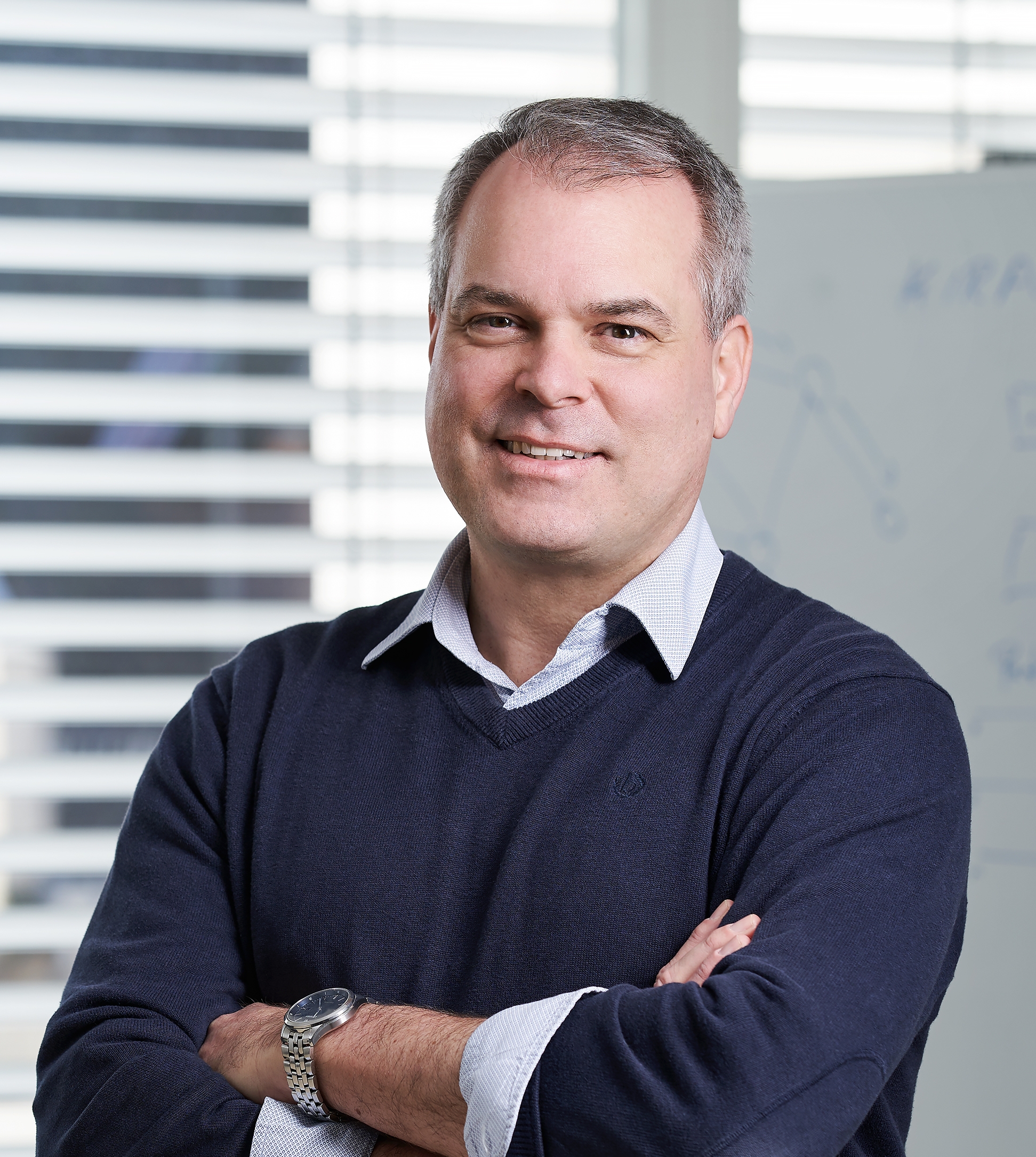}}]{Roland Bless}
is an Assistant Professor and senior researcher at Karlsruhe Institute of Technology (KIT) in the group of Prof. Zitterbart at the Institute of Telematics. He received his  diploma in Computer Science at the University of Karlsruhe in 1996 and got his PhD (Dr.-Ing.) in the realm of Quality of Service Management in 2002. In 2009 he finished his Habilitation at the Computer Science Department of Informatics at KIT. Since 1998 he contributes to Internet Standardization in the IETF. His current fields of research include autonomous control planes for 6G and future networks, scalable zero-touch routing protocols, and congestion control. Dr.~Bless is member of Gesellschaft für Informatik, ACM SIGCOMM, IEEE ComSoc (IEEE Member since 1997), and ISOC.
\end{IEEEbiography}
\begin{IEEEbiography}[{\includegraphics[width=1in,height=1.25in,clip,keepaspectratio]{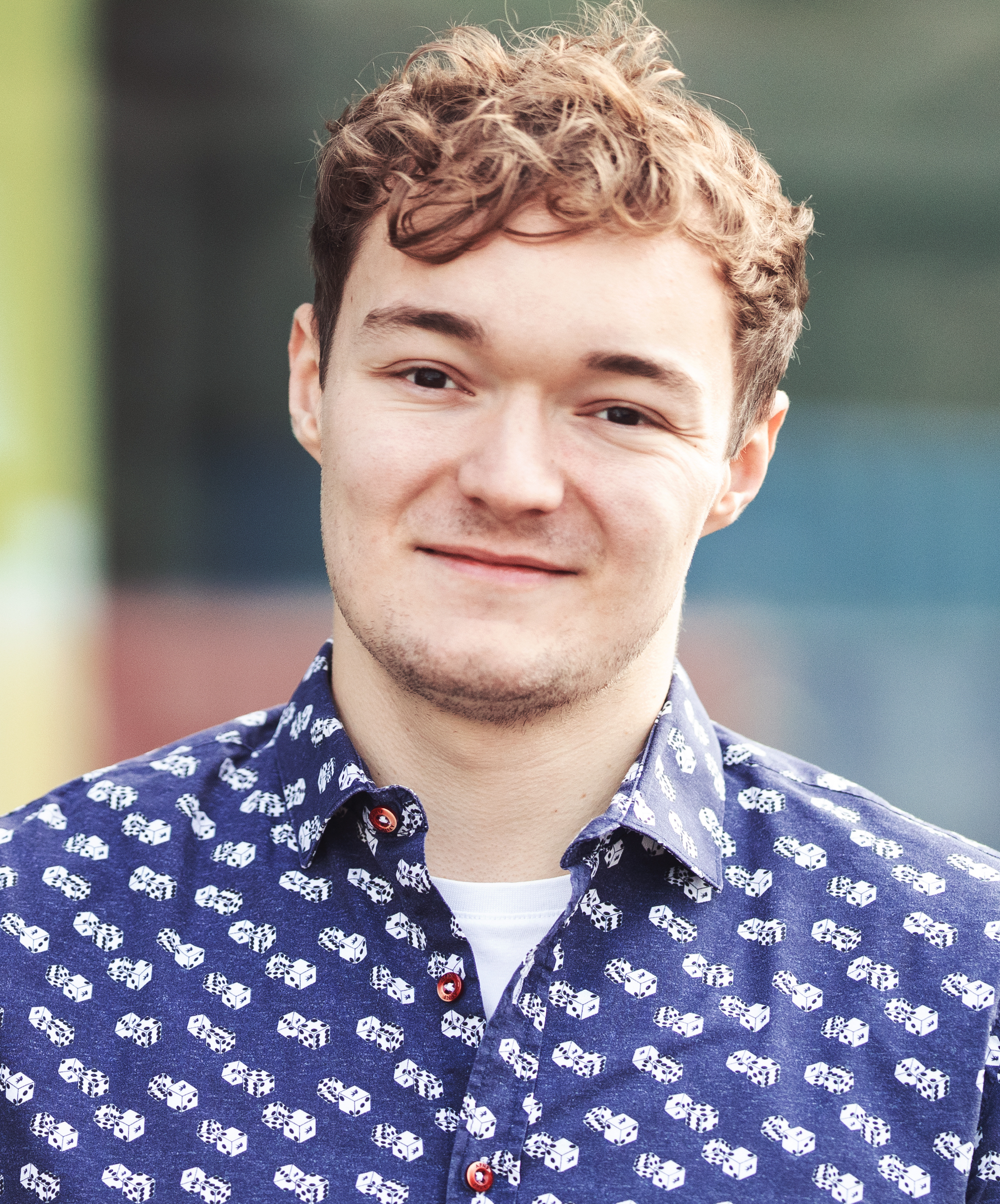}}]{Leon \textcolor{\hlcolor}{Janzen}}
completed his B.Sc. and M.Sc. degrees in computer science and IT security (2019, 2021) at the Technical University of Darmstadt (TUDa). He is currently pursuing the Ph.D. degree in computer science at TUDa. His research focuses on the resilience and security of future cellular networks. He received the Best Paper Award at the ACM CHI Conference on Human Factors in Computing Systems in 2023.
\end{IEEEbiography}
\begin{IEEEbiography}
[{\includegraphics[width=1in,height=1.25in,clip,keepaspectratio]{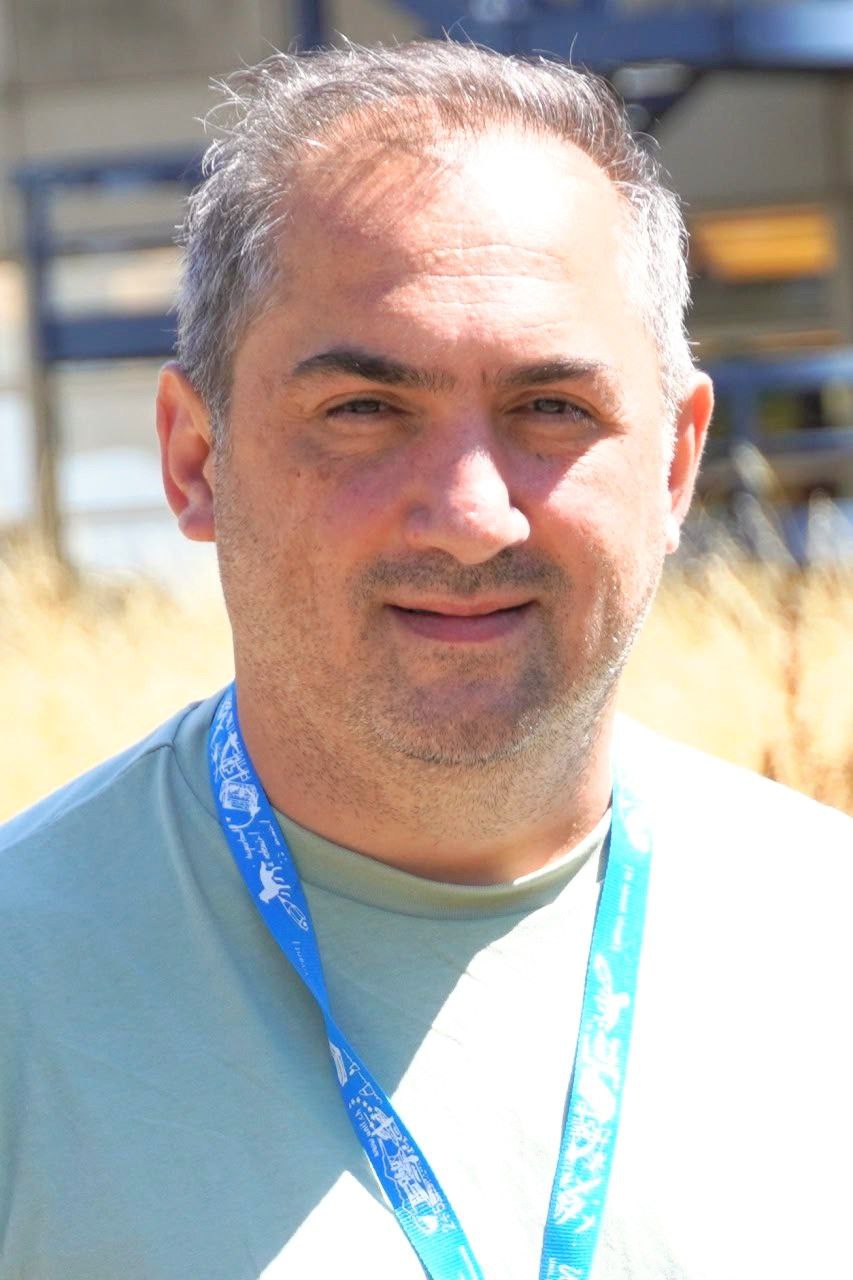}}]
{Kamyar Abedi} is Ph.D. researcher at Karlsruhe Institute of Technology (KIT) under the supervision of Prof. Strufe at the Practical IT Security group.
He earned his master's degree in data network and security from Birmingham City University (2011,2012). Since then, he has been involved in many mobile network security projects from 2G to 4G. He is currently working on 5G networks and beyond with a focus on privacy and security. 
\end{IEEEbiography}
\begin{IEEEbiography}[{\includegraphics[width=1in,height=1.25in,clip,keepaspectratio]{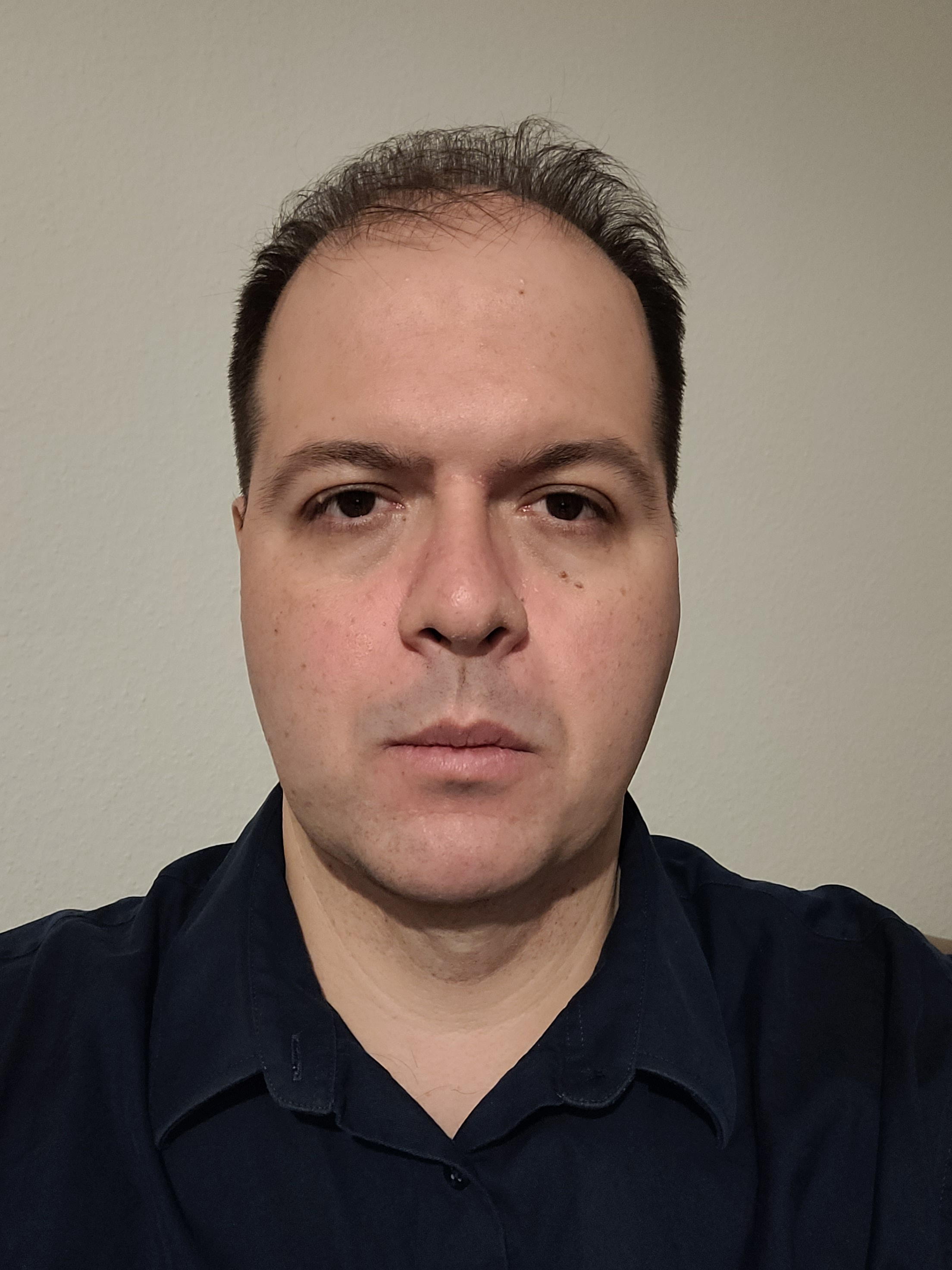}}]{Marko Andjelkovic} received his Dipl.-Ing. degree in Electronics from the Faculty of Electronic Engineering, University of Nis, Serbia, in 2008. In 2021, he received a Dr.-Ing. degree from the University of Potsdam, Germany, with the thesis entitled “ A Methodology for Characterization, Modeling, and Mitigation of Single Event Transient Effects in CMOS Standard Combinational Cells”. From 2010 to 2016 he was working as a research associate at the University of Nis. Since 2016, he is with  IHP - Leibniz Institut für innovative Mikroelektronik, where he is employed as a research associate at the System Architectures Department. His current research is focused on characterization and modeling of fault effects in digital electronic circuits, on-chip fault sensors, and design of fault-tolerant circuits and systems. He has published over 70 journal and conference papers.
\end{IEEEbiography}
\begin{IEEEbiography}[{\includegraphics[width=1in,height=1.25in,clip,keepaspectratio]{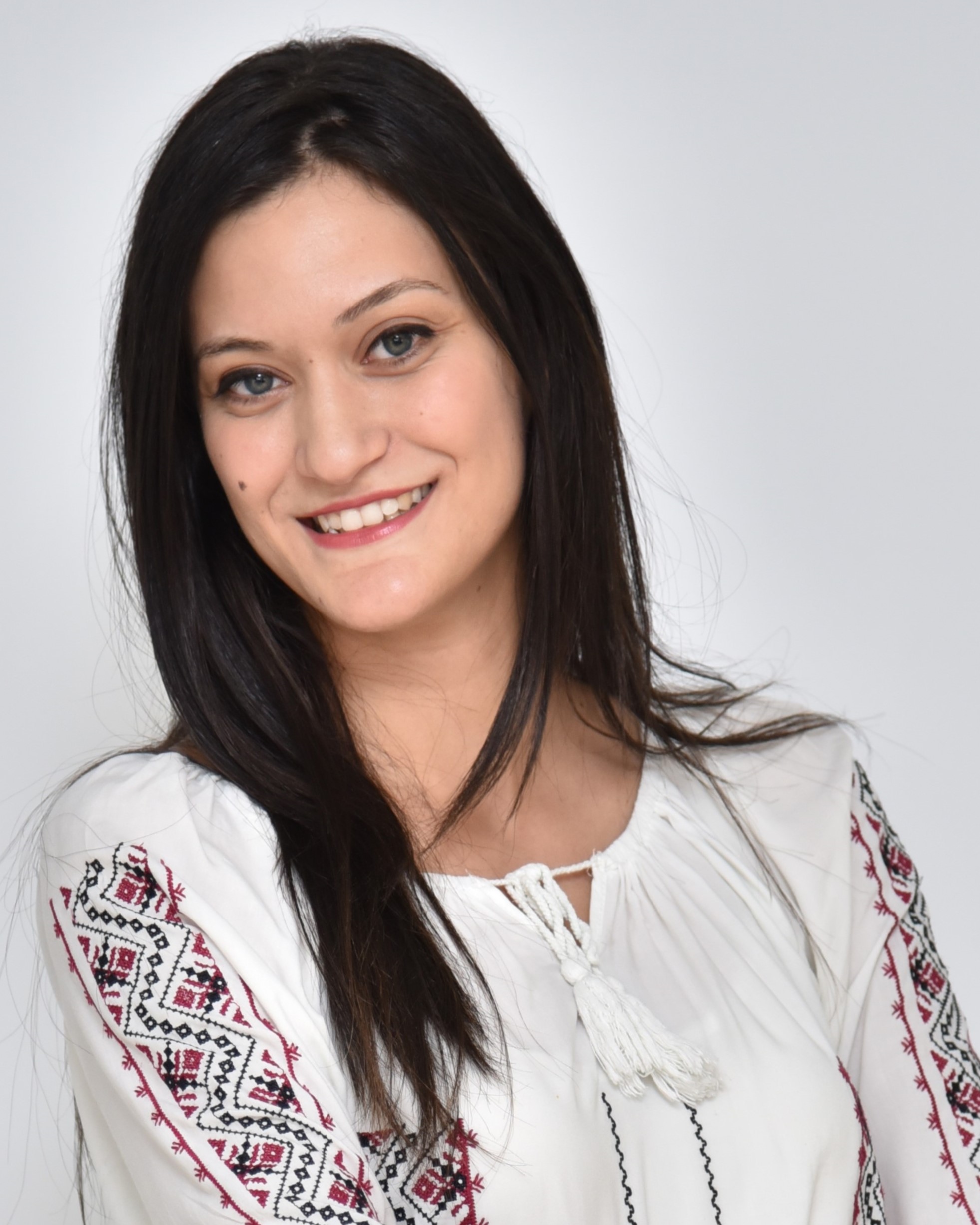}}]{Hekma Chaari}
is graduated with a PhD degree in Telecommunications in April 2017. The topic of her research activities focused on the video quality transmission evaluation over radio communication networks. Before that, she obtained her master’s degree in New Technologies for dedicated systems and her engineering’s degree in Electronics in the National School of Engineers in Sfax – Tunisia.
From 2010 to 2019, she worked as an academic assistant and the telecommunications department coordinator at the National School of Engineers in Telecommunications and Electronics Sfax – Tunisia and in the International Institute of Technologies also in Sfax.
Since January 2024, she joined the Chair of Electrical Smart City Systems as a researcher in trustworthiness, resilience and real-time.
\end{IEEEbiography}
\begin{IEEEbiography}
[{\includegraphics[width=1in,height=1.25in,clip,keepaspectratio]{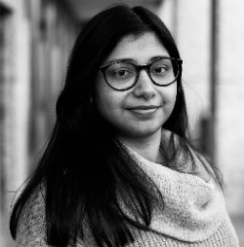}}]{Pousali Chakraborty} is a researcher at the Fraunhofer FOKUS Institute in the Software-based Networks department since 2020. She completed her Master’s degree in Computer Science from the Technical University of Berlin, Germany in 2019. She is working with the Open5GCore toolkit to develop different features within software 5G core networks. Her research work is in the direction of private networks, roaming support and anomaly detection in software core networks to improve resilience.
\end{IEEEbiography}
\begin{IEEEbiography}
[{\includegraphics[width=1in,height=1.25in,clip,keepaspectratio]{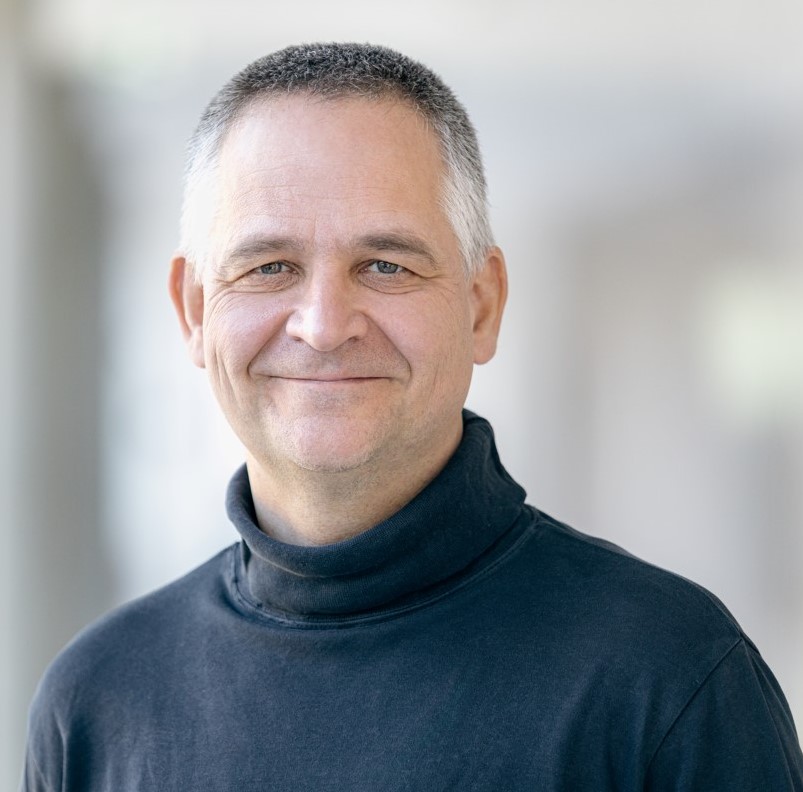}}]{Michael Kreutzer}
earned his diploma in computer science at Saarland University and in 2006 he completed his doctorate at the Albert-Ludwigs-University of Freiburg with a thesis on “Scalable and robust service discovery in mobile ad hoc networks”. From 2008 to 2015, he was Managing Director of the Darmstadt Center for Advanced Security Research (CASED), and from 2011 to 2015 he was Research Coordinator of the BMBF Competence Center European Center for Security and Privacy by Design (EC SPRIDE). Michael Kreutzer has been a scientist at Fraunhofer SIT since the end of 2015 and has been acting head of the Advanced Cryptographic Engineering department since fall 2020. 
Michael Kreutzer's areas of expertise are network security, disinformation detection, cybersecurity governance, and post-quantum cryptography. He is involved in advising start-ups on the topic of cyber security. 
\end{IEEEbiography}
\begin{IEEEbiography}
[{\includegraphics[width=1in,height=1.25in,clip,keepaspectratio]{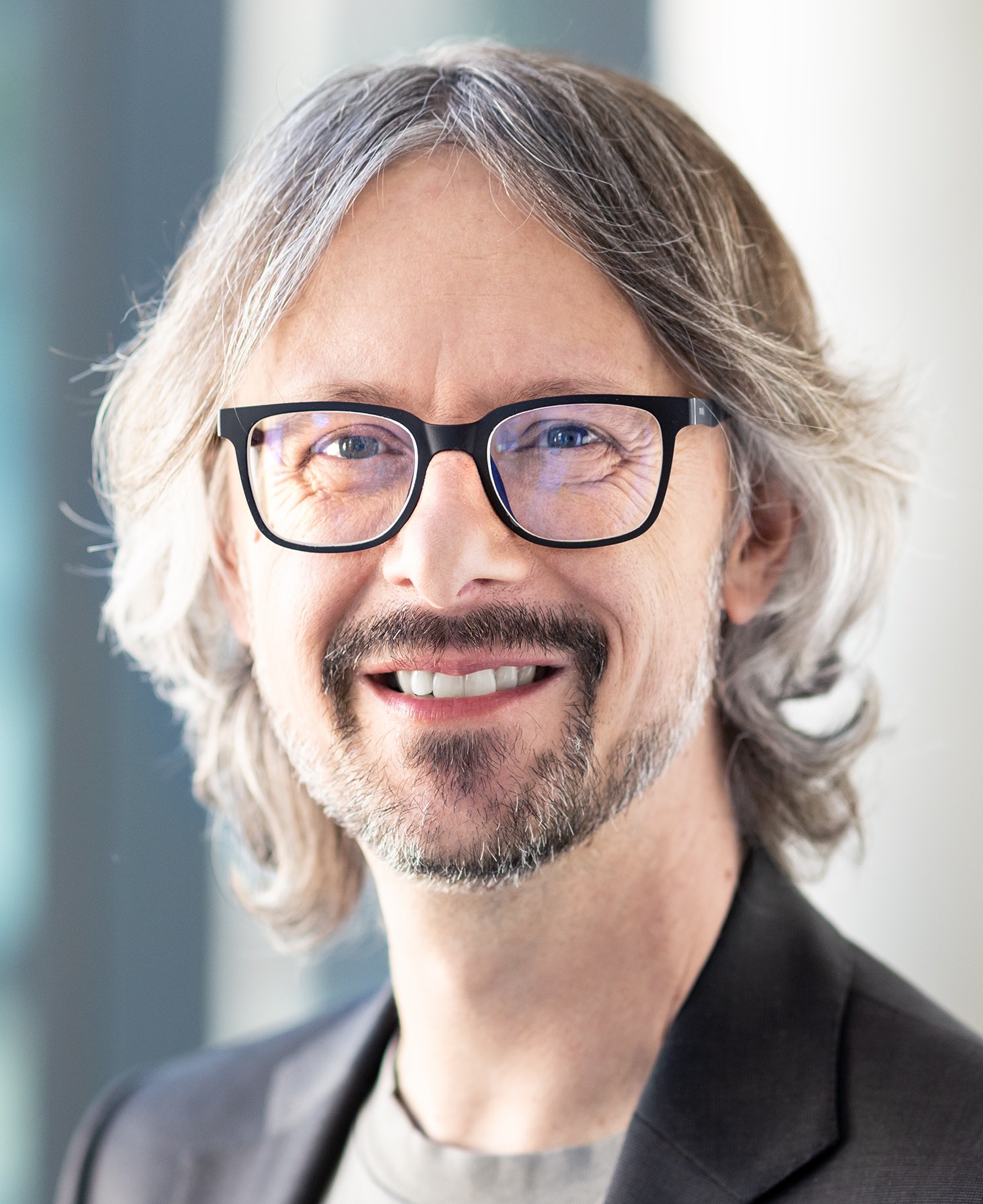}}]
{Matthias Hollick} received the Ph.D. degree from Technical University of Darmstadt in 2004. He is currently the Head of the Secure Mobile Networking Lab, Department of Computer Science, Technical University of Darmstadt, Germany. He has been researching and teaching at Technical University of Darmstadt, Universidad Carlos III de Madrid, and the University of Illinois at Urbana–Champaign. His research interests include resilient, secure, privacy-preserving, and quality-of-service-aware communication for mobile and wireless systems and networks.
\end{IEEEbiography}
\begin{IEEEbiography} 
[{\includegraphics[width=1in,height=1.25in,clip,keepaspectratio]{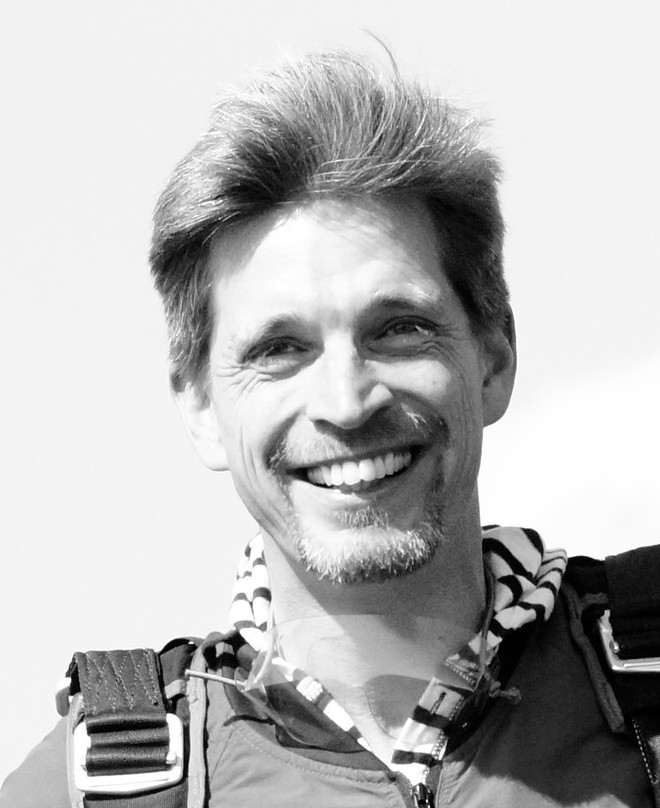}}]
{Thorsten Strufe} is Professor of Privacy and IT Security at KIT, adjunct professor for Privacy and Network Security at TU Dresden, Deputy Speaker of the Cluster of Excellence CeTI (Centre for Tactile Internet with Human-in-the-Loop), and PI in the German Competence Center for IT Security KASTEL, the Research Training Group RoSI (TU Dresden), and the 5G-Lab Germany.
His research interests lie in the areas of privacy and resilience. More recently, he has focused on studying user behavior and security of behavioral data, as collected by ubiquitous computing infrastructures and both wearables and AR devices. One of the challenges that drives him is how to create competitive  services and apps without extensive collection of personal information, thus respecting the privacy of their users. 
\end{IEEEbiography}
\begin{IEEEbiography}[{\includegraphics[width=1in,height=1.25in,clip,keepaspectratio]{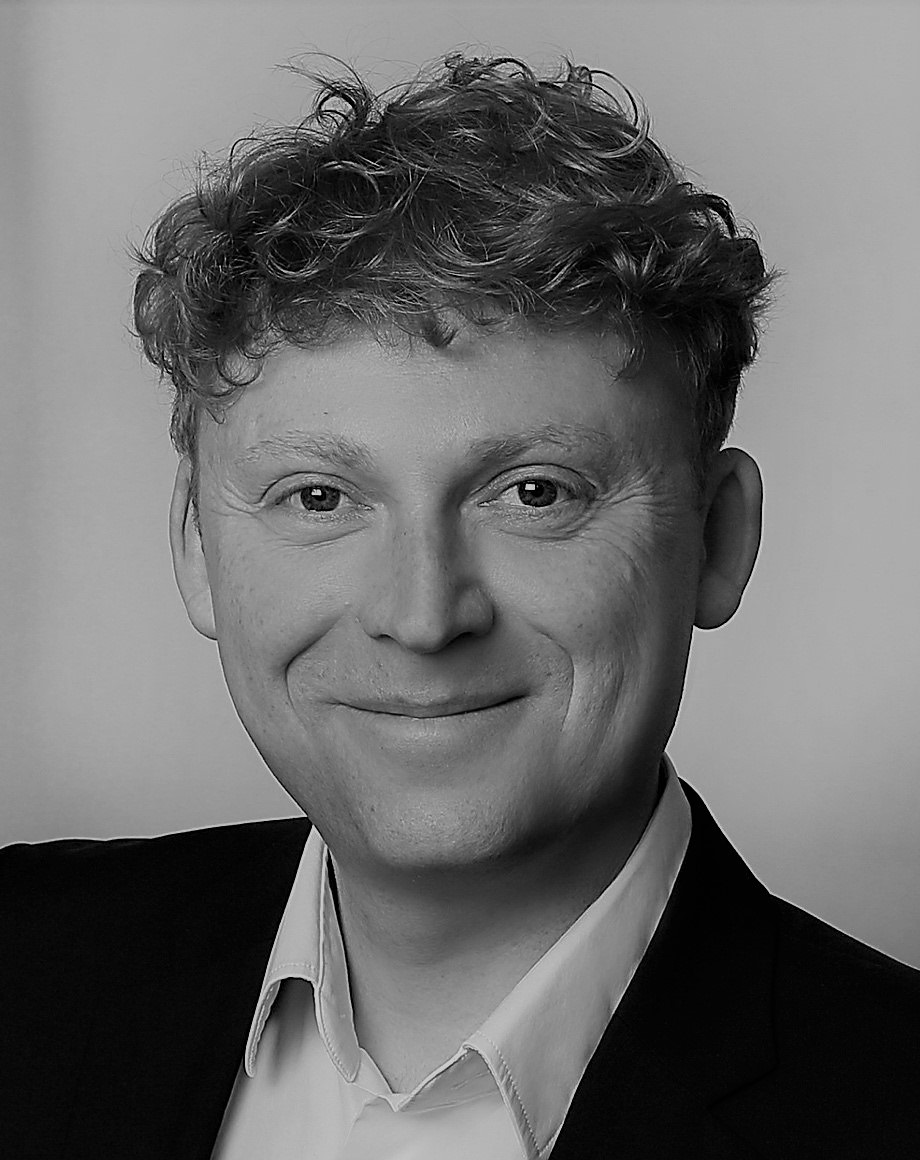}}]{Norman Franchi} (Member, IEEE) received the Dr.-Ing. (Ph.D.E.E.) and Dipl.-Ing. (M.S.E.E.) degrees in electrical, electronic and communications engineering (EEI), in 2015 and 2007, respectively. From 2007 to 2011, he was with the Automotive Research and Development Sector as a System and Application Engineer in advanced networked control system design. From 2012 to 2015, he was a Research Associate with the Institute for Electronics Engineering, FAU, focused on software-defined radio-based V2X communications. From 2015 to 2021, he was the Gerhard Fettweis’ Vodafone Chair with Dresden University of Technology (TU Dresden), where he was the Senior Research Group Leader of Resilient Mobile Communications Systems and 5G Industrial Campus Networks. From 2019 to 2020, he was the Managing Director of the 5G Laboratory GmbH, Germany. In 2020, he founded the company Advancing Individual Networks (AIN) GmbH, Germany, a technology start-up for design, optimization, and operation of the IIoT networks. He is currently a Full Professor (W3) with Friedrich-Alexander Universität Erlangen-Nürnberg, Germany, where he heads the Institute for Electrical Smart City Systems. His research interests include 6G, joint communications and sensing, resilient and secure systems, the IIoT, open RAN, V2X, and green ICT for smart cities. He is a member of the IEEE ISAC Initiative, Open 6G Hub Germany, 5G Laboratory Germany, and 6G Platform Germany. Furthermore, he is an Advisory Board Member of the Industrial Radio Laboratory Germany (IRLG) and the KI Park Deutschland.
\end{IEEEbiography}

\begin{IEEEbiography}[{\includegraphics[width=1in,height=1.25in,clip,keepaspectratio]{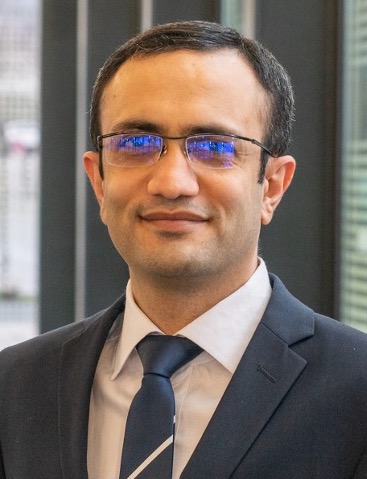}}]{Vahid Jamali} (Senior Member, IEEE) received the doctoral degree (with distinctions) from Friedrich-Alexander University Erlangen-Nürnberg (FAU) in 2019. He has been an Assistant Professor with the Technical University of Darmstadt (TUDa), since 2022, leading the Resilient Communication Systems Laboratory. Prior to joining TUDa, he held academic appointments at Princeton University (2021–2022) and FAU (2019– 2021), as a Post-Doctoral Researcher; and at Stanford University as a Visiting Researcher in 2017. His research interests include wireless and molecular communications. He is currently serving as an Associate Editor of the \textsc{IEEE Transactions on Communications}, \textsc{IEEE Communications Letters}, and \textsc{IEEE Open Journal of the Communications Society} as well as acting as a Vice-Chair for the IEEE ComSoc -- German chapter. He has received several awards for his publications including the Best Paper Awards from the IEEE ICC in 2016, the ACM NanoCOM in 2019, the Asilomar CSSC in 2020, and the IEEE WCNC in 2021; and the Best Journal Paper Award (Literaturpreis) from the German Information Technology Society (ITG) in 2020.
\end{IEEEbiography}

\EOD

\begin{thebibliography}{00}
	\bibitem{wang2023road}
	C.-X. Wang, X. You, X. Gao, et al., ``On the road to 6G: Visions, requirements, key technologies, and testbeds,'' \emph{ IEEE Communications Surveys \& Tutorials}, vol. 25, no. 2, pp. 905--974, 2023.
	
	\bibitem{Hexa-X}
	``Project Hexa-X'', Accessed date: Oct. 5, 2023. [Online]. Available: https://hexa-x.eu/.
	
	\bibitem{RISE-6G}
	``Project RISE-6G'', Accessed date: Oct. 5, 2023. [Online]. Available: https://rise-6g.eu/.
	
	\bibitem{REINDEER}
	``REINDEER --- Resilient interactive applications through hyper diversity in energy efficient RadioWeaves technology’’, Accessed date: Oct. 5, 2023. [Online]. Available: https://reindeer-project.eu/. 
	
	
	\bibitem{6GStart}
	``6GStart: Starting the sustainable 6G SNS initiative for Europe'', Accessed date: Oct. 5, 2023. [Online]. Available: https://5g-ppp.eu/6gstart/.
	
	\bibitem{6GTandem}
	``6GTandem: A dual-frequency distributed MIMO approach for future 6G applications'', Accessed date: Oct. 5, 2023. [Online]. Available: https://security-link.se/6gtandem/.
	
	\bibitem{TERA6G}
	``TERA6G'', Accessed date: Oct. 5, 2023. [Online]. Available: https: //www.hhi.fraunhofer.de/en/departments/pc/projects/tera6g.html. 
	
	\bibitem{Open6GHub}
	‘‘Open6GHub -- 6G for Society and Sustainability’’, Accessed date: Oct. 5, 2023. [Online]. Available: https://www.open6ghub.de/en/. 
	
	\bibitem{6G-RIC}
	‘‘6G-RIC -- 6G research and innovation cluster’’, Accessed date: Oct. 5, 2023. [Online]. Available: https://6g-ric.de/.
	
	\bibitem{6G-life}
	‘‘6G-life -- Digital transformation and sovereignty of future communication networks’’, Accessed date: Oct. 5, 2023. [Online]. Available: https://6g-life.de/.
	
	\bibitem{6GEM}
	‘‘6GEM -- 6G research hub for open, efficient and secure mobile communications systems’’, Accessed date: Oct. 5, 2023. [Online]. Available: https://www.6gem.de/en/.
	
	\bibitem{6G-ANNA}
	‘‘6G-ANNA -- Holistic approaches for 6th generation mobile networks’’, Accessed date: Apr. 10, 2024. [Online]. Available: https://6g-anna.de/en/.
	
	\bibitem{Oulu}
	‘‘6G flagship, key drivers and research challenges for 6G ubiquitous wireless intelligence, white paper, Sep. 2019’’. [Online]. Available: https://www.6gflagship.com/key-drivers-and-research-challenges-for-6g-ubiquitous-wireless-intelligence/.
	
	\bibitem{RINGS}
	‘‘RINGS’’, Accessed date: Oct. 5, 2023. [Online]. Available: https://beta.nsf.gov/funding/opportunities/resilient-intelligent-nextg-systems-rings.
	
	\bibitem{China_Project2019}
	‘‘Key R\&D projects 2019: Wideband communications and new networks’’, Accessed date: Oct. 5, 2023. [Online]. Available: https://service.most.gov.cn/u/cms/static/201812/12164952skqa.pdf.
	
	\bibitem{Space_Japan}
	‘‘Research and development on satellite-terrestrial integration technology for beyond 5G’’, Accessed date: Oct. 5, 2023. [Online]. Available: https://www2.nict.go.jp/spacelab/en/pj\_stit.html.
	
	\bibitem{KT6G}
	‘‘KT and Hanwha systems are jointly developing 6G quantum cryptography technology’’, Accessed date: Oct. 5, 2023. [Online]. Available: https://www.ajudaily.com/view/20220712111419400.
	
	\bibitem{de2021survey}
	C. De Alwis, A. Kalla, Q.-V. Pham, et al., “Survey on 6G frontiers: Trends, applications, requirements, technologies and future research,” \emph{IEEE Open Journal of the Communications Society}, vol. 2, pp. 836–-886, 2021.
	
	\bibitem{alsabah20216g}
	M. Alsabah, M. A. Naser, B. M. Mahmmod, et al., “6G wireless communications networks: A comprehensive survey,” \emph{IEEE Access}, vol. 9, pp. 148191--148243, 2021.
	
	\bibitem{cassottana2023resilience}
	B. Cassottana, M. M. Roomi, D. Mashima, and G. Sansavini, “Resilience analysis of cyber-physical systems: A review of models and methods,” \emph{Risk Analysis}, 2023.
	
	\bibitem{gribble2001robustness}
	S. D. Gribble, “Robustness in complex systems,” \emph{in Proceedings Eighth Workshop on Hot Topics in Operating Systems, IEEE}, Elmau, Germany, 2001, pp. 21–26.
	
	\bibitem{nistCSF}
	NIST, The NIST cybersecurity framework, https://www.nist.gov/cyberframework.
	
	\bibitem{smith2011network}
	P. Smith, D. Hutchison, J. P. Sterbenz, et al., “Network resilience: A systematic approach,” \emph{IEEE Communications Magazine}, vol. 49, no. 7, pp. 88--97, 2011.
	
	\bibitem{maklachkova2022analysis}
	V. Maklachkova, A. Shvedov, and S. Alyev, “Analysis of resilience indicators in corporate networks and possible ways to improve it,” \emph{in Systems of Signals Generating and Processing in the Field of on Board Communications, IEEE}, 2022, pp. 1--5.
	
	\bibitem{pajic2017design}
	M. Pajic, J. Weimer, N. Bezzo, O. Sokolsky, G. J. Pappas, and I. Lee, “Design and implementation of attack-resilient cyberphysical systems: With a focus on attack-resilient state estimators,” \emph{IEEE Control Systems Magazine}, vol. 37, no. 2, pp. 66--81, 2017.
	
	\bibitem{yang2020adversary}
	Z. Yang, A. Gang, and W. U. Bajwa, “Adversary-resilient distributed and decentralized statistical inference and machine learning: An overview of recent advances under the Byzantine threat model,” \emph{IEEE Signal Processing Magazine}, vol. 37, no. 3, pp. 146--159, 2020.
	
		\bibitem{kantarci2015resilient}
	B. Kantarci and H. T. Mouftah, “Resilient design of a cloud system over an optical backbone,” \emph{IEEE Network}, vol. 29, no. 4, pp. 80--87, 2015.
	
	\bibitem{sterbenz2017smart}
	J. P. Sterbenz, “Smart city and IoT resilience, survivability, and disruption tolerance: Challenges, modelling, and a survey of research opportunities,” \emph{in 9th International Workshop on Resilient Networks Design and Modeling (RNDM)}, IEEE, Alghero, Italy, 2017, pp. 1--6.
	
	\bibitem{hutchison2018architecture}
	D. Hutchison and J. P. Sterbenz, “Architecture and design for resilient networked systems,” \emph{Computer Communications}, vol. 131, pp. 13--21, 2018.
	
	\bibitem{pradhan2018chariot}
	S. Pradhan, A. Dubey, S. Khare, et al., “Chariot: Goal-driven orchestration middleware for resilient iot systems,” \emph{ACM Transactions on Cyber-Physical Systems}, vol. 2, no. 3, pp. 1--37, 2018.
	
	\bibitem{benkhelifa2018critical}
	E. Benkhelifa, T. Welsh, and W. Hamouda, “A critical review of practices and challenges in intrusion detection systems for IoT: Toward universal and resilient systems,” \emph{IEEE Communications Surveys \& Tutorials}, vol. 20, no. 4, pp. 3496--3509, 2018.
	
	\bibitem{rak2020guide}
	J. Rak and D. Hutchison, Guide to disaster-resilient communication networks. \emph{Springer Nature}, 2020.
	
	\bibitem{khodaei2020scalable}
	M. Khodaei and P. Papadimitratos, “Scalable \& resilient vehicle-centric certificate revocation list distribution in vehicular communication systems,” \emph{IEEE Transactions on Mobile Computing}, vol. 20, no. 7, pp. 2473--2489, 2020.
	
	\bibitem{althobaiti2020cybersecurity}
	O. S. Althobaiti and M. Dohler, ``Cybersecurity challenges associated with the internet of things in a post-quantum world.'' \emph{IEEE Access}, vol. 8, pp. 157356--157381, 2020.
	
	\bibitem{saki2021survey}
	A. A. Saki, M. Alam, K. Phalak, A. Suresh, R. O. Topaloglu, and S. Ghosh, ``A survey and tutorial on security and resilience of quantum computing.'' \emph{in 2021 IEEE European Test Symposium (ETS)}, IEEE, 2021, pp. 1--10.
	
	
	\bibitem{banerjee2021introducing}
	I. Banerjee, M. Warnier, F. M. Brazier, and D. Helbing, “Introducing participatory fairness in emergency communication can support self-organization for survival,” \emph{Scientific Reports}, vol. 11, no. 1, p. 7209, 2021.
	
	\bibitem{berger2021survey}
	C. Berger, P. Eichhammer, H. P. Reiser, J. Domaschka, F. J. Hauck, and G. Habiger, “A survey on resilience in the IoT: Taxonomy, classification, and discussion of resilience mechanisms,” \emph{ACM Computing Surveys (CSUR)}, vol. 54, no. 7, pp. 1--39, 2021.
	
	\bibitem{banerjee2022designing}
	I. Banerjee, M. Warnier, and F. M. Brazier, “Designing inclusion and continuity for resilient communication during disasters,” \emph{Sustainable and Resilient Infrastructure}, vol. 7, no. 6, pp. 955--970, 2022.
	
	\bibitem{hutchison2023importance}
	D. Hutchison, D. Pezaros, J. Rak, and P. Smith, ''On the importance of resilience engineering for networked systems in a changing world.'' \emph{IEEE Communications Magazine}, 2023.
	
	\bibitem{dietzel2010resilient}
	S. Dietzel, E. Schoch, B. Konings, M. Weber, and F. Kargl, “Resilient secure aggregation for vehicular networks,” \emph{IEEE Network}, vol. 24, no. 1, pp. 26--31, 2010.
	
	\bibitem{stefanovic2017resilient}
	C. Stefanovic, M. Angjelichinoski, P. Danzi, and P. Popovski, “Resilient and secure low-rate connectivity for smart energy applications through power talk in DC microgrids,” \emph{IEEE Communications Magazine}, vol. 55, no. 10, pp. 83--89, 2017.
	
	\bibitem{liu2017resilient}
	J. Liu, H. Guo, and L. Zhao, “Resilient and low-latency information acquisition for FiWi enhanced smart grid,” \emph{IEEE Network}, vol. 31, no. 5, pp. 80--86, 2017.
	
	\bibitem{ye2015energy}
	Y. Ye, F. J. Arribas, J. Elmirghani, et al., “Energy-efficient resilient optical networks: Challenges and trade-offs,” \emph{IEEE Communications Magazine}, vol. 53, no. 2, pp. 144--150, 2015.
	

	
	\bibitem{deepak2019overview}
	G. Deepak, A. Ladas, Y. A. Sambo, H. Pervaiz, C. Politis, and M. A. Imran, “An overview of post-disaster emergency communication systems in the future networks,” \emph{IEEE Wireless Communications}, vol. 26, no. 6, pp. 132--139, 2019.
	
	\bibitem{kaleem2019uav}
	Z. Kaleem, M. Yousaf, A. Qamar, et al., “UAV-empowered disaster-resilient edge architecture for delay-sensitive communication,” \emph{IEEE Network}, vol. 33, no. 6, pp. 124--132, 2019.
	
	\bibitem{naqvi2018drone}
	S. A. R. Naqvi, S. A. Hassan, H. Pervaiz, and Q. Ni, “Drone-aided communication as a key enabler for 5G and resilient public safety networks,” \emph{IEEE Communications Magazine}, vol. 56, no. 1, pp. 36--42, 2018.
	
	\bibitem{hu2021building}
	N. Hu, Z. Tian, Y. Sun, et al., “Building agile and resilient UAV networks based on SDN and blockchain,” \emph{IEEE Network}, vol. 35, no. 1, pp. 57--63, 2021.
	
	\bibitem{miranda2016survey}
	K. Miranda, A. Molinaro, and T. Razafindralambo, “A survey on rapidly deployable solutions for post-disaster networks,” \emph{IEEE Communications Magazine}, vol. 54, no. 4, pp. 117--123, 2016.
	
	\bibitem{zhang2016self}
	H. Zhang, C. Jiang, R. Q. Hu, and Y. Qian, “Self-organization in disaster-resilient heterogeneous small cell networks,” \emph{IEEE Network}, vol. 30, no. 2, pp. 116--121, 2016.
	
	\bibitem{tyson2014beyond}
	G. Tyson, E. Bodanese, J. Bigham, and A. Mauthe, “Beyond content delivery: Can icns help emergency scenarios?” \emph{IEEE Network}, vol. 28, no. 3, pp. 44–49, 2014.
	
	\bibitem{sakano2013disaster}
	T. Sakano, Z. M. Fadlullah, T. Ngo, et al., “Disaster-resilient networking: A new vision based on movable and deployable resource units,” \emph{IEEE Network}, vol. 27, no. 4, pp. 40--46, 2013.
	
	\bibitem{sterbenz2010resilience}
	J. P. Sterbenz, D. Hutchison, E. K. Çetinkaya, et al., “Resilience and survivability in communication networks: Strategies, principles, and survey of disciplines,” \emph{Computer Networks}, vol. 54, no. 8, pp. 1245--1265, 2010.
	
	
	
	\bibitem{khan2017trust}
	Z. A. Khan, J. Ullrich, A. G. Voyiatzis, and P. Herrmann, “A trust-based resilient routing mechanism for the internet of things,” \emph{in Proceedings of the 12th International Conference on Availability, Reliability and Security}, Reggio Calabria, Italy, 2017, pp. 1--6.
	
	\bibitem{modarresi2017multilevel}
	A. Modarresi and J. P. Sterbenz, “Multilevel IoT model for smart cities resilience,” \emph{in Proceedings of the 12th International Conference on Future Internet Technologies}, Fukuoka, Japan, 2017, pp. 1--7.
	
	\bibitem{laprie1985dependable}
	J.-C. Laprie, “Dependable computing and fault-tolerance,” \emph{Digest of Papers FTCS-15}, vol. 10, no. 2, p. 124, 1985.
	
	\bibitem{laprie2008dependability}
	J.-C. Laprie, “From dependability to resilience,” \emph{in 38th IEEE/IFIP Int. Conf. on Dependable Systems and Networks}, Anchorage, Alaska, USA, 2008, G8--G9.
	
	\bibitem{delic2016resilience}
	K. A. Delic, “On resilience of IoT systems: The internet of things (ubiquity symposium),” \emph{Ubiquity}, vol. 2016, no. February, pp. 1--7, 2016.
	
	\bibitem{us2010dhs}
	U. D. of Homeland Security Risk Steering Committee et al., “DHS risk lexicon: 2010 edition,” \emph{Washington, DC (www. dhs. gov/xlibrary/assets/dhs-risk-lexicon-2010. pdf)}, vol. 348, 2010.
	
	\bibitem{thompson2016new}
	M. A. Thompson, M. J. Ryan, J. Slay, and A. C. McLucas, “A new resilience taxonomy,” \emph{in INCOSE International Symposium, Wiley Online Library}, vol. 26, Edinburgh, Scotland, 2016, pp. 1318--1330.
	
	\bibitem{witti2018secure}
	M. Witti and D. Konstantas, “A secure and privacy-preserving internet of things framework for smart city,” \emph{in Proceedings of the 6th International Conference on Information Technology: IoT and Smart City}, Hong Kong, China, 2018, pp. 145--150.
	
	\bibitem{vugrin2011resilience}
	E. D. Vugrin, D. E. Warren, and M. A. Ehlen, “A resilience assessment framework for infrastructure and economic systems: Quantitative and qualitative resilience analysis of petrochemical supply chains to a hurricane,” \emph{Process Safety Progress}, vol. 30, no. 3, pp. 280--290, 2011.
	
	\bibitem{avizienis2001fundamental}
	A. Avizienis, J.-C. Laprie, and B. Randell, “Fundamental concepts of dependability,” \emph{Department of Computing Science Technical Report Series}, 2001.
	
	\bibitem{avizienis2004basic}
	A. Avizienis, J.-C. Laprie, B. Randell, and C. Landwehr, “Basic concepts and taxonomy of dependable and secure computing,” \emph{IEEE Transactions on Dependable and Secure Computing}, vol. 1, no. 1, pp. 11–33, 2004.
	
	\bibitem{brinkmeier09methods}
	M. Brinkmeier, M. Fischer, S. Grau, G. Schäfer, and T. Strufe, “Methods for improving resilience in communication networks and P2P overlays,” \emph{Praxis der Informationsverarbeitung und Kommunikation: PIK: Fachzeitschrift für den Einsatz von Informationssystemen}, vol. 32, no. 1, pp. 64–78, 2009.
	
	\bibitem{bishop2011resilience}
	M. Bishop, M. Carvalho, R. Ford, and L. M. Mayron, “Resilience is more than availability,” \emph{in Proceedings of the 2011 New Security Paradigms Workshop}, Marin County, California, USA, 2011, pp. 95--104.
	
	\bibitem{erdene2012new}
	O. Erdene-Ochir, A. Kountouris, M. Minier, and F. Valois, “A new metric to quantify resiliency in networking,” \emph{IEEE Communications Letters}, vol. 16, no. 10, pp. 1699--1702, 2012.
	
	\bibitem{sterbenz2014redundancy}
	J. P. Sterbenz, D. Hutchison, E. K. Çetinkaya, et al., “Redundancy, diversity, and connectivity to achieve multilevel network resilience, survivability, and disruption tolerance invited paper,” \emph{Telecommunication Systems}, vol. 56, no. 1, pp. 17--31, 2014.
	
	\bibitem{chen09sensor}
	X. Chen, K. Makki, K. Yen, and N. Pissinou, “Sensor network security: A survey,” \emph{IEEE Communications Surveys \& Tutorials}, vol. 11, no. 2, pp. 52--73, 2009.
	
	\bibitem{csrc2024resilience}
	Information system resilience, https://csrc.nist.gov/glossary/term/ information\_system\_resilience, Accessed: 2024-01-15.
	
	\bibitem{EU2020resilience}
	“Strategic foresight report -- Charting the course towards a more resilient Europe,” European Commission, 2020. [Online]. Available: https://commission.europa.eu/system/files/2021-04/strategic\_foresight\_report\_2020\_1\_0.pdf.
	
	\bibitem{landwehr2001computer}
	C. E. Landwehr, “Computer security,” \emph{International journal of information security}, vol. 1, no. 1, pp. 3--13, 2001.
	
	\bibitem{rai2007temperature}
	V. K. Rai, “Temperature sensors and optical sensors,” \emph{Applied Physics B}, vol. 88, pp. 297--303, 2007.
	
	\bibitem{kuzubasoglu2020flexible}
	B. A. Kuzubasoglu and S. K. Bahadir, “Flexible temperature sensors: A review,” \emph{Sensors and Actuators A: Physical}, vol. 315, pp. 112--282, 2020.
	
	\bibitem{catterall2010ion}
	W. A. Catterall, “Ion channel voltage sensors: Structure, function, and pathophysiology,” \emph{Neuron}, vol. 67, no. 6, pp. 915--928, 2010.
	
	\bibitem{rajachandrasekar2012monitoring}
	R. Rajachandrasekar, X. Besseron, and D. K. Panda, “Monitoring and predicting hardware failures in HPC clusters with FTB-IPMI,” \emph{in IEEE 26th International Parallel and Distributed Processing Symposium Workshops \& PhD Forum}, IEEE, Shanghai, China, 2012, pp. 1136--1143.
	
	\bibitem{amouri2011}
	A. Amouri and M. Tahoori, “A low-cost sensor for aging and late transitions detection in modern FPGAs,” \emph{in 21st Int. Conference on Field Programmable Logic and Applications}, Chania, Greece, 2011.
	
	\bibitem{anik2020}
	M. T. H. Anik et al., “On-chip voltage and temperature digital sensor for security, reliability, and portability,” \emph{in Int. Conference on Computer-Aided Design (ICCAD)}, San Diego, CA, USA, 2020.
	
	\bibitem{rahmanikia2017}
	N. Rahmanikia, A. Amiri, H. Noori, and F. Mehdipour, “Performance evaluation metrics for ring-oscillator-based temperature sensors on FPGAs: A quality factor,” \emph{Integration, the VLSI Journal}, vol. 57, pp. 81--100, 2017.
	
	\bibitem{keedy1979structuring}
	J. L. Keedy, “On structuring operating systems with monitors,” \emph{ACM SIGOPS Operating Systems Review}, vol. 13, no. 1, pp. 5--9, 1979.
	
	\bibitem{xu2019indoor}
	Q. Xu, Y. Han, B. Wang, M. Wu, and K. R. Liu, “Indoor events monitoring using channel state information time series,” \emph{IEEE Internet of Things Journal}, vol. 6, no. 3, pp. 4977--4990, 2019.
	
	\bibitem{cai2017proactive}
	H. Cai, Q. Zhang, Q. Li, and J. Qin, “Proactive monitoring via jamming for rate maximization over MIMO Rayleigh fading channels,” \emph{IEEE Communications Letters}, vol. 21, no. 9, pp. 2021--2024, 2017.
	
	\bibitem{hu2017proactive}
	D. Hu, Q. Zhang, P. Yang, and J. Qin, “Proactive monitoring via jamming in amplify-and-forward relay networks,” \emph{IEEE Signal Processing Letters}, vol. 24, no. 11, pp. 1714--1718, 2017.
	
	\bibitem{housh2018model}
	M. Housh and Z. Ohar, “Model-based approach for cyber-physical attack detection in water distribution systems,” \emph{Water research}, vol. 139, pp. 132--143, 2018.
	
	\bibitem{zhang2013time}
	Z. Zhang, S. Gong, A. D. Dimitrovski, and H. Li, “Time synchronization attack in smart grid: Impact and analysis,” \emph{IEEE Transactions on Smart Grid}, vol. 4, no. 1, pp. 87--98, 2013.
	
	\bibitem{uwagbole2017applied}
	S. O. Uwagbole, W. J. Buchanan, and L. Fan, “Applied machine learning predictive analytics to SQL injection attack detection and prevention,” \emph{in IFIP/IEEE Symposium on Integrated Network and Service Management (IM)}, IEEE, Lisbon, Portugal, 2017, pp. 1087--1090.
	
	\bibitem{silva2014prbs}
	A. Silva, E. Pontes, F. Zhou, A. Guelf, and S. Kofuji, “PRBS/EWMA based model for predicting burst attacks (Brute Froce, DoS) in computer networks,” \emph{in Ninth International Conference on Digital Information Management (ICDIM 2014)}, IEEE, Phitsanulok, Thailand, 2014, pp. 194--200.
	
	\bibitem{zhan2015predicting}
	Z. Zhan, M. Xu, and S. Xu, “Predicting cyber attack rates with extreme values,” \emph{IEEE Transactions on Information Forensics and Security}, vol. 10, no. 8, pp. 1666--1677, 2015.
	
	\bibitem{lundberg2006resilience}
	J. Lundberg and B. Johansson, “Resilience, stability and requisite interpretation in accident investigations,” \emph{in Proceedings of the Second Resilience Engineering Symposium}, Juan-les-Pins, France, 2006, pp. 191--198.
	
	\bibitem{amrouch2014}
	H. Amrouch, V. M. van Santen, T. Ebi, V. Wenzel, and J. Henkel, “Towards interdependencies of aging mechanisms,” \emph{in Int. Conference on Computer-Aided Design (ICCAD)}, San Jose, California, USA, 2014. R. C.
	
	\bibitem{baumann2005}
	R. C. Baumann, “Radiation-induced soft errors in advanced semiconductor Technologies,” \emph{IEEE Transactions on Device and Materials Reliability}, vol. 100--101, 2005.
	
	\bibitem{Goerl2019}
	R. Goerl et al., “Combined ionizing radiation electromagnetic interference test procedure to achieve reliable integrated circuits,” \emph{Microelectronics Reliability}, vol. 100--101, 2019.
	
	\bibitem{sethumadhavan15trustworthy}
	S. Sethumadhavan, A. Waksman, M. Suozzo, Y. Huang, and J. Eum, “Trustworthy hardware from untrusted components,” \emph{Communications of the ACM}, vol. 58, no. 9, pp. 60--71, 2015.
	
	\bibitem{moriam18protecting}
	S. Moriam et al., “Protecting communication in many-core systems against active attackers,”\emph{ in Proceedings of the 2018 on Great Lakes Symposium on VLSI}, Chicago, IL, USA, 2018, pp. 45--50.
	
	\bibitem{walther18improving}
	P. Walther, C. Janda, E. Franz, et al., “Improving quantization for channel reciprocity based key generation,” \emph{in 43rd IEEE Conference on Local Computer Networks (LCN)}, Chicago, IL, USA, 2018, pp. 545--552.
	
	\bibitem{walther19blind}
	P. Walther, E. Franz, and T. Strufe, “Blind synchronization of channel impulse responses for channel reciprocity-based key generation,” \emph{in IEEE 44th Conference on Local Computer Networks (LCN)}, Osnabrueck, Germany, 2019, pp. 76--83.
	
	\bibitem{walther21ray}
	P. Walther, M. Richter, and T. Strufe, “Ray-tracing based inference attacks on physical layer security,” \emph{Electronic Communications of the EASST}, vol. 80, 2021.
	
	\bibitem{mpitziopoulos2009survey}
	A. Mpitziopoulos, D. Gavalas, C. Konstantopoulos, and G. Pantziou, “A survey on jamming attacks and countermeasures in WSNs,” \emph{IEEE communications surveys \& tutorials}, vol. 11, no. 4, pp. 42--56, 2009.
	
	\bibitem{pandey2022security}
	G. K. Pandey, D. S. Gurjar, H. H. Nguyen, and S. Yadav, “Security threats and mitigation techniques in UAV communications: A comprehensive survey,” \emph{IEEE Access}, vol. 10, pp. 112 858--112 897, 2022. 
	
	\bibitem{kapetanovic2015physical}
	D. Kapetanovic, G. Zheng, and F. Rusek, “Physical layer security for massive MIMO: An overview on passive eavesdropping and active attacks,” \emph{IEEE Communications Magazine}, vol. 53, no. 6, pp. 21--27, 2015.
	
	\bibitem{faruk2022review}
	M. J. H. Faruk, S. Tahora, M. Tasnim, H. Shahriar, and N. Sakib, ``A review of quantum cybersecurity: threats, risks and opportunities.'' in \emph{2022 1st International Conference on AI in Cybersecurity (ICAIC)}, pp. 1--8, IEEE, 2022.
	
	\bibitem{nejatollahi2019post}
	H. Nejatollahi, N. Dutt, S. Ray, F. Regazzoni, I. Banerjee, and R. Cammarota, ``Post-quantum lattice-based cryptography implementations: A survey.'' \emph{ACM Computing Surveys (CSUR)}, vol. 51, no. 6, pp. 1--41, 2019.
	
	\bibitem{pirayesh2022jamming}
	H. Pirayesh and H. Zeng, “Jamming attacks and anti-jamming strategies in wireless networks: A comprehensive survey,” \emph{IEEE communications surveys \& tutorials}, vol. 24, no. 2, pp. 767--809, 2022.
	
	\bibitem{grover2014jamming}
	K. Grover, A. Lim, and Q. Yang, “Jamming and anti–jamming techniques in wireless networks: A survey,” \emph{International Journal of Ad Hoc and Ubiquitous Computing}, vol. 17, no. 4, pp. 197--215, 2014. 
	
	\bibitem{shakiba2021physical}
	M. Shakiba-Herfeh, A. Chorti, and H. Vincent Poor, “Physical layer security: Authentication, integrity, and confidentiality,” \emph{Physical layer security}, pp. 129--150, 2021.
	
	\bibitem{jeung2011adaptive}
	J. Jeung, S. Jeong, and J. Lim, “Adaptive rapid channel-hopping scheme mitigating smart jammer attacks in secure WLAN,” \emph{in 2011-MILCOM 2011 Military Communications Conference}, IEEE, Baltimore, MD, USA, 2011, pp. 1231--1236.
	
	\bibitem{pinola2012experimental}
	J. Pinola, J. Prokkola, and E. Piri, “An experimental study on jamming
	tolerance of 3G/WCDMA,” \emph{in MILCOM 2012-2012 IEEE Military
		Communications Conf.}, IEEE, Orlando, FL, USA, 2012, pp. 1--7.
	
	\bibitem{davaslioglu2019deepwifi}
	K. Davaslioglu, S. Soltani, T. Erpek, and Y. E. Sagduyu, “DeepWiFi: Cognitive WiFi with deep learning,” \emph{IEEE Transactions on Mobile
		Computing}, vol. 20, no. 2, pp. 429--444, 2019.
	
	\bibitem{yan2014mimo}
	Q. Yan, H. Zeng, T. Jiang, M. Li, W. Lou, and Y. T. Hou, “MIMO-
	based jamming resilient communication in wireless networks,” \emph{in
		IEEE INFOCOM 2014-IEEE Conference on Computer Communications}, IEEE, Toronto, ON, Canada, 2014, pp. 2697--2706.
	
	\bibitem{yan2016jamming}
	Q. Yan, H. Zeng, T. Jiang, M. Li, W. Lou, and Y. T. Hou, “Jamming resilient communication using MIMO interference cancellation,” \emph{IEEE Transactions on Information Forensics and Security}, vol. 11, no. 7,
	pp. 1486--1499, 2016.
	
	\bibitem{rezki2017secret}
	Z. Rezki, M. Zorgui, B. Alomair, and M.-S. Alouini, “Secret key
	agreement: Fundamental limits and practical challenges,” \emph{IEEE Wireless Communications}, vol. 24, no. 3, pp. 72--79, 2017.
	
	\bibitem{wang2015survey}
	T. Wang, Y. Liu, and A. V. Vasilakos, “Survey on channel reciprocity based key establishment techniques for wireless systems,” \emph{Wireless
		Networks}, vol. 21, pp. 1835--1846, 2015.
	
	\bibitem{alleaume2014using}
	R. Alléaume, C. Branciard, J. Bouda, et al., “Using quantum key
	distribution for cryptographic purposes: A survey,” \emph{Theoretical Computer Science}, vol. 560, pp. 62--81, 2014.
	
	\bibitem{huang2021quantum}
	X. Huang, S.-B. Zhang, Y. Chang, C. Qiu, D.-M. Liu, and M. Hou,
	“Quantum key agreement protocol based on quantum search algorithm,” \emph{International Journal of Theoretical Physics}, vol. 60, pp. 838--847, 2021.
	
	\bibitem{amer2021introduction}
	O. Amer, V. Garg, and W. O. Krawec, “An introduction to practical quantum key distribution,” \emph{IEEE Aerospace and Electronic Systems Magazine}, vol. 36, no. 3, pp. 30--55, 2021.
	
	\bibitem{hoang2021physical}
	T. M. Hoang, T. Q. Duong, H. D. Tuan, S. Lambotharan, and L. Hanzo, “Physical layer security: Detection of active eavesdropping attacks by support vector machines,” \emph{IEEE Access}, vol. 9, pp. 31 595--31 607, 2021.
	
	\bibitem{renzo2019smart}
	M. D. Renzo, M. Debbah, D.-T. Phan-Huy, et al., “Smart radio environments empowered by reconfigurable AI meta-surfaces: An idea whose time has come,” \emph{EURASIP Journal on Wireless Communications and Networking}, vol. 2019, no. 1, pp. 1--20, 2019.
	
	\bibitem{wu2019towards}
	Q. Wu and R. Zhang, “Towards smart and reconfigurable environment: Intelligent reflecting surface aided wireless network,” \emph{IEEE Communications Magazine}, vol. 58, no. 1, pp. 106--112, 2019.
	
	\bibitem{yu2021smart}
	X. Yu, V. Jamali, D. Xu, D. W. K. Ng, and R. Schober, “Smart and reconfigurable wireless communications: From IRS modeling to algorithm design,” \emph{IEEE Wireless Communications}, vol. 28, no. 6, pp. 118--125, 2021.
	
	\bibitem{ji2020secret}
	Z. Ji, P. L. Yeoh, D. Zhang, et al., “Secret key generation for intelligent reflecting surface assisted wireless communication networks,” \emph{IEEE Transactions on Vehicular Technology}, vol. 70, no. 1, pp. 1030--1034, 2020.
	
	\bibitem{yu2020robust}
	X. Yu, D. Xu, Y. Sun, D. W. K. Ng, and R. Schober, “Robust and secure wireless communications via intelligent reflecting surfaces,” \emph{IEEE Journal on Selected Areas in Communications}, vol. 38, no. 11, pp. 2637--2652, 2020.
	
	\bibitem{ji2021random}
	Z. Ji, P. L. Yeoh, G. Chen, et al., “Random shifting intelligent reflecting surface for OTP encrypted data transmission,” \emph{IEEE Wireless Communications Letters}, vol. 10, no. 6, pp. 1192--1196, 2021.
	
	\bibitem{abedi24improving}
	K. Abedi, G. T. Nguyen, and T. Strufe, “Improving resilience of future mobile network generations implementing zero trust paradigm,” \emph{in Proceedins of IEEE NOMS}, 2024.
	
	\bibitem{osman21mitigating}
	A. Osman, J. Born, and T. Strufe, “Mitigating internal, stealthy DoS attacks in microservice networks,” \emph{in Stabilization, Safety, and Security of Distributed Systems: 23rd International Symposium, SSS 2021}, Virtual Event, November 17–20, 2021, Proceedings 23, Springer, 2021, pp. 500--504.
	
	\bibitem{li19infas}
	T. Li et al., “INFAS: In-network flow management scheme for SDN control plane protection,” \emph{in IFIP/IEEE Symposium on Integrated Network and Service Management (IM)}, IEEE, Arlington, VA, USA, 2019, pp. 367--373.
	
	\bibitem{schuchard10losing}
	M. Schuchard, A. Mohaisen, D. Foo Kune, N. Hopper, Y. Kim, and E. Y. Vasserman, “Losing control of the internet: Using the data plane to attack the control plane,” \emph{in Proceedings of the 17th ACM Conference on Computer and Communications Security}, Chicago, Illinois, USA, 2010, pp. 726--728.
	
	\bibitem{freitas2022graph}
	S. Freitas, D. Yang, S. Kumar, H. Tong, and D. H. Chau, “Graph vulnerability and robustness: A survey,” \emph{IEEE Transactions on Knowledge and Data Engineering}, vol. 35, no. 6, pp. 5915--5934, 2022.
	
	\bibitem{cohen15small}
	R. Cohen, R. Hess-Green, and G. Nakibly, “Small lies, lots of damage: a partition attack on link-state routing protocols,” \emph{in IEEE Conference on Communications and Network Security (CNS)}, IEEE, Florence, Italy, 2015, pp. 397--405.
	
	\bibitem{cert97bind}
	P. Vixie et al., “CERT advisory CA-1997-22 BIND the Berkeley internet name daemon,” \emph{CERT, Tech. Rep.}, 1997.
	
	\bibitem{alexiou10formal}
	N. Alexiou, S. Basagiannis, P. Katsaros, T. Dashpande, and S. A. Smolka, “Formal analysis of the kaminsky DNS cache-poisoning attack using probabilistic model checking,” \emph{in IEEE 12th International Symposium on High Assurance Systems Engineering}, IEEE, San Jose, CA, USA, 2010, pp. 94--103.
	
	\bibitem{herzberg12security}
	A. Herzberg and H. Shulman, “Security of patched DNS,” \emph{in Computer Security–ESORICS 2012: 17th European Symposium on Research in Computer Security}, Pisa, Italy, September 10-12, 2012. Proceedings 17, Springer, 2012, pp. 271--288.
	
	\bibitem{man20dns}
	K. Man, Z. Qian, Z. Wang, X. Zheng, Y. Huang, and H. Duan, “Dns cache poisoning attack reloaded: Revolutions with side channels,” \emph{in Proceedings of the 2020 ACM SIGSAC Conference on Computer and Communications Security}, Virtual Event USA, 2020, pp. 1337--1350. 
	
	\bibitem{BORBOR201996}
	D. Borbor, L. Wang, S. Jajodia, and A. Singhal, “Optimizing the network diversity to improve the resilience of networks against unknown attacks,” \emph{Computer Communications}, vol. 145, pp. 96--112, 2019.
	
	\bibitem{STERBENZ20101245}
	J. P. Sterbenz, D. Hutchison, E. K. Çetinkaya, et al., “Resilience and survivability in communication networks: Strategies, principles, and survey of disciplines,” \emph{Computer Networks}, vol. 54, no. 8, pp. 1245--1265, 2010, Resilient and Survivable networks.
	
	\bibitem{rescue2022}
	M. Stute, F. Kohnhäuser, L. Baumgärtner, et al., “RESCUE: A resilient and secure device-to-device communication framework for emergencies,” \emph{IEEE Transactions on Dependable and Secure Computing}, vol. 19, no. 3, pp. 1722--1734, 2022.
	
	\bibitem{rose_zero_2020}
	V. Stafford, “Zero trust architecture,” \emph{NIST special publication}, vol. 800, p. 207, 2020.
	
	\bibitem{Organic6G-IEEEAccess:2023}
	M.-I. Corici, F. Eichhorn, R. Bless, et al., “Organic 6G networks: Vision, requirements, and research approaches,” \emph{IEEE Access}, vol. 11, pp. 70 698--70 715, 2023.
	
	\bibitem{BlessZitterbartDespotovic2022_1000148953}
	R. Bless, M. Zitterbart, Z. Despotovic, and A. Hecker, “KIRA: Distributed scalable ID-based routing with fast forwarding,” \emph{in 21st IFIP Networking Conference}, IEEE, Catania, Italy, 2022, pp. 1--9.
	
	\bibitem{bhusal2020power}
	N. Bhusal, M. Abdelmalak, M. Kamruzzaman, and M. Benidris, “Power system resilience: Current practices, challenges, and future directions,” \emph{IEEE Access}, vol. 8, pp. 18 064--18 086, 2020.
	
	\bibitem{blaabjerg2017distributed}
	F. Blaabjerg, Y. Yang, D. Yang, and X. Wang, “Distributed power-generation systems and protection,” \emph{Proceedings of the IEEE}, vol. 105, no. 7, pp. 1311--1331, 2017.
	
	\bibitem{bie2017battling}
	Z. Bie, Y. Lin, G. Li, and F. Li, “Battling the extreme: A study on the power system resilience,” \emph{Proceedings of the IEEE}, vol. 105, no. 7, pp. 1253--1266, 2017.
	
	\bibitem{yu2017survey}
	W. Yu, F. Liang, X. He, et al., ``A survey on the edge computing for the Internet of Things.'' \emph{IEEE Access}, vol. 6, pp. 6900--6919, 2017.
	
	\bibitem{pan2017future}
	J. Pan and J. McElhannon, ``Future edge cloud and edge computing for internet of things applications.'' \emph{IEEE Internet of Things Journal}, vol. 5, no. 1, pp. 439--449, 2017.
	
	\bibitem{javed2022future}
	A. R. Javed, F. Shahzad, S. ur Rehman, et al., ``Future smart cities: Requirements, emerging technologies, applications, challenges, and future aspects,'' \emph{Cities}, vol. 129, p. 103794, 2022.
	
	\bibitem{band2022smart}
	S. S. Band, S. Ardabili, M. Sookhak, et al., ``When smart cities get smarter via machine learning: An in-depth literature review.'' \emph{IEEE Access}, vol. 10, pp. 60985--61015, 2022.
	
	\bibitem{letaief2021edge}
	K. B. Letaief, Y. Shi, J. Lu, and J. Lu, “Edge artificial intelligence for 6G: Vision, enabling technologies, and applications,” \emph{IEEE Journal on Selected Areas in Communications}, vol. 40, no. 1, pp. 5--36, 2021. 
	
	\bibitem{mahmood2022comprehensive}
	M. R. Mahmood, M. A. Matin, P. Sarigiannidis, and S. K. Goudos, “A comprehensive review on artificial intelligence/machine learning algorithms for empowering the future IoT toward 6G era,” \emph{IEEE Access}, vol. 10, pp. 87 535--87 562, 2022.
	
	\bibitem{khan2022digital}
	L. U. Khan, W. Saad, D. Niyato, Z. Han, and C. S. Hong, “Digital-twin-enabled 6G: Vision, architectural trends, and future directions,” \emph{IEEE Communications Magazine}, vol. 60, no. 1, pp. 74--80, 2022.
	
	\bibitem{serodio20236g}
	C. Ser{\^o}dio, J. Cunha, G. Candela, S. Rodriguez, X. R. Sousa, and F. Branco, ``The 6G ecosystem as support for IoE and private networks: Vision, requirements, and challenges.'' \emph{Future Internet}, vol. 15, no. 11, p. 348, 2023.
	
	\bibitem{mohammed2024enabling}
	S. M. Mohammed, A. Al-Barrak, and N. T. Mahmood, ``Enabling Technologies for Ultra-Low Latency and High-Reliability Communication in 6G Networks.'' \emph{Ing{\'e}nierie des Syst{\`e}mes d'Information}, vol. 29, no. 3, 2024.
	
	\bibitem{dressler2022v}
	F. Dressler, C. F. Chiasserini, F. H. Fitzek, et al., “V-Edge: Virtual edge computing as an enabler for novel microservices and cooperative computing,” \emph{IEEE Network}, vol. 36, no. 3, pp. 24--31, 2022.
	
	\bibitem{aguiar2023inter}
	A. Aguiar, O. Altintas, F. Dressler, G. Karlsson, and F. Klingler, “Inter-vehicular communication-from edge support to vulnerable road users II (Dagstuhl seminar 22512),” \emph{in Dagstuhl Reports, Schloss Dagstuhl-Leibniz-Zentrum für Informatik}, vol. 12, 2023.
	
	\bibitem{pannu2023data}
	G. S. Pannu, S. Ucar, T. Higuchi, O. Altintas, and F. Dressler, “Data sharing in virtual edge computing using coded caching,” \emph{in IEEE Vehicular Networking Conference (VNC)}, IEEE, Istanbul, Turkiye, 2023, pp. 104--111.
	
	\bibitem{yonis2023performance}
	A. Z. Yonis, ``Performance of ultra reliable low latency communication (URLLC) in 5G wireless networks.'' \emph{system (BMS)}, vol. 10, p. 13, 2023.
	
	\bibitem{7751164}
	J. Rak, M. Jonsson, and A. Vinel, “A taxonomy of challenges to resilient message dissemination in VANETs,” \emph{in 17th International Telecommunications Network Strategy and Planning Symposium (Networks)}, Montreal, QC, Canada, 2016, pp. 127--132.
	
	\bibitem{5473884}
	K. Pelechrinis, M. Iliofotou, and S. V. Krishnamurthy, “Denial of service attacks in wireless networks: The case of jammers,” \emph{IEEE Communications Surveys \& Tutorials}, vol. 13, no. 2, pp. 245--257, 2011.

	\bibitem{boccella2020evaluating}
	A. R. Boccella, P. Centobelli, R. Cerchione, T. Murino, and R. Riedel, ``Evaluating centralized and heterarchical control of smart manufacturing systems in the era of Industry 4.0.'' \emph{Applied Sciences}, vol. 10, no. 3, p. 755, 2020.
	
	\bibitem{fortoul2023smart}
	J. A. Fortoul-Diaz, L. A. Carrillo-Martinez, A. Centeno-Tellez, F. Cortes-Santacruz, I. Olmos-Pineda, and R. R. Flores-Quintero, ``A smart factory architecture based on industry 4.0 technologies: open-source software implementation.'' \emph{IEEE Access}, 2023.
	
	\bibitem{peng2022resource}
	Q. Peng, H. Ren, C. Pan, N. Liu, and M. Elkashlan, ``Resource allocation for uplink cell-free massive MIMO enabled URLLC in a smart factory.'' \emph{IEEE Transactions on Communications}, vol. 71, no. 1, pp. 553--568, 2022.
	
	\bibitem{fu2022systematic}
	Y. Fu, Y. Hu, and V. Sundstedt, ``A systematic literature review of virtual, augmented, and mixed reality game applications in healthcare.'' \emph{ACM Transactions on Computing for Healthcare (HEALTH)}, vol. 3, no. 2, pp. 1--27, 2022.
	
	\bibitem{akyildiz2022wireless}
	I. F. Akyildiz and H. Guo, ``Wireless communication research challenges for extended reality (XR).'' \emph{ITU Journal on Future and Evolving Technologies}, vol. 3, no. 1, pp. 1--15, 2022.
	
	\bibitem{hu2020cellular}
	F. Hu, Y. Deng, W. Saad, M. Bennis, and A. H. Aghvami, ``Cellular-connected wireless virtual reality: Requirements, challenges, and solutions.'' \emph{IEEE Communications Magazine}, vol. 58, no. 5, pp. 105--111, 2020.
	
	\bibitem{chattopadhyay2020autonomous}
	A. Chattopadhyay, K.-Y. Lam, and Y. Tavva, “Autonomous vehicle: Security by design,” \emph{IEEE Transactions on Intelligent Transportation Systems}, vol. 22, no. 11, pp. 7015--7029, 2020.
	
	\bibitem{casola2016security}
	V. Casola, A. De Benedictis, M. Rak, and E. Rios, “Security-by-design in clouds: A security-SLA driven methodology to build secure cloud applications,” \emph{Procedia Computer Science}, vol. 97, pp. 53--62, 2016.
	
	\bibitem{simoiu2020secure}
	C. Simoiu, ``Secure by default: A behavioral approach to cyber security''. \emph{Stanford University}, 2020.
	
	
	
	
	
	
	
	
	
	
	

	
	
\end{thebibliography}
\end{document}